\documentclass[aps,prb,twocolumn,superscriptaddress,longbibliography,floatfix]{revtex4-2}
\usepackage{amsmath,amssymb,graphicx,bm,xcolor,dsfont,booktabs,multirow,array,longtable,float,placeins}
\usepackage[colorlinks=true,linkcolor=blue,citecolor=blue,urlcolor=blue]{hyperref}
\newcommand{\Tr}{\mathrm{Tr}}
\newcommand{\ii}{\mathrm{i}}
\newcommand{\one}{\mathds{1}}
\newcommand{\Z}{\mathbb{Z}}
\newcommand{\T}{\Theta}
\newcommand{\IGG}{\mathrm{IGG}}
\newcommand{\SMtext}[2]{#1}
\begin{document}

\title{Melting noncoplanar order on the square-kagome lattice: Chiral, flat-band, and staggered spin liquids from projective symmetry}

\author{Harsh Raj}
\affiliation{Department of Physics, Indian Institute of Technology Madras, Chennai 600036, India}
\author{Atanu Maity}
\affiliation{Institut f\"ur Theoretische Physik und Astrophysik and W\"urzburg-Dresden Cluster of Excellence ctd.qmat, Julius-Maximilians-Universit\"at W\"urzburg, Am Hubland, Campus S\"ud, W\"urzburg 97074, Germany}
\affiliation{Department of Physics, Indian Institute of Technology Madras, Chennai 600036, India}
\author{Yasir Iqbal}
\email{yiqbal@physics.iitm.ac.in}
\affiliation{Department of Physics, Indian Institute of Technology Madras, Chennai 600036, India}

\begin{abstract}
Noncoplanar magnetic orders carry a scalar spin chirality, and their melting by quantum fluctuations has been proposed as a route to chiral spin liquids. We ask what this route can produce on the square-kagome lattice, where couplings across the octagons stabilize noncoplanar cuboc and conical orders. We classify the fermionic spin liquids of this lattice by their projective symmetry groups, identify the symmetry class into which each noncoplanar order would melt, and compare candidate states with Gutzwiller-projected wave functions. A magnetic selection rule constrains the route: the chirality of every cuboc and conical order alternates in space, so a liquid that keeps the symmetry of its parent order has zero Chern number and no spontaneous thermal Hall response. A Chern state can descend only from uniformly chiral orders, and the regular ones we find need ferromagnetic bow-tie couplings. The energetics do not favor a chiral liquid either: among the projected wave functions we compare, a cuboc1-ordered state is lowest for strong cross-octagon couplings, and nonchiral states, a valence-bond crystal or a time-reversal-symmetric liquid, for weaker ones. The two inequivalent site orbits shape the spinons: a relative time-reversal sign between them pins flat bands to the Fermi level, and because the translations of $p4mm$ are not commutators, staggered chiral liquids exist, which are absent on sixfold-symmetric lattices.
\end{abstract}

\maketitle
\tableofcontents

\section{Introduction}

Noncoplanar magnetic orders carry a scalar spin chirality, and when quantum fluctuations melt such an order the chirality may survive in a chiral spin liquid, a state without magnetic order that breaks time reversal and can carry chiral edge modes and a thermal Hall response~\cite{KalmeyerLaughlin1987,WenWilczekZee1989}. Quantum melting of noncoplanar orders has been studied as a general route to such liquids~\cite{HickeyCincioPapicParamekanti2017}. We call the route in which the liquid keeps the magnetic symmetry of its parent order the inherited route, and the parent the melted order. On the kagome lattice, further-neighbor couplings stabilize cuboctahedral (cuboc) orders~\cite{DomengeSindzingreLhuillierPierre2005,MessioLhuillierMisguich2011}, and a chiral liquid was proposed as their melted form~\cite{MessioBernuLhuillier2012}. Bosonic and fermionic mean-field and variational studies~\cite{MessioLhuillierMisguich2013,BieriMessioBernuLhuillier2015,BieriLhuillierMessio2016,HuZhuZhangGongBeccaSheng2015} as well as density-matrix renormalization group and exact-diagonalization studies~\cite{GongZhuSheng2014,HeShengChen2014,GongZhuBalentsSheng2015,WietekSterdyniakLauchli2015} find chiral spin liquids near these orders, in the regions where classical ordered phases meet~\cite{WietekSterdyniakLauchli2015}. On the square-kagome lattice, couplings across the octagons stabilize cuboc and conical noncoplanar orders, and Gemb\'e \textit{et al.}\ proposed that their quantum melting could give chiral spin liquids~\cite{Gembe2023}. Whether such a liquid keeps the symmetry of its melted order decides how it can be observed: a liquid whose chirality alternates in space has no spontaneous thermal Hall response, whereas a uniformly chiral liquid can have one. In this work, we address this question with the projective symmetry groups of fermionic spin liquids on the square-kagome lattice and with Gutzwiller-projected wave functions.

The square-kagome lattice, also called the squagome or shuriken lattice, is a network of corner-sharing triangles with coordination four~\cite{SiddharthanGeorges2001,TomczakRichter2003,NakanoSakai2013}. It is the medial (line-graph) lattice of the square-octagon tiling~\cite{NakanoSakai2013}, in the same way as the kagome lattice is the medial lattice of the honeycomb lattice, and its unit cell contains four triangles, one square, and one octagon. Its spin-$1/2$ Heisenberg antiferromagnet has a nonmagnetic ground state with a spin gap and many low-lying singlets~\cite{SiddharthanGeorges2001,Richter2009CMP,RichterDerzhkoSchnack2022}. Exact diagonalization of clusters up to $36$ sites gives a ground-state energy of about $-0.440J$ per site and magnetization plateaus at $1/3$ and $2/3$ of saturation~\cite{NakanoSakai2013}. Numerical studies agree that it is a valence-bond crystal but differ on which one. Quantum dimer and resonating-valence-bond descriptions favor valence-bond crystals with resonating building blocks~\cite{Rousochatzakis2013,RalkoRousochatzakis2015}, and variational Monte Carlo finds a pinwheel valence-bond crystal as an instability of a $U(1)$ Dirac spin liquid~\cite{Astrakhantsev2021}. Tensor-network calculations find a loop-six valence-bond crystal at the isotropic point and a sequence of valence-bond crystals and Lieb ferrimagnets as the ratio of the two nearest-neighbor couplings varies~\cite{Schmoll2023,JahromiIqbal2026}. Schwinger-boson mean-field theory of the model with inequivalent bow-tie edges finds, besides ordered phases, a topological spin liquid with weak nematicity~\cite{LuganJaubertRalko2019}.

In a magnetic field the lattice shows magnetization plateaus and jumps~\cite{MoritaTohyama2018,HasegawaNakanoSakai2018} and, near saturation, localized magnons on its flat magnon band~\cite{RichterDerzhkoSchulenburg2004,Derzhko2013}, and its classical Ising and Heisenberg models show reentrant disorder~\cite{PohleBentonJaubert2016} and cuboc ground states with unusual magnetization curves~\cite{SchmidtRichter2024,RichterSchmidtSchnack2025}. The lattice is realized, in decorated and distorted form, in copper minerals and their synthetic analogs. KCu$_6$AlBiO$_4$(SO$_4$)$_5$Cl shows no order down to $58$~mK and a gapless continuum~\cite{Fujihala2020}, although nuclear magnetic resonance suggests a magnetic ground state with a small gap~\cite{Goto2024}. Na$_6$Cu$_7$BiO$_4$(PO$_4$)$_4$[Cl,(OH)]$_3$ shows no order down to $50$~mK~\cite{Yakubovich2021,Liu2022} and lies close to a magnetic instability in \textit{ab initio} models~\cite{Niggemann2023,Taran2026}, whereas KCu$_7$TeO$_4$(SO$_4$)$_5$Cl orders at $T_N=3.2$~K~\cite{Murtazoev2023,Markina2024,Jing2026}.

In the parton description of a frustrated Heisenberg magnet the spin is represented by fermionic spinons coupled to an emergent gauge field~\cite{BaskaranZouAnderson1987,BaskaranAnderson1988,AffleckZouHsuAnderson1988,Wen2002,Wen2004}. Symmetric spin liquids are then distinguished by their projective symmetry group (PSG), the projective representation under which the spinons transform~\cite{Wen2002}, and chiral spin liquids, in which time reversal and some lattice symmetries are broken while their products survive, are classified in the same framework~\cite{MessioLhuillierMisguich2013,BieriLhuillierMessio2016}. PSG classifications of fermionic spin liquids exist for many two-dimensional~\cite{Wen2002,ReutherLeeAlicea2014,ZhengMeiQi2015,Lu2016,BieriLhuillierMessio2016,LuRan2011,LuRanLee2011,BieriMessioBernuLhuillier2015,ChoyKim2009,YangParamekantiKim2010,MaitySamajdarIqbal2024,MaityIqbalSamajdar2026,SahooFlint2020,Maity2023,Sonnenschein2024,Chatterjee2026,Maity2025SS} and three-dimensional~\cite{Sonnenschein2020,Chauhan2023,LiuHalaszBalents2019,LiuHalaszBalents2021,SchneiderHalimehPunk2022,ChernKimCastelnovo2022,LiuBalentsIqbal2026,HuangKimLu2017,JinZhou2020,HuangChoiKimLu2018,ChernKim2021,LiBiswasParameswaran2025} lattices, and the same framework classifies bosonic spinons in Schwinger-boson theory~\cite{WangVishwanath2006}. For the square-kagome lattice, the fermionic mean-field study of Astrakhantsev \textit{et al.}\ classified $U(1)$ states by their nearest-neighbor fluxes, among them chiral states with flux $\pi/2$ per triangle, and listed a PSG classification as an open problem~\cite{Astrakhantsev2021}.

We work in Abrikosov-fermion mean-field theory with spin-singlet and spin-triplet amplitudes on all twelve bond orbits up to the couplings $J_\times$ and $J_+$ across the octagons~\cite{Gembe2023}, where the triplet amplitudes describe the effect of spin--orbit coupling~\cite{DoddsBhattacharjeeKim2013,LiuHalaszBalents2021}. The mean-field amplitudes are first treated as free parameters constrained only by the PSG, so that spectral types refer to regions of amplitude space. In a second step we solve the self-consistency equations of the Heisenberg model of Ref.~\cite{Gembe2023} within each PSG manifold at representative points of its noncoplanar phases, and we compare Gutzwiller-projected wave functions.

Our main results are as follows. Because the translations of $p4mm$ are not commutators, a chiral state may combine a translation with time reversal, and staggered chiral classes appear, which are absent for $p6mm$~\cite{MessioLhuillierMisguich2013}. Only one of the eight classes, with unitary rotations and mirrors combined with time reversal, admits a Chern number (Secs.~\ref{sec:chiral-psg} and \ref{sec:chern}). The scalar chirality of every cuboc and conical order alternates under a translation or the fourfold rotation, so a liquid that keeps the magnetic group of such an order has no Chern number and no spontaneous thermal Hall response (Sec.~\ref{sec:order-classes}). This selection rule needs only the magnetic group, but it assumes that the liquid inherits that group, which for cuboc1 is the nematic group $p2mm$, whose PSGs we classify separately (Sec.~\ref{sec:nematic}). Uniformly chiral orders, from which a Chern state could be inherited, exist among the regular orders, but those we find need ferromagnetic bow-tie couplings and sizable further-neighbor couplings (Sec.~\ref{sec:regular}), and the regular orders of two phases of Ref.~\cite{Gembe2023} give way to long-wavelength modulations when $J_\times\neq0$ (Sec.~\ref{sec:classical}). Unprojected self-consistent mean-field theory selects the uniform class-$\mathcal{B}$ chiral state, a gapped Chern state (Secs.~\ref{sec:kl} and \ref{sec:mf}), but projection reverses this ranking, as at the nearest-neighbor point~\cite{Astrakhantsev2021}. Among the projected wave functions we compare (Sec.~\ref{sec:vmc}), the cuboc1-ordered state is lowest for strong cross-octagon couplings, a pinwheel valence-bond crystal for weaker ones, and nonchiral states close to decoupled squares and chains near the boundary of the cuboc1 phase. No chiral state, including the nematic chiral state that would inherit the symmetry of cuboc1, is lowest at any coupling considered. This differs from the kagome lattice~\cite{BieriMessioBernuLhuillier2015,BieriLhuillierMessio2016,HuZhuZhangGongBeccaSheng2015,GongZhuBalentsSheng2015}. Beyond the chiral question, the two site orbits, both with nontrivial site symmetry, reduce the cohomology classes to a much smaller set of algebraic PSGs (Sec.~\ref{sec:algebraic}). A relative time-reversal sign $\zeta$ between the orbits decides realizability and, for $\zeta=-$, pins sublattice-imbalance flat bands~\cite{Sutherland1986,Lieb1989} to the Fermi level for any amplitudes (Secs.~\ref{sec:alg-zeta} and \ref{sec:flatband}). Because the nearest-neighbor graph contains triangles, nearest-neighbor ans\"atze can already break time reversal (Sec.~\ref{sec:chiral}).

Spin--orbit coupling allows spin-dependent spinon hopping whose Dzyaloshinskii--Moriya (DM)-like vectors follow Moriya's rules relative to the singlet hopping in every PSG, and projective signs invert them only on bonds that carry no singlet hopping (Sec.~\ref{sec:moriya}). It turns $U(1)$ states into mean-field spinon quantum spin Hall insulators (Sec.~\ref{sec:qsh}). With conserved $S^z$ a charge-conjugation theorem confines them to the classes in which no symmetry conjugates the spinon charge, and with generic spin--orbit coupling they also occur in time-reversal-symmetric (TRS) PSGs whose translations conjugate the spinon charge and, rarely, in other particle--hole classes. These states are not stable spin liquids: with conserved $S^z$ the physical state orders, and with generic spin--orbit coupling the gauge field confines. This fermionic analysis complements a Schwinger-boson study of the effect of \textit{ab initio} DM interactions on decorated square-kagome magnets~\cite{Taran2026}.

The paper is organized as follows. Section~\ref{sec:lattice} introduces the lattice and its symmetries, and Sec.~\ref{sec:parton} the parton construction and the working definitions. Section~\ref{sec:algebraic} classifies the projective symmetry groups and derives the flat bands. Sections~\ref{sec:ansatz} and \ref{sec:chiral} construct the symmetric and chiral spin liquids, and Sec.~\ref{sec:spectra} discusses their spectra. Section~\ref{sec:soc} adds spin--orbit coupling. Section~\ref{sec:orders} relates the spin liquids to the noncoplanar orders through classical, self-consistent mean-field, and projected variational calculations, and Sec.~\ref{sec:lsm} discusses the stability of the mean-field states and a Lieb--Schultz--Mattis-type constraint. Section~\ref{sec:conclusion} summarizes the results and their limitations. Conventions and derivations are given in the appendices, and method and survey details and all tabulated data in the Supplemental Material (SM)~\cite{RajMaityIqbalZenodo}.
\section{Lattice, Symmetries, and Bonds}
\label{sec:lattice}

\begin{figure*}
\centering
\includegraphics[width=\textwidth]{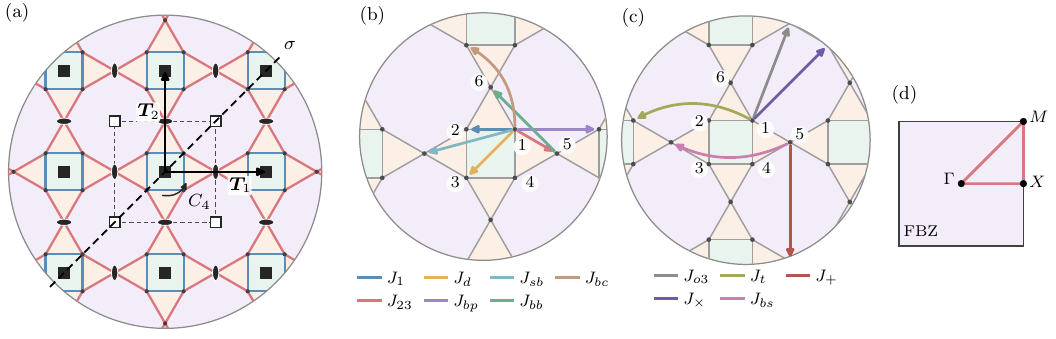}
\caption{(a) Square-kagome lattice with the elements of $p4mm$: the primitive translations $\bm{T}_1$ and $\bm{T}_2$, the unit cell centered on a square (dashed), the fourfold rotation $C_4$ about the square center, and the diagonal mirror $\sigma:(X,Y)\to(Y,X)$ (dashed line). Filled and open squares mark the fourfold rotation centers at the square centers (Wyckoff position $1a$) and octagon centers ($1b$), and lenses mark the twofold rotation centers at the bow-tie sites ($2c$, site symmetry $2mm$). The square sites ($4d$, site symmetry $..m$) lie on the diagonal mirror lines. Nearest-neighbor bonds are colored by orbit: square edges ($J_1$, blue) and bow-tie edges ($J_{23}$, rose). (b), (c) Reference cell with the sublattice labels $\mu=1,\dots,6$ and the reference bonds $i\to j$ of the twelve bond orbits up to $J_+$: (b) the nearest-neighbor orbits and the five orbits at graph distance two, (c) the five longer orbits, among them the couplings $J_\times$ and $J_+$ of Ref.~\cite{Gembe2023} across the octagons. Curved arrows mark bonds that pass through a site ($J_{bc}$, $J_t$) or overlap another reference bond ($J_{bs}$). (d) First Brillouin zone (FBZ) with the path $\Gamma\to X\to M\to\Gamma$ used for band structures, with $X=(1/2,0)$ and $M=(1/2,1/2)$ in units of $2\pi/a$. The bond orbits are listed in \SMtext{Table~\ref{sm:tab:bonds} of the SM}{the SM}.}
\label{fig:lattice}
\end{figure*}

The square-kagome lattice consists of squares joined at their corners by pairs of triangles, the bow-ties, so that every site is shared by two triangles. Each unit cell contains six sites, twelve nearest-neighbor bonds, four triangles, one square, and one octagon, and every site has four neighbors. Four sites of a cell belong to its square (square sites) and two join neighboring squares (bow-tie sites). We use the ideal geometry, in which all nearest-neighbor bonds have unit length, the triangles are equilateral, and the squares are parallel to the Bravais vectors, the geometry of Fig.~1 of Ref.~\cite{Gembe2023}. With the origin at a square center, the Bravais lattice is spanned by
\begin{equation}
\bm{T}_1=(a,0),\qquad \bm{T}_2=(0,a),\qquad a=1+\sqrt3,
\label{eq:T}
\end{equation}
and a site is labeled by $i=(x,y,\mu)$, with $\bm{r}_i=x\bm{T}_1+y\bm{T}_2+\bm{d}_\mu$ and
\begin{align}
\bm{d}_1&=\left(\tfrac12,\tfrac12\right), & \bm{d}_2&=\left(-\tfrac12,\tfrac12\right), & \bm{d}_3&=\left(-\tfrac12,-\tfrac12\right),\nonumber\\
\bm{d}_4&=\left(\tfrac12,-\tfrac12\right), & \bm{d}_5&=\left(\tfrac{a}{2},0\right), & \bm{d}_6&=\left(0,\tfrac{a}{2}\right).
\end{align}
The square sites $\mu=1,\dots,4$ are numbered counterclockwise, and $\mu=5$, $6$ are the bow-tie sites between the square at the origin and its neighbors along $\bm{T}_1$ and $\bm{T}_2$ (Fig.~\ref{fig:lattice}). Every triangle has one square edge and two bow-tie edges, so that the nearest-neighbor graph is not bipartite.

The full isometry group of the ideal lattice is the symmorphic wallpaper group $p4mm$ (see also Ref.~\cite{Issing2026}), generated by the translations $T_1$, $T_2$, the rotation $C_4$ by $\pi/2$ about the square center at the origin, and the diagonal mirror $\sigma:(X,Y)\to(Y,X)$. The bow-tie sites occupy the Wyckoff position $2c$ with site symmetry $2mm$, and the square sites the position $4d$, $(\pm x,\pm x)$ with $x=(\sqrt3-1)/4$ in units of $a$, with site symmetry $..m$. The generators act on the site labels as
\begin{align}
T_1&:(x,y,\mu)\mapsto(x+1,y,\mu),\nonumber\\
T_2&:(x,y,\mu)\mapsto(x,y+1,\mu),\nonumber\\
C_4&:(x,y,\mu)\mapsto(-y,x,[\mu+1])\quad(\mu\le4),\nonumber\\
&\phantom{:}\ \ (x,y,5)\mapsto(-y,x,6),\quad (x,y,6)\mapsto(-y-1,x,5),\nonumber\\
\sigma&:(x,y,\mu)\mapsto(y,x,\mu_\sigma),
\label{eq:action}
\end{align}
where $[k]=((k-1)\bmod4)+1$ and the mirror fixes $\mu=1$, $3$ and exchanges $2\leftrightarrow4$ and $5\leftrightarrow6$. Time reversal $\T$ acts trivially on the sites. The group $p4mm\times\Z_2^\T$ is defined by the thirteen relations
\begin{align}
&T_2T_1T_2^{-1}T_1^{-1}=C_4T_1C_4^{-1}T_2^{-1}=C_4T_2C_4^{-1}T_1=e,\nonumber\\
&\sigma T_1\sigma^{-1}T_2^{-1}=\sigma T_2\sigma^{-1}T_1^{-1}=e,\nonumber\\
&C_4^4=\sigma^2=(\sigma C_4)^2=\T^2=e,\nonumber\\
&O\,\T\,O^{-1}\T^{-1}=e\quad (O=T_1,T_2,C_4,\sigma),
\label{eq:relations}
\end{align}
with words read from right to left. Every site has a nontrivial stabilizer: the bow-tie site $(0,0,5)$ is fixed by the twofold rotation $T_1C_4^2$ about it, by the mirror $\sigma C_4:(X,Y)\to(X,-Y)$, and by $\T$, which generate $\Z_2^3$, and the square site $(0,0,1)$ is fixed by $\sigma$ and $\T$, which generate $\Z_2^2$. These stabilizers, through the on-site representations they must carry, shape the whole classification (Sec.~\ref{sec:algebraic}).

We include the twelve bond orbits up to the length $1+\sqrt3$ of the couplings $J_\times$ and $J_+$ across the octagons~\cite{Gembe2023}, drawn in Fig.~\ref{fig:lattice}(b) and (c). Their lengths, reference bonds, and stabilizers are listed in \SMtext{Sec.~\ref{sm:sec:bonds} of the SM}{the SM}. The nearest-neighbor orbits are the square edges $J_1$ and the bow-tie edges $J_{23}$, whose eight bonds per cell form a single orbit of $p4mm$ and comprise the couplings $J_2$ and $J_3$ of Ref.~\cite{Gembe2023}. At graph distance two lie the square diagonals $J_d$ and the orbits $J_{bp}$, $J_{sb}$, $J_{bb}$, and $J_{bc}$, and we call the range that includes them GD2. At graph distance three lie $J_{o3}$, $J_t$, and $J_{bs}$, and at graph distance four $J_\times$ and $J_+$. The bonds $J_{23}$, $J_{sb}$, and $J_{o3}$ join a square site to a bow-tie site, and all other orbits join two sites of the same orbit. This distinction controls the time-reversal selection rules of Sec.~\ref{sec:alg-zeta}. The mirror $\sigma$ fixes the reference bonds of $J_d$ and $J_\times$ with both endpoints, and axial mirrors fix those of $J_{bs}$ and $J_+$. These elements constrain the reference amplitudes in addition to the bond-reversing elements. Inequivalent bow-tie edges, as in the $J_1$-$J_2$-$J_3$ model of Refs.~\cite{Fujihala2020,Gembe2023}, lower the symmetry to $p4$, which we do not treat here.
\section{Parton Mean-Field Theory and Projective Symmetry Groups}
\label{sec:parton}

\subsection{Abrikosov Fermions and Mean-Field Ans\"atze}

We use the standard fermionic parton construction~\cite{Abrikosov1965,BaskaranZouAnderson1987,AffleckZouHsuAnderson1988,Wen2002} in the conventions of Ref.~\cite{LiuHalaszBalents2021}, summarized in Appendix~\ref{app:bilinears}. The spin is written as $\bm{S}_i=\frac12 f^\dagger_{i\alpha}\bm{\sigma}_{\alpha\beta}f_{i\beta}$ with the constraint $f^\dagger_{i\alpha}f_{i\alpha}=1$, and the fermions of site $i$ form the matrix
\begin{equation}
\psi_i=\begin{pmatrix} f_{i\uparrow} & f^\dagger_{i\downarrow}\\ f_{i\downarrow} & -f^\dagger_{i\uparrow}\end{pmatrix},
\label{eq:psi}
\end{equation}
on which spin rotations act from the left and $SU(2)$ gauge transformations from the right, $\psi_i\to\psi_iW_i^\dagger$. Pauli matrices $\sigma^a$ act in spin space and $\tau^a$ in gauge space. The mean-field Hamiltonian is
\begin{equation}
H_{\rm MF}=\sum_{i,j}\sum_{\alpha=0,x,y,z}\Tr\left[\sigma^\alpha\psi_i u^{(\alpha)}_{ij}\psi_j^\dagger\right]+\sum_i\Tr\left[\psi_i a_i\psi_i^\dagger\right],
\label{eq:HMF}
\end{equation}
summed over ordered pairs, with the singlet link $u^{(0)}_{ij}$ and the triplet links $u^{(a)}_{ij}$,
\begin{align}
u^{(0)}_{ij}&=\ii\chi^0_{ij}\tau^0+\chi^3_{ij}\tau^3+\Delta^1_{ij}\tau^1+\Delta^2_{ij}\tau^2,\nonumber\\
u^{(a)}_{ij}&=t^{a0}_{ij}\tau^0+\ii\left(t^{a1}_{ij}\tau^1+t^{a2}_{ij}\tau^2+t^{a3}_{ij}\tau^3\right),
\label{eq:links}
\end{align}
real coefficients, and $u^{(\alpha)}_{ji}=u^{(\alpha)\dagger}_{ij}$. Here $\chi^0$ and $\chi^3$ give the spin-independent hopping, $\Delta^{1,2}$ the singlet pairing, and the triplet coefficients spin-dependent hopping and pairing, while the multipliers $a_i=a^k_i\tau^k$ impose the constraint on average. Triplet links are allowed when spin--orbit coupling locks spin rotations to lattice operations, and describe at mean-field level the effect of Dzyaloshinskii--Moriya and other anisotropic exchange~\cite{Dzyaloshinsky1958,Moriya1960,DoddsBhattacharjeeKim2013,LiuHalaszBalents2021}. A gauge transformation acts as
\begin{equation}
u^{(\alpha)}_{ij}\to W_i\, u^{(\alpha)}_{ij}W_j^\dagger,\qquad a_i\to W_i\, a_iW_i^\dagger,
\end{equation}
and gauge-related ans\"atze describe the same projected state.

\subsection{Projective Symmetry and Invariant Gauge Group}

A lattice operation $O$ maps an ansatz $u_{ij}$ onto $u_{O(i)O(j)}$. Since gauge-related ans\"atze describe the same spin state, the ansatz is symmetric under $O$ if the two agree up to a gauge transformation $G_O(i)$~\cite{Wen2002},
\begin{align}
u^{(0)}_{O(i)O(j)}&=G_O(O(i))\,u^{(0)}_{ij}\,G_O^\dagger(O(j)),\nonumber\\
u^{(b)}_{O(i)O(j)}&=\sum_a\left[R_O\right]_{ba}G_O(O(i))\,u^{(a)}_{ij}\,G_O^\dagger(O(j)),
\label{eq:psg-spatial}
\end{align}
with $a_{O(i)}=G_O(O(i))\,a_i\,G_O^\dagger(O(i))$. Here $R_O\in SO(3)$ is the spin rotation that accompanies $O$ in the presence of spin--orbit coupling (Appendix~\ref{app:rotations}), and without it only the singlet links occur. Time reversal reverses every link and multiplier~\cite{Wen2002},
\begin{equation}
u^{(\alpha)}_{ij}=-G_\T(i)\,u^{(\alpha)}_{ij}\,G_\T^\dagger(j),\qquad a_i=-G_\T(i)\,a_i\,G_\T^\dagger(i).
\label{eq:psg-T}
\end{equation}
The combined operations $G_OO$ form the projective symmetry group (PSG) of the ansatz. Ans\"atze with gauge-inequivalent PSGs describe distinct spin liquids, so the PSG plays for a spin liquid the role that the space group plays for a crystal.

The PSG elements that belong to the identity operation, $u_{ij}=G(i)\,u_{ij}\,G^\dagger(j)$, form a special subgroup, the invariant gauge group (IGG). It always contains $\{\pm\tau^0\}\cong\Z_2$ and fixes the low-energy gauge theory~\cite{Wen2002}. In the canonical gauge the IGG is global. An ansatz with only imaginary hopping, $u^{(0)}_{ij}=\ii\chi^0_{ij}\tau^0$ and $a_i=0$, has the IGG $SU(2)$, real hopping $\chi^3_{ij}\tau^3$ lowers it to $U(1)=\{e^{\ii\theta\tau^3}\}$, and pairing lowers it to $\Z_2$.

PSG elements compose as
\begin{equation}
G_{O_1O_2}(i)=G_{O_1}(i)\,G_{O_2}\big(O_1^{-1}(i)\big),
\label{eq:compose}
\end{equation}
so that every relation~\eqref{eq:relations}, which equals the identity for the spins, becomes an element of the IGG for the spinons. The algebraic PSGs are the solutions of these conditions, up to gauge transformations $G_O(i)\to W_iG_O(i)W^\dagger_{O^{-1}(i)}$ that preserve the IGG and up to re-liftings, which multiply a generator by an element of the IGG. Because $R_O\in SO(3)$, the classification is the same with and without spin--orbit coupling. Formulations with $SU(2)$ spin rotations, as in Ref.~\cite{ReutherLeeAlicea2014}, carry in addition the double-cover signs, for instance $U_{C_4}^4=-1$. Since $-1$ in spin space acts on the spinons as $-\tau^0\in\IGG$, their labels differ from ours by these fixed signs and describe the same PSGs.

On the square-kagome lattice one further condition enters. An element $h$ of the stabilizer of a site $i$ maps the site onto itself, so $G_h(i)$ is an on-site $SU(2)$ matrix, and these matrices must form a representation of the stabilizer that is compatible with the relation values. A PSG is therefore a set of relation values together with an on-site representation for each site orbit. On a lattice whose sites lie at the general position this condition is always met. Here it removes most cohomology classes and splits others (Sec.~\ref{sec:algebraic}).

\subsection{Gauge-Invariant Fluxes}

Gauge-inequivalent ans\"atze are distinguished by their fluxes. For a closed loop $C$ starting and ending at a base site $i$, the flux operator built from the singlet links is~\cite{Wen2002}
\begin{equation}
P_C(i)=u_{ij}u_{jk}\cdots u_{li}.
\end{equation}
A gauge transformation changes it only by a rotation at the base site, $P_C(i)\to W_iP_C(i)W_i^\dagger$. For a loop of length $L$ it takes the form $P_C=\ii^L Q_C$ with $Q_C\propto\cos\phi_C\,\tau^0+\ii\sin\phi_C\,\hat{\bm n}_C\cdot\bm\tau$. For hopping ans\"atze in the canonical gauge, $\hat{\bm n}_C=\hat{\bm z}$, and $\phi_C=L\pi/2-\Phi_C$ (mod $2\pi$), where $\Phi_C=\arg(t_{ij}t_{jk}\cdots t_{li})$ is the hopping flux and $t_{ij}$ is the coefficient of $f^\dagger_{i\alpha}f_{j\alpha}$. The fluxes also reveal the IGG, which consists of the $SU(2)$ matrices that commute with all flux operators and on-site terms at the base site. It is $SU(2)$ if all $P_C\propto\tau^0$, $U(1)$ if all $\hat{\bm n}_C$ are collinear, and $\Z_2$ otherwise.

\begin{figure}
\centering
\includegraphics[width=\columnwidth]{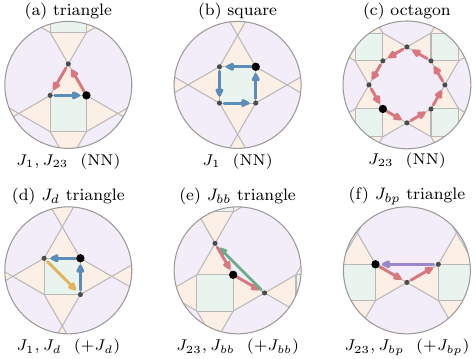}
\caption{Elementary gauge-invariant loops of the square-kagome lattice, one representative per symmetry orbit, all based at the reference square site $(0,0,1)$ (black dot). Arrows give the order of the links in the flux operator $P_C$ (counterclockwise). Loop bonds are colored by orbit as in Fig.~\ref{fig:lattice}, and the faint background shows the nearest-neighbor bonds. (a) Triangle ($J_1$, $J_{23}$), (b) square ($J_1$), and (c) octagon ($J_{23}$) generate all loops for nearest-neighbor bonds. Further-neighbor bonds close triangles such as (d) the half-square triangle ($J_1$, $J_d$), (e) the triangle ($J_{23}$, $J_{bb}$) at a square-site corner of an octagon, and (f) the triangle ($J_{23}$, $J_{bp}$) at a bow-tie corner of an octagon. The range needed for each loop is indicated below the panel.}
\label{fig:loops}
\end{figure}

With nearest-neighbor bonds, all loops are generated by the faces of the lattice: the four triangles of a cell, the square, and the octagon [Fig.~\ref{fig:loops}(a)--(c)]. Further-neighbor bonds add the triangles of Fig.~\ref{fig:loops}(d)--(f). For $U(1)$ and $SU(2)$ ans\"atze, the counterclockwise hopping fluxes through the faces of a unit cell add up to the flux per cell,
\begin{equation}
\sum_{\triangle}\Phi_\triangle+\Phi_s+\Phi_o=\Phi\pmod{2\pi},
\label{eq:fluxsum}
\end{equation}
where $\Phi_s$ and $\Phi_o$ are the square and octagon fluxes. The PSG fixes $\Phi$: it is $0$ or $\pi$ for $\eta_{12}=\pm1$, and continuous in the flux families of Sec.~\ref{sec:alg-u1}. In the staggered gauge sectors ($w_T=1$) the translations conjugate the $U(1)$ charge, so the face fluxes alternate in sign between neighboring cells and Eq.~\eqref{eq:fluxsum} holds with the sign $(-1)^{x+y}$.

Time reversal acts on the flux operators as $P_C(i)\to(-1)^LG_\T(i)P_C(i)G_\T^\dagger(i)$, so loops of odd length are the ones that detect it. In a TRS ansatz, every odd loop has a traceless $P_C$, that is, a hopping flux $0$ or $\pi$. In an $SU(2)$ ansatz, by contrast, every $P_C\propto\tau^0$, so an odd loop carries the flux $\pm\pi/2$, which is odd under time reversal. A TRS $SU(2)$ state therefore has no odd loop. Every triangle of the lattice has one square edge and two bow-tie edges, so such a state cannot have amplitudes on both nearest-neighbor orbits.
\subsection{Chiral Projective Symmetry Groups}
\label{sec:chiral-psg}

A chiral spin liquid breaks time reversal, but in a restricted way: every lattice symmetry $g$ is kept, either on its own or combined with time reversal as $g\T$~\cite{KalmeyerLaughlin1987,WenWilczekZee1989,BieriLhuillierMessio2016}. We record this choice by a sign $\epsilon(g)=\pm1$, with $\epsilon(g)=+1$ if $g$ is kept on its own and $\epsilon(g)=-1$ if only $g\T$ is a symmetry. Since products of symmetries must behave consistently, $\epsilon(g_1g_2)=\epsilon(g_1)\epsilon(g_2)$, so $\epsilon$ is a character of $p4mm$. It is therefore fixed by its values on the generators, which must respect the relations~\eqref{eq:relations}.

These relations impose only one condition. Because $C_4$ turns $T_1$ into $T_2$, the two translations must carry the same sign, $\epsilon(T_1)=\epsilon(T_2)\equiv\epsilon_T$. The signs $\epsilon_T$, $\epsilon(C_4)$, and $\epsilon(\sigma)$ are otherwise free, which gives the eight chiral classes of Table~\ref{tab:classes}. In four of them, $\epsilon_T=-1$: a translation is a symmetry only together with time reversal, so time-reversal-odd quantities, such as the scalar chirality, change sign from one unit cell to the next. We call these classes staggered and mark them with a prime. They exist on every lattice with the symmetry $p4mm$ (or $p4$)~\cite{MessioLhuillierMisguich2013}. They do not exist for $p6mm$, where every translation can be written as a product of rotations and therefore always has $\epsilon=+1$, nor for $p4g$, where every translation is the square of a glide reflection.

\begin{table}[!htb]
\caption{Chiral classes of $p4mm$. A class is labeled by the signs $(\epsilon_T,\epsilon(C_4),\epsilon(\sigma))$, and the last column lists the generators that are symmetries only in combination with time reversal. Here $\sigma$ is the diagonal mirror, and the axial mirror $\sigma C_4^{-1}$ has the sign $\epsilon(\sigma)\epsilon(C_4)$. The primed classes are staggered.}
\label{tab:classes}
\begin{ruledtabular}
\begin{tabular}{ccc}
class & $(\epsilon_T,\epsilon(C_4),\epsilon(\sigma))$ & elements combined with $\T$\\\hline
$\mathcal{A}$ & $(+,+,+)$ & --\\
$\mathcal{B}$ & $(+,+,-)$ & $\sigma\T$\\
$\mathcal{C}$ & $(+,-,+)$ & $C_4\T$\\
$\mathcal{D}$ & $(+,-,-)$ & $C_4\T$, $\sigma\T$\\
$\mathcal{A}'$ & $(-,+,+)$ & $T_{1,2}\T$\\
$\mathcal{B}'$ & $(-,+,-)$ & $T_{1,2}\T$, $\sigma\T$\\
$\mathcal{C}'$ & $(-,-,+)$ & $T_{1,2}\T$, $C_4\T$\\
$\mathcal{D}'$ & $(-,-,-)$ & $T_{1,2}\T$, $C_4\T$, $\sigma\T$
\end{tabular}
\end{ruledtabular}
\end{table}

Class $\mathcal{B}$ is the analog of the Kalmeyer--Laughlin state: translations and rotations are kept, and the mirrors are kept only together with time reversal. As we show in Sec.~\ref{sec:chern}, it is the only class that allows a Chern number.

The PSGs of a chiral class are easy to obtain. The symmetries of each class, some of them combined with $\T$, form a group with the same multiplication rules as $p4mm$ itself. The gauge matrices $G_g$ therefore satisfy the same equations, and the algebraic PSGs of every chiral class are those of the space group $p4mm$ alone (Sec.~\ref{sec:alg-sg}). The class enters only when an ansatz is built: for an element with $\epsilon(g)=-1$, the symmetry condition~\eqref{eq:psg-spatial} acquires the extra minus sign of time reversal~\cite{BieriLhuillierMessio2016},
\begin{equation}
u^{(\alpha)}_{g(i)g(j)}=-\,G_{g\T}(g(i))\,\big[R_g\,u_{ij}\big]^{(\alpha)}\,G^\dagger_{g\T}(g(j)).
\end{equation}

\subsection{Working Definitions and Units}
\label{sec:defs}

\textit{Realized PSG.} An ansatz built from a PSG can accidentally be invariant under a larger gauge group, and it then belongs to a different PSG. We call a PSG \textit{realized} at a given range if a generic ansatz with all amplitudes it allows up to that range has exactly the IGG of the PSG, and a connected bond graph. A chiral PSG must, in addition, break time reversal.

\textit{Other terms.} A state \textit{descends} from a parent if its PSG is a subgroup of the parent PSG and it has a smaller IGG. A descendant is \textit{genuine} if it is realized with bonds up to $J_+$. The bow-tie sites and the square sites form the two \textit{site orbits}, and symmetry-related bonds form \textit{bond orbits}, named by their couplings. The \textit{extended stabilizer} of a site is its stabilizer in $p4mm\times\Z_2^\T$, extended by the IGG. A \textit{sector} of $U(1)$ PSGs collects the PSGs in which the same symmetries reverse the $U(1)$ rotation (Sec.~\ref{sec:alg-u1}).

\textit{Sampled results.} Spectral types, Dirac flavor numbers, and the Chern numbers of our surveys are computed at fixed amplitudes. They describe regions of amplitude space, not the PSG as a whole. Only the flat bands (Sec.~\ref{sec:flatband}) and the symmetry selection rules hold for all amplitudes.

\textit{Units.} Spinon energies are given in units of the hopping $|t|$ on a reference bond, with $t_{ij}=2(\ii\chi^0_{ij}-\chi^3_{ij})$ the coefficient of $f^\dagger_{i\alpha}f_{j\alpha}$. Energies of Heisenberg models are per site, in units of $|J_1|$, for $S=1/2$. The gap $\Delta$ is the single-spinon gap, in units of $|t|$ for free amplitudes and of $|J_1|$ for the self-consistent states of Sec.~\ref{sec:mf}. The two-spinon continuum starts at $2\Delta$.

\section{Algebraic PSGs}
\label{sec:algebraic}

We now solve the PSGs [the derivations are outlined in Appendix~\ref{app:psg}]. For the IGG $\Z_2$, a PSG is fixed by two pieces of data. The first is its set of relation values, one sign $\pm1$ for each relation~\eqref{eq:relations}. Gauge-inequivalent sets of signs are classified by the cohomology group $H^2(p4mm\times\Z_2^\T,\Z_2)=\Z_2^{10}$, so there are $2^{10}=1024$ classes. The second is how the symmetries that leave a site fixed act on that site. These elements form the site stabilizer, and their gauge matrices $G_h(i)$ are on-site $SU(2)$ matrices that must be consistent with the relation values.

On a lattice whose sites have no symmetry of their own, the second condition is always satisfied, and every class gives exactly one PSG. On the square-kagome lattice both site orbits have nontrivial stabilizers. A class can therefore give no PSG, if one of the orbits admits no consistent on-site matrices, or more than one, if the two orbits admit several inequivalent combinations~\cite{Mackey1952,Brown1982}.

\subsection{$\Z_2$ Projective Symmetry Groups}
\label{sec:alg-z2}

In a suitable gauge, every $\Z_2$ PSG of the square-kagome lattice takes the form
\begin{align}
G_{T_1}(x,y,\mu)&=\tau^0,\nonumber\\
G_{T_2}(x,y,\mu)&=\eta_{12}^{x}\,\tau^0,\nonumber\\
G_{C_4}(x,y,\mu)&=\eta_{C_4T}^{\,x}\,\eta_{12}^{\,xy}\,g_{C_4}(\mu),\nonumber\\
G_\sigma(x,y,\mu)&=\eta_{\sigma T}^{\,x+y}\,\eta_{12}^{\,xy}\,g_\sigma(\mu),\nonumber\\
G_\T(x,y,\mu)&=\eta_{\T T}^{\,x+y}\,g_\T(\mu).
\label{eq:z2psg}
\end{align}
The dependence on the cell $(x,y)$ is a set of signs, and the sublattice matrices $g_O(\mu)$ can be chosen from the quaternion group $\{\pm\tau^0,\pm\ii\tau^1,\pm\ii\tau^2,\pm\ii\tau^3\}$, with $g_{C_4}(\mu)=\tau^0$ for $\mu=2,3,4,6$. Restricting to the quaternion group loses nothing: every site stabilizer here is a product of $\Z_2$ factors, whose $SU(2)$ representations can always be rotated into this group.

A $\Z_2$ class is labeled by ten signs, the values of the following relations:
\begin{align}
&\eta_{12}:\ T_2T_1T_2^{-1}T_1^{-1}, && \eta_{\T\sigma}:\ \sigma\T\sigma^{-1}\T^{-1},\nonumber\\
&\eta_{C_4T}:\ C_4T_2C_4^{-1}T_1, && \eta_{\T C_4}:\ C_4\T C_4^{-1}\T^{-1},\nonumber\\
&\eta_{\sigma T}:\ \sigma T_1\sigma^{-1}T_2^{-1}, && \eta_\sigma:\ \sigma^2,\quad \eta_{\sigma C_4}:\ (\sigma C_4)^2,\nonumber\\
&\eta_{\T T}:\ T_1\T T_1^{-1}\T^{-1}, && \eta_{C_4}:\ C_4^4,\quad \eta_\T:\ \T^2.
\label{eq:z2labels}
\end{align}
The other relations add nothing new. We set $C_4T_1C_4^{-1}T_2^{-1}=+1$ by multiplying $G_{T_1}$ by $-1$ if needed, and the relations for $T_2$ repeat $\eta_{\sigma T}$ and $\eta_{\T T}$. Six of the ten signs belong to the space group, three tell whether time reversal commutes with $T_1$, $\sigma$, and $C_4$, and one is $\eta_\T$.

Three signs have no counterpart on lattices with $p6mm$ symmetry, such as the kagome lattice. With $\eta_{C_4T}=-1$ or $\eta_{\sigma T}=-1$, the rotation or the mirror commutes with the translations only up to a sign. With $\eta_{\T T}=-1$, time reversal carries the alternating sign $(-1)^{x+y}$ on the spinons.

We label a $\Z_2$ PSG by these signs, $\mathrm{Z}n_{T_2}n_{C_4T}n_{\sigma T}$-$n_{\T T}n_{\T\sigma}n_{\T C_4}$-$n_\sigma n_{\sigma C_4}n_{C_4}(n_\T)$, with the digit $0$ for $+1$ and $1$ for $-1$, for example Z000-011-100(1). Here $n_{T_2}$ stands for $\eta_{12}$. A sign $\zeta=\pm$ is appended when a class has two PSGs (Sec.~\ref{sec:alg-zeta}). The sublattice matrices of all PSGs are given in the SM.

The on-site condition reduces the $1024$ classes to $200$ PSGs. Consistent on-site matrices exist on the bow-tie sites, with stabilizer $2mm\times\Z_2^\T\cong\Z_2^3$, for $240$ classes, and on the square sites, with stabilizer $m\times\Z_2^\T\cong\Z_2^2$, for $640$. Only the $160$ classes allowed on both orbits survive. Of these, $120$ give one PSG and $40$ give two, for a total of $120+2\times40=200$ $\Z_2$ PSGs (Appendix~\ref{app:psg} and \SMtext{Sec.~\ref{sm:sec:counting} of the SM}{the SM}). A script in the data files~\cite{RajMaityIqbalZenodo} checks the classification independently: it reconstructs every PSG from its canonical form, evaluates all group relations, distinguishes the $200$ $\Z_2$ PSGs by gauge invariants, and recomputes the realized PSGs of Sec.~\ref{sec:ansatz}. The two orbits alone form a checkerboard lattice and a lattice of squares. How their PSGs combine into those of the square-kagome lattice, and the counts for other $p4mm$ lattices, are given in Sec.~\ref{sm:sec:relationsm} of the SM.

\subsection{The Relative Time-Reversal Sign}
\label{sec:alg-zeta}

The $40$ classes with two PSGs are those in which time reversal acts on every site as a plain sign, $G_\T=h_b\tau^0$ on the bow-tie sites and $G_\T=h_s\tau^0$ on the square sites. All four time-reversal signs of these classes, $\eta_\T$, $\eta_{\T T}$, $\eta_{\T\sigma}$, and $\eta_{\T C_4}$, are $+1$. The two PSGs of such a class differ only in the relative sign
\begin{equation}
\zeta=h_bh_s=\tfrac14\Tr G_\T(0,0,5)\,\Tr G_\T(0,0,1),
\end{equation}
which no gauge transformation can change.

This sign decides which bonds are allowed. With $G_\T(i)=h_i\tau^0$, the time-reversal condition~\eqref{eq:psg-T} becomes $u_{ij}=-h_ih_j\,u_{ij}$ and $a_i=-a_i$. On-site terms therefore always vanish, and a bond survives only if $h_ih_j=-1$:
\begin{itemize}
\item For $\zeta=+$, all $h_i$ are equal, and every amplitude vanishes, on every bond at any range. These PSGs describe no state at all.
\item For $\zeta=-$, bonds within a site orbit vanish, and only the bonds between square and bow-tie sites survive ($J_{23}$, $J_{sb}$, and $J_{o3}$ up to $J_+$).
\end{itemize}
The same holds for $U(1)$ and $SU(2)$ PSGs in which time reversal acts as a sign. The PSGs with no amplitude are exactly the $\zeta=+$ ones: $40$ $\Z_2$, $25$ $U(1)$, and $4$ $SU(2)$ PSGs. Every other PSG is realized by bonds up to $J_+$ (Sec.~\ref{sec:ansatz}).

\subsection{Flat Bands at the Fermi Level}
\label{sec:flatband}

The sign $\zeta$ has a direct effect on the spinon bands. In a PSG with $\zeta=-$, only bonds between square and bow-tie sites survive (Sec.~\ref{sec:alg-zeta}), for singlet and triplet amplitudes alike. The Bloch Hamiltonian therefore connects only the two site orbits,
\begin{equation}
h(\bm{k})=\begin{pmatrix}0&B(\bm{k})\\B^\dagger(\bm{k})&0\end{pmatrix},
\label{eq:hflat}
\end{equation}
where $B(\bm{k})$ is a $4\times2$ matrix between the four square sites and the two bow-tie sites of a cell, per spin ($8\times4$ in Nambu space). A $4\times2$ matrix has rank at most two, so $h(\bm{k})$ has at least two zero modes per spin at every momentum. These modes form exactly flat bands at $E=0$. At half filling they are half filled and pinned to the Fermi level, for any amplitudes on $J_{23}$, $J_{sb}$, and $J_{o3}$.

These are the familiar zero modes of a bipartite hopping problem with unequal sublattices~\cite{Sutherland1986,Lieb1989}. The new element is that here they are imposed by the projective action of time reversal, not by the lattice geometry. The same mechanism works on any lattice with two site orbits of unequal size that admits a PSG with $\zeta=-$. On the square-kagome lattice this covers $40$ $\Z_2$, $25$ $U(1)$ (including one flux family), and $4$ $SU(2)$ PSGs. Only for these PSGs are the bands flat for all amplitudes. The flat bands carry an extensive degeneracy at half filling, which interactions, gauge fluctuations, or the projection will lift, and we do not study their fate here.

\subsection{$SU(2)$ Projective Symmetry Groups}

An $SU(2)$ ansatz has only imaginary hopping, $u^{(0)}_{ij}=\ii\chi_{ij}\tau^0$, and its IGG contains every uniform $SU(2)$ matrix. Any uniform part of a gauge matrix $G_O(i)$ can therefore be absorbed into the IGG, and only a site-dependent sign is left. An $SU(2)$ PSG is thus a pattern of signs on the lattice: a $\Z_2$ PSG in which all on-site matrices are $\pm\tau^0$. The sign $\eta_\T$ no longer distinguishes PSGs, since it can be absorbed in the same way. We find $8$ $SU(2)$ PSGs, four classes each with $\zeta=\pm$,
\begin{equation}
\begin{gathered}
\mathrm{SU}000\text{-}000\text{-}000\pm,\quad \mathrm{SU}011\text{-}000\text{-}001\pm,\\
\mathrm{SU}100\text{-}000\text{-}000\pm,\quad \mathrm{SU}111\text{-}000\text{-}001\pm,
\end{gathered}
\end{equation}
labeled like the $\Z_2$ PSGs without the digit $n_\T$. Only the four PSGs with $\zeta=-$ give states, all with bonds between square and bow-tie sites only.

\subsection{$U(1)$ Projective Symmetry Groups}
\label{sec:alg-u1}

For the IGG $U(1)=\{e^{\ii\theta\tau^3}\}$, every gauge matrix either keeps or reverses the $U(1)$ rotation,
\begin{equation}
G_O(i)=e^{\ii\phi_O(i)\tau^3}(\ii\tau^1)^{w_O},
\end{equation}
with $w_O=0$ if $O$ keeps it and $w_O=1$ if $O$ reverses it, that is, if $O$ exchanges spinon particles and holes. As for the chiral classes, $C_4$ forces $w_{T_1}=w_{T_2}\equiv w_T$, but nothing forces $w_T=0$. The choices $w=(w_T,w_{C_4},w_\sigma,w_\T)$ give $16$ sectors, twice as many as for $p6mm$. In the eight staggered sectors, $w_T=1$, the translations exchange spinon particles and holes. These sectors still describe TRS states. They should not be confused with the staggered chiral classes of Sec.~\ref{sec:chiral-psg}, in which a translation is combined with time reversal.

\begin{table}
\caption{$U(1)$ projective symmetry groups of $p4mm\times\Z_2^\T$ by sector $w=(w_T,w_{C_4},w_\sigma,w_\T)$: abstract cohomology classes ($H^2$), relation-value classes realizable on the square-kagome sites ($\eta$), and PSGs. ``$\times U(1)$'' marks one-parameter families with a continuous flux $\Phi$ per unit cell. Sectors with $w_T=1$ are staggered.}
\label{tab:u1}
\begin{ruledtabular}
\begin{tabular}{cccc}
$w$ & $H^2$ & $\eta$ & PSG\\
\hline
$(0,0,0,0)$ & 64 & 4 & 8\\
$(0,0,0,1)$ & 128 & 4 & 4\\
$(0,0,1,0)$ & $64\times U(1)$ & $1\times U(1)$ & $2\times U(1)$\\
$(0,0,1,1)$ & 128 & 4 & 4\\
$(0,1,0,0)$ & 128 & 4 & 8\\
$(0,1,0,1)$ & 128 & 4 & 4\\
$(0,1,1,0)$ & 128 & 4 & 8\\
$(0,1,1,1)$ & 128 & 4 & 4\\
$(1,0,0,0)$ & 128 & 4 & 8\\
$(1,0,0,1)$ & 128 & 4 & 4\\
$(1,0,1,0)$ & 128 & 2 & 4\\
$(1,0,1,1)$ & 128 & 4 & 4\\
$(1,1,0,0)$ & 128 & 4 & 8\\
$(1,1,0,1)$ & 128 & 4 & 4\\
$(1,1,1,0)$ & 128 & 2 & 4\\
$(1,1,1,1)$ & 128 & 4 & 4\\
\hline
total & $1856+64\times U(1)$ & $56+1\times U(1)$ & $80+2\times U(1)$
\end{tabular}
\end{ruledtabular}
\end{table}

Table~\ref{tab:u1} gives three counts for each sector, which follow the same steps as for $\Z_2$. The first column counts all cohomology classes, $1856$ in total. The on-site conditions on the two site orbits keep $56$ of them (second column). As for $\Z_2$, some classes then split into two PSGs that differ only in the relative on-site data of the two orbits, which gives $80$ discrete $U(1)$ PSGs (third column). The sectors with $w_\T=0$ contain $48$ discrete PSGs and the two flux families, exactly the $25$ $U(1)$ PSGs with $\zeta=+$ and the $25$ with $\zeta=-$ (Sec.~\ref{sec:alg-zeta}). Every $U(1)$ PSG with $w_\T=0$ is therefore either empty or a flat-band PSG, and the $32$ $U(1)$ states with bonds within the site orbits are exactly those with $w_\T=1$, the class-AII PSGs of Sec.~\ref{sec:qsh}.

One sector is special. In $w=(0,0,1,0)$, where only the mirror reverses the $U(1)$ rotation, the translations need to commute only up to a phase, $T_2T_1T_2^{-1}T_1^{-1}=e^{\ii\Phi\tau^3}$, with any flux $\Phi\in[0,\pi]$ per unit cell. This gives two continuous families of PSGs. Only the family with $\zeta=-$ allows amplitudes, on square--bow-tie bonds only, and its octagon flux varies continuously with $\Phi$.

We label the discrete $U(1)$ PSGs U$w$-$n$, with the sector $w=w_Tw_{C_4}w_\sigma w_\T$ and an index $n$ within the sector, and the two families U0010-F0 and U0010-F1. Their forms are given in the SM.

\subsection{The Space Group Alone}
\label{sec:alg-sg}

Without time reversal we obtain the PSGs of the space group $p4mm$ alone. These are the PSGs of every chiral class (Sec.~\ref{sec:chiral-psg}): $40$ $\Z_2$, $24$ discrete $U(1)$ PSGs and one flux family, and $4$ $SU(2)$ PSGs (Table~\ref{tab:counts}). Their labels drop the time-reversal digits and $\zeta$, and carry the chiral class as a prefix, for example $\mathcal{B}$:SU011-001.

All PSG labels in this paper follow the same scheme:
\begin{itemize}
\item The letters Z, U, and SU give the IGG.
\item The digits give the relation signs, with $0$ for $+1$ and $1$ for $-1$. For $U(1)$ they give the sector $w$ followed by an index $n$.
\item A label without a prefix, such as Z000-011-100(1), is a TRS PSG of $p4mm\times\Z_2^\T$.
\item A prefix $\mathcal{A}$--$\mathcal{D}$ or $\mathcal{A}'$--$\mathcal{D}'$, such as $\mathcal{B}$:SU011-001, gives the chiral class of a time-reversal-breaking PSG.
\item A prefix N, such as N:U1000-0 or N:Z00000-110, marks a PSG of the nematic group of cuboc1 (Sec.~\ref{sec:nematic}).
\end{itemize}
Table~\ref{tab:named} lists the states that recur in the text.

\begin{table}
\caption{Number of algebraic projective symmetry groups on the square-kagome lattice. $U(1)$ entries give discrete PSGs plus one-parameter families with a continuous flux. The second column gives the dimension of $H^2(G,\Z_2)$ over $\Z_2$, of which only part is realized on the lattice.}
\label{tab:counts}
\begin{ruledtabular}
\begin{tabular}{lcccc}
symmetry group & $\dim H^2$ & $\Z_2$ & $U(1)$ & $SU(2)$\\
\hline
$p4mm\times\Z_2^\T$ & 10 & 200 & $80+2\times U(1)$ & 8\\
$p4mm$ & 6 & 40 & $24+1\times U(1)$ & 4
\end{tabular}
\end{ruledtabular}
\end{table}

\begin{table*}
\caption{Named states that recur in the text, with their PSG, symmetry class (TRS: time-reversal-symmetric, nematic: the class of cuboc1, Sec.~\ref{sec:nematic}, SOC: with spin--orbit coupling), spinon spectrum at half filling at the amplitudes of the section given, with the single-spinon gap $\Delta$, and Chern number $C$ per spin of the occupied bands. The fluxes $(\Phi_{\rm ud},\Phi_{\rm lr},\Phi_s)$ are defined in Sec.~\ref{sec:flux-states}.}
\label{tab:named}
\begin{ruledtabular}
\begin{tabular}{lllllc}
state & PSG & class & spectrum & $C$ & Sec.\\
\hline
flux state $(0,0,0)$ & U0001-2 & TRS & zero-energy lines & -- & \ref{sec:flux-states}\\
flux state $(0,0,\pi)$ & U0001-0 & TRS & Dirac & -- & \ref{sec:flux-states}\\
flux state $(\pi,0,0)$ & U0111-2 & TRS & flat band at $E_F$ & -- & \ref{sec:flux-states}\\
flux state $(\pi,0,\pi)$ & U0111-0 & TRS & gapped, $\Delta=\sqrt3-1$ & 0 & \ref{sec:flux-states}\\
class-$\mathcal{B}$ state $(\pi/2,\pi/2,\pi)$ & $\mathcal{B}$:SU011-001 & $\mathcal{B}$ & gapped, $\Delta=2-\sqrt2$ & $-1$ & \ref{sec:kl}\\
best projected spin liquid ($\lambda\le0.4$, C1b) & Z000-011-100(1) & TRS & (projected) & -- & \ref{sec:vmc}\\
nematic candidate & N:U1000-0 & nematic & gapped, $\Delta=0.156$ & 0 & \ref{sec:nematic}\\
projected nematic chiral state & N:Z00000-110 & nematic & (projected) & -- & \ref{sec:vmc}\\
conical candidate & $\mathcal{D}'$:U100-0 & $\mathcal{D}'$ & gapped, $\Delta\approx0.51$ & 0 & \ref{sec:order-classes}\\
spinon quantum spin Hall Q1, Q2 & U0001-2, U0001-0 & TRS, SOC & gapped, $\Z_2$-odd & -- & \ref{sec:qsh}
\end{tabular}
\end{ruledtabular}
\end{table*}

\section{Symmetric Spin Liquids}
\label{sec:ansatz}

\subsection{Construction}

To build an ansatz from a PSG, we fix the amplitudes on one reference bond of each bond orbit, and the PSG then gives them on all other bonds of the orbit. The reference amplitudes are restricted by time reversal and by every symmetry that maps the bond onto itself. These are the elements that reverse the bond and, for $J_d$, $J_\times$, $J_{bs}$, and $J_+$, also mirrors that fix both of its ends. Equations~\eqref{eq:psg-spatial} and \eqref{eq:psg-T} then become linear equations for the real coefficients in Eq.~\eqref{eq:links}, and their solutions are the allowed amplitudes. On-site terms are restricted in the same way by the site stabilizers.

For each PSG we then check whether it is realized (Sec.~\ref{sec:defs}), by computing the IGG of a generic allowed ansatz on a finite patch. Self-consistent multipliers cannot reduce this IGG, since they commute with it.\footnote{The mean-field free energy at $T>0$ is strictly concave in the multipliers and invariant under the IGG of the bonds, so its unique maximum commutes with that IGG. $T=0$ is reached as the limit $T\to0^+$.} A realized PSG need not solve the self-consistency equations of a given Hamiltonian, a question we turn to in Sec.~\ref{sec:orders}.

\subsection{Realized Spin Liquids}

Every PSG except the $\zeta=+$ ones of Sec.~\ref{sec:alg-zeta} is realized with bonds up to $J_+$. This gives $160$ $\Z_2$, $57$ $U(1)$, and $4$ $SU(2)$ spin liquids, $221$ in total. They fall into two groups:
\begin{itemize}
\item The $69$ states with $\zeta=-$ have bonds only between square and bow-tie sites, and the flat bands of Sec.~\ref{sec:flatband}. They include all four $SU(2)$ states. Since they forbid the square-edge amplitude, they are not candidates for a model with a dominant $J_1$.
\item The other $152$ states, $120$ $\Z_2$ and $32$ $U(1)$, also have bonds within the site orbits.
\end{itemize}
With singlet and triplet amplitudes, all $221$ PSGs are realized already with nearest-neighbor bonds. With singlet amplitudes only, $40$ $\Z_2$, $21$ $U(1)$, and the $4$ $SU(2)$ PSGs are realized with nearest-neighbor bonds, $156$, $57$, and $4$ with all orbits at graph distance two, and all $221$ once $J_{o3}$ is added. The couplings $\{J_1,J_{23},J_\times,J_+\}$ of Ref.~\cite{Gembe2023} realize $66$, $27$, and $4$ (\SMtext{Table~\ref{sm:tab:realized} of the SM}{the SM}).

The triangles of the lattice matter here. On a lattice whose nearest-neighbor bonds form a bipartite graph, a TRS nearest-neighbor ansatz always has a larger IGG when $\T^2=-1$, so such $\Z_2$ PSGs cannot be realized at that range~\cite{RajIqbalSHD}. The triangles of the square-kagome lattice remove this restriction: $36$ of the $40$ $\Z_2$ PSGs realized with nearest-neighbor singlets have $\eta_\T=-1$. The amplitudes and shortest realizing range of every spin liquid are given in the SM.

The realized PSGs are linked by a hierarchy: each $SU(2)$ state has $U(1)$ and $\Z_2$ descendants obtained by adding symmetry-allowed amplitudes. The full map is given in Appendix~\ref{app:hierarchy}.

\subsection{Flux States of Ref.~\cite{Astrakhantsev2021}}
\label{sec:flux-states}

\begin{table}
\caption{Nearest-neighbor flux states of Ref.~\cite{Astrakhantsev2021}, labeled by the hopping fluxes $(\Phi_{\rm ud},\Phi_{\rm lr},\Phi_s)$, with their PSG, class, and spinon spectrum at half filling. The chiral states have flux $\pi/2$ per triangle and the IGG $SU(2)$, and their full label is class:PSG, for example $\mathcal{B}$:SU011-001. Spectra are for equal amplitudes on $J_1$ and $J_{23}$ and zero multipliers (Figs.~\ref{fig:chiral} and \ref{fig:nn-spectra}), with $k_i=\bm{k}\cdot\bm{T}_i$. The types of the TRS states hold for amplitude ratios between $0.02$ and $50$. The single-spinon gaps of the gapped states are $\Delta=\sqrt3-1$ $(\pi,0,\pi)$ and $\Delta=2-\sqrt2$ $(\pi/2,\pi/2,\pi)$ in units of the hopping, and $C$ is the Chern number per spin.}
\label{tab:flux}
\begin{ruledtabular}
\begin{tabular}{llll}
fluxes & PSG & class & spectrum\\
\hline
$(0,0,0)$ & U0001-2 & TRS & zero-energy lines $k_{1,2}=0$\\
$(0,0,\pi)$ & U0001-0 & TRS & Dirac, two nodes\\
$(\pi,0,0)$ & U0111-2 & TRS & flat band at $E_F$\\
$(\pi,0,\pi)$ & U0111-0 & TRS & gapped\\
$(\pi/2,\pi/2,0)$ & SU000-000 & $\mathcal{B}$ & two flat bands at $E_F$\\
$(\pi/2,\pi/2,\pi)$ & SU011-001 & $\mathcal{B}$ & gapped, $C=-1$\\
$(\pi/2,-\pi/2,0)$ & SU000-000 & $\mathcal{C}$ & zero-energy lines $k_{1,2}=0$\\
$(\pi/2,-\pi/2,\pi)$ & SU011-001 & $\mathcal{C}$ & Fermi pockets
\end{tabular}
\end{ruledtabular}
\end{table}

Reference~\cite{Astrakhantsev2021} labels its nearest-neighbor states by the hopping fluxes $(\Phi_{\rm ud},\Phi_{\rm lr},\Phi_s)$ through the up and down triangles, the left and right triangles, and the square. Table~\ref{tab:flux} gives the PSGs of its eight states, and the four chiral states are discussed in Sec.~\ref{sec:kl}. Two remarks help in reading the table. First, $(\pi,0,0)$ and $(\pi,0,\pi)$ are fully symmetric even though up/down and left/right triangles carry different fluxes. The reason is that $C_4$ and the mirror act together with the particle--hole transformation $\ii\tau^1$, which turns a triangle flux $0$ into $\pi$ and vice versa. Second, changing the sign of all hoppings shifts every triangle flux by $\pi$ but leaves squares and octagons unchanged, so, for example, $(0,0,0)$ and $(\pi,\pi,0)$ are the same state.

Our spectra agree with Ref.~\cite{Astrakhantsev2021} for seven of its eight states. The exception is $(\pi,0,\pi)$, which that work lists with a Fermi surface. We find it gapped, with $\Delta=\sqrt3-1$ at equal amplitudes and a nonzero gap for all amplitude ratios. Its on-site constraints require a staggered multiplier on the bow-tie sites (Sec.~\ref{sec:nn-u1}).

\section{Chiral Spin Liquids}
\label{sec:chiral}

\subsection{Realization}

\begin{figure*}
\centering
\includegraphics[width=\textwidth]{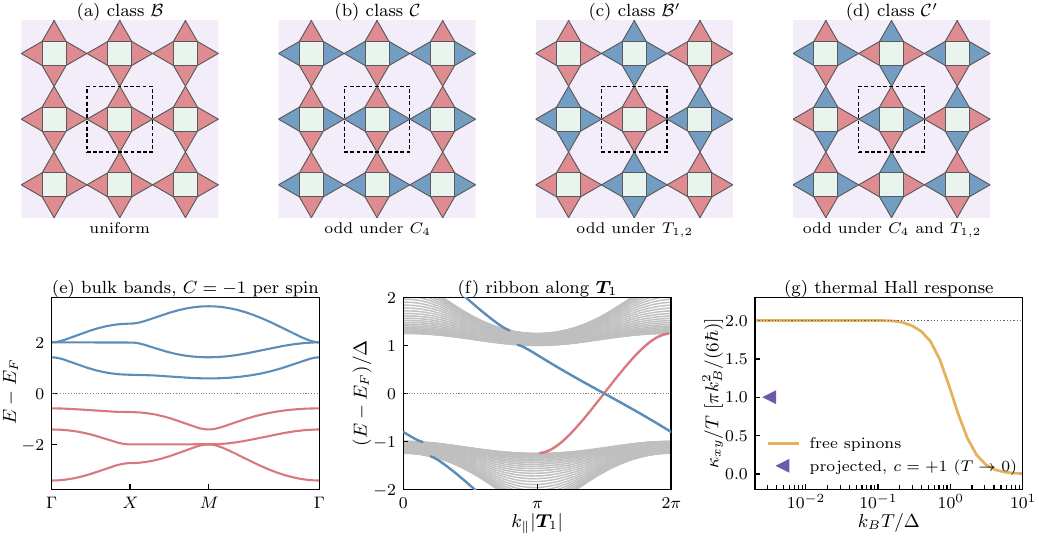}
\caption{Chiral classes and the class-$\mathcal{B}$ state of the square-kagome lattice. (a)--(d) Sign of a time-reversal-odd flux through the nearest-neighbor triangles (red positive, blue negative) allowed by the classes $\mathcal{B}$, $\mathcal{C}$, $\mathcal{B}'$, and $\mathcal{C}'$, from $\Phi(gP)=\epsilon(g)\det(g)\Phi(P)$. The dashed square is the unit cell. In the classes $\mathcal{A}$, $\mathcal{D}$, $\mathcal{A}'$, and $\mathcal{D}'$ this flux vanishes. (e) Spinon bands per spin of the nearest-neighbor $SU(2)$ state $\mathcal{B}$:SU011-001, with flux $+\pi/2$ through every triangle (counterclockwise) and $\pi$ through squares and octagons, and equal hopping amplitudes $|t|=1$ on $J_1$ and $J_{23}$. Occupied bands are rose, empty bands blue. (f) Spectrum of a ribbon of $30$ unit cells, periodic along $\bm{T}_1$, in units of $\Delta$. States with more than half of their weight near the lower (upper) edge are rose (blue), and bulk states gray. Each edge carries one chiral mode per spin, the lower-edge mode moving along $+x$. (g) Thermal Hall conductivity of the free spinons, Eq.~\eqref{eq:kxy}, with the free-spinon plateau (dotted) and, schematically, the zero-temperature value expected after projection (triangle, drawn at the low-temperature end of the axis). The free-spinon curve at finite temperature is a mean-field result, which gauge fluctuations modify.}
\label{fig:chiral}
\end{figure*}

On a bipartite lattice, a nearest-neighbor ansatz is always gauge-equivalent to its time-reversed partner. Consequently, realizing a genuinely chiral ansatz requires longer-range bonds. The triangular plaquettes of the square-kagome lattice remove this restriction, allowing chiral ansätze to be realized even with nearest-neighbor bonds. Table~\ref{tab:chiralcounts} summarizes the number of realized chiral PSGs (Sec.~\ref{sec:defs}) in each class. With singlet amplitudes on all orbits up to $J_+$, every chiral PSG of every class is realized. The four $SU(2)$ PSGs of class $\mathcal{A}$ need orbits beyond graph distance two and therefore appear in no column of Table~\ref{tab:chiralcounts}. With triplet amplitudes, every chiral PSG is realized already with nearest-neighbor bonds.

\begin{table}
\caption{Number of realized chiral PSGs in each class with singlet amplitudes, for nearest-neighbor bonds (NN), bonds up to graph distance two (GD2), and the couplings $\{J_1,J_{23},J_\times,J_+\}$ of Ref.~\cite{Gembe2023} (CO). Each class has $40$ $\Z_2$, $25$ $U(1)$ (including the flux family), and $4$ $SU(2)$ PSGs.}
\label{tab:chiralcounts}
\begin{ruledtabular}
\begin{tabular}{lccccccccc}
 & \multicolumn{3}{c}{$\Z_2$} & \multicolumn{3}{c}{$U(1)$} & \multicolumn{3}{c}{$SU(2)$}\\
\cline{2-4}\cline{5-7}\cline{8-10}
class & NN & GD2 & CO & NN & GD2 & CO & NN & GD2 & CO\\
\hline
$\mathcal{A}$ & 8 & 40 & 20 & 2 & 25 & 3 & 0 & 0 & 0\\
$\mathcal{B}$ & 16 & 40 & 20 & 13 & 25 & 17 & 4 & 4 & 4\\
$\mathcal{C}$ & 16 & 40 & 24 & 13 & 25 & 17 & 4 & 4 & 4\\
$\mathcal{D}$ & 8 & 40 & 22 & 2 & 25 & 11 & 0 & 4 & 2\\
$\mathcal{A}'$ & 8 & 40 & 24 & 2 & 25 & 10 & 0 & 4 & 2\\
$\mathcal{B}'$ & 16 & 40 & 24 & 13 & 25 & 17 & 4 & 4 & 4\\
$\mathcal{C}'$ & 16 & 40 & 28 & 13 & 25 & 19 & 4 & 4 & 4\\
$\mathcal{D}'$ & 8 & 40 & 24 & 2 & 25 & 11 & 0 & 4 & 2
\end{tabular}
\end{ruledtabular}
\end{table}

The classes $\mathcal{B}$, $\mathcal{C}$, $\mathcal{B}'$, and $\mathcal{C}'$ realize more PSGs with nearest-neighbor bonds because they allow a time-reversal-odd flux through the triangles. Each triangle is mapped onto itself by the mirror through its bow-tie corner, which reverses its orientation and hence its flux. A nonzero flux can survive only if this mirror is combined with time reversal, which reverses the flux once more. This is the case in $\mathcal{B}$, $\mathcal{C}$, $\mathcal{B}'$, and $\mathcal{C}'$, and not in $\mathcal{A}$, $\mathcal{D}$, $\mathcal{A}'$, and $\mathcal{D}'$. The sign pattern of the flux then follows from the class [Fig.~\ref{fig:chiral}(a)--(d)]: uniform in $\mathcal{B}$, alternating under $C_4$ in $\mathcal{C}$, alternating between neighboring cells in $\mathcal{B}'$, and both in $\mathcal{C}'$. In general, a symmetry $g$ maps the flux of a loop $P$ onto $\Phi(gP)=\epsilon(g)\det(g)\Phi(P)$. 

\subsection{Chern Numbers and Selection Rule}
\label{sec:chern}

The same rule applies to the Chern number $C$ of the occupied spinon bands. A unitary mirror and time reversal each reverse $C$, so a symmetry $g$ maps $C\to\epsilon(g)\det(g)\,C$. A nonzero Chern number therefore needs $\epsilon(g)=\det(g)$ for every $g$: translations and rotations must be unitary, and mirrors must be combined with time reversal. Among the eight classes, only class $\mathcal{B}$ satisfies this. In all others, at least one symmetry reverses $C$ and forces $C=0$: a translation combined with $\T$ in the staggered classes, $C_4\T$ in $\mathcal{C}$, $\mathcal{D}$, $\mathcal{C}'$, and $\mathcal{D}'$, and a unitary mirror in $\mathcal{A}$ and $\mathcal{C}$.

Two rotations will matter later. The twofold rotation $C_4^2$ about the square center is unitary in every class, since it is the square of $C_4$. The twofold rotations $T_1C_4^2$ and $T_2C_4^2$ about the bow-tie sites carry the sign $\epsilon_T$, so they are combined with $\T$ in all staggered classes.

We computed Chern numbers for all realized chiral PSGs at three amplitude points, with the Fukui--Hatsugai--Suzuki method~\cite{FukuiHatsugaiSuzuki2005}. For $\Z_2$ states we quote the Kitaev index $\nu$ of the Bogoliubov--de Gennes bands, normalized so that $\nu=-4C$ for a state with Chern number $C$ per spin. As required, every gapped state outside class $\mathcal{B}$ has $C=0$. In class $\mathcal{B}$ we find $C=\pm1$ and $\pm2$ per spin for $U(1)$ states, $C=\pm1$ for $SU(2)$ states (all from $\mathcal{B}$:SU011-001), and $\nu=\pm4$ and $\pm8$ for $\Z_2$ states. These values are consistent with the eigenvalues of the fourfold rotation at its fixed momenta, which fix $C$ modulo $4$~\cite{FangGilbertBernevig2012}.

\subsection{Chiral Flux States and the Class-$\mathcal{B}$ State}
\label{sec:kl}

The four chiral flux states of Ref.~\cite{Astrakhantsev2021} have flux $\pi/2$ through every triangle. They belong to class $\mathcal{B}$ if this flux is the same on all triangles, and to class $\mathcal{C}$ if it alternates between the up/down and left/right triangles, for either square flux $0$ or $\pi$ (Table~\ref{tab:flux}). Only one of them has a Chern number: $\mathcal{B}$:SU011-001, with flux $\pi/2$ through every triangle and $\pi$ through squares and octagons [Fig.~\ref{fig:chiral}(e)]. We call it the class-$\mathcal{B}$ state.

At equal amplitudes, the class-$\mathcal{B}$ state has the single-spinon gap $\Delta=2-\sqrt2\approx0.586$ and Chern number $C=-1$ per spin, and its time-reversed partner has $C=+1$. Ribbons along $\bm{T}_1$ and $\bm{T}_1+\bm{T}_2$ carry one chiral edge mode per spin on each edge [Fig.~\ref{fig:chiral}(f)], as the Chern number requires. Sign conventions are fixed in Appendix~\ref{app:conventions}.

The chiral edge modes give a thermal Hall response. For free spinons~\cite{SumiyoshiFujimoto2013},
\begin{equation}
\frac{\kappa_{xy}}{T}=-d_s\,\frac{3}{\pi^2}\sum_n\int_{\rm BZ}\frac{d^2k}{2\pi}\,\Omega_n(\bm{k})\,g_2\!\left(\frac{\varepsilon_n(\bm{k})-\mu}{k_BT}\right),
\label{eq:kxy}
\end{equation}
in units of $\pi k_B^2/(6\hbar)$, where $\Omega_n$ is the Berry curvature of band $n$, $g_2(x)=\int_x^\infty y^2(-\partial_yf)\,dy$, and $d_s=2$ counts the spin species. For $k_BT\lesssim0.15\Delta$ it reaches the plateau $-d_sC=+2$ [Fig.~\ref{fig:chiral}(g)]. This is a mean-field value. After projection, a gapped state with $C=\pm1$ per spin is expected to become a Kalmeyer-Laughlin spin liquid~\cite{KalmeyerLaughlin1987,WenWilczekZee1989}, with chiral central charge $c=\mp1$, half the free-spinon value, because the gauge field removes one chiral mode. For the state of Fig.~\ref{fig:chiral}, with $C=-1$, this gives $c=+1$. More generally, a $U(1)$ state with $|C|\ge2$ gives $|c|=2|C|-1$, and a $\Z_2$ state with Kitaev index $\nu$ gives $c=\nu/2$~\cite{Kitaev2006}. A class-$\mathcal{B}$ $\Z_2$ state with $\nu=+4$ therefore has $c=+2$, whereas an $SU(2)$ or $U(1)$ state with the same $\nu$, that is $C=-1$, has $c=+1$: the two describe different topological orders.

\section{Spinon Spectra}
\label{sec:spectra}

We now look at the spinon bands of the realized states, with the mean-field amplitudes treated as free parameters. At half filling a state can be gapped, have Dirac nodes or nodal lines, have Fermi pockets, or, on this lattice, have exactly flat bands at the Fermi level. Except for the flat bands of Sec.~\ref{sec:flatband} and nine $U(1)$ classes with forced zero modes (Sec.~\ref{sec:qsh}), the PSG alone does not fix which of these occurs.

Spinon momenta refer to the canonical gauge of Eq.~\eqref{eq:z2psg}. A different choice of gauge shifts them by $(\pi,0)$, $(0,\pi)$, or $(\pi,\pi)$ for $\Z_2$ states, and by any amount for $U(1)$ and $SU(2)$ states, without changing spectral types, gaps, or Chern numbers.

To compare all states on an equal footing, we evaluate each ansatz at a fixed point in its amplitude space, denoted as \textit{Point A}. At this point, all singlet amplitudes allowed up to $J_+$ are included, with magnitudes set to $1$ ($J_1$), $0.9$ ($J_{23}$), $0.5$ ($J_d$), $0.45$ ($J_{bp}$), $0.4$ ($J_{sb}$, $J_{bb}$), $0.35$ ($J_{bc}$), $0.3$ ($J_{o3}$, $J_\times$), and $0.25$ ($J_t$, $J_{bs}$, $J_+$). Their directions are chosen randomly within the corresponding symmetry-allowed spaces, and the Lagrange multipliers are determined self-consistently. Including all symmetry-allowed bonds helps avoid accidental nodal lines that may arise when certain amplitudes are omitted. \textit{Point B} uses the same amplitude magnitudes but different random directions.

\subsection{Nearest-Neighbor $SU(2)$ States}

\begin{figure*}
\centering
\includegraphics[width=\textwidth]{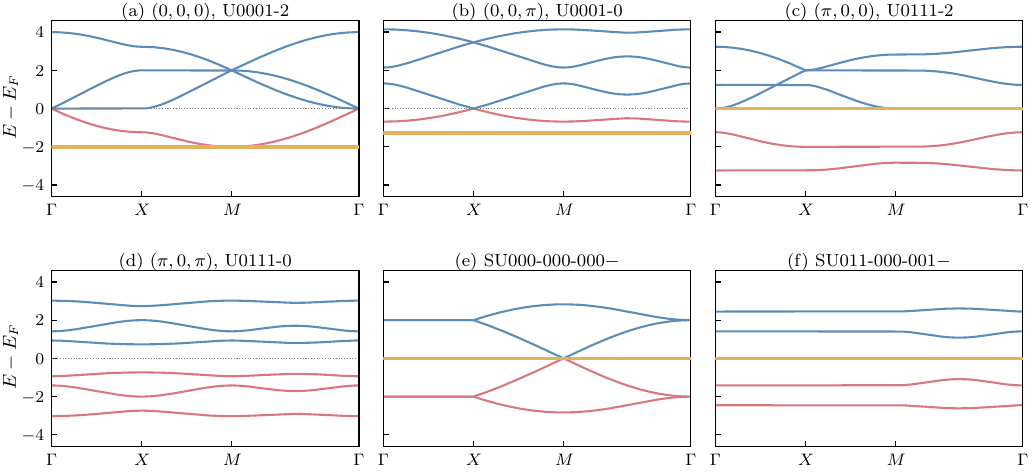}
\caption{Spinon bands per spin of the nearest-neighbor symmetric states along $\Gamma\to X\to M\to\Gamma$, measured from the Fermi level at half filling. (a)--(d) The four TRS flux states of Table~\ref{tab:flux}, with real hopping of unit magnitude on $J_1$ and $J_{23}$ and zero on-site multipliers. In (a) the highest occupied band lies at $E=0$ along the lines $k_1=0$ and $k_2=0$, and hence along $\Gamma\to X$. (e), (f) The $SU(2)$ states SU000-000-000$-$ and SU011-000-001$-$, Eqs.~\eqref{eq:su2a} and \eqref{eq:su2b}, with unit hopping on the bow-tie edges. Occupied bands are rose, empty bands blue, and bands that are flat at these amplitudes amber.}
\label{fig:nn-spectra}
\end{figure*}

The two nearest-neighbor $SU(2)$ states with bonds on the bow-tie edges have simple closed-form bands. With hopping $t_b$, both have two flat bands at $E=0$. The other bands are
\begin{equation}
E^2=4t_b^2\pm2t_b^2\left[(1-\cos k_1)(1-\cos k_2)\right]^{1/2}
\label{eq:su2a}
\end{equation}
for SU000-000-000$-$, which touch the flat bands at $M$, and
\begin{equation}
E^2=4t_b^2\pm2t_b^2\left(1+\sin k_1\sin k_2\right)^{1/2}
\label{eq:su2b}
\end{equation}
for SU011-000-001$-$, which stay at least $|t_b|(4-2\sqrt2)^{1/2}\approx1.08|t_b|$ away from them [Fig.~\ref{fig:nn-spectra}(e), (f)]. Here $k_i=\bm{k}\cdot\bm{T}_i$.

\subsection{Nearest-Neighbor $U(1)$ States}
\label{sec:nn-u1}

Figure~\ref{fig:nn-spectra}(a)--(d) shows the band structures of the four TRS flux states listed in Table~\ref{tab:flux}. Two of these states exhibit a flat band for arbitrary ratios of the $J_1$ and $J_{23}$ hopping amplitudes. For the $(0,0,0)$ state, the flat band occurs at energy $\pm 2|t_1|$, where $t_1$ denotes the hopping amplitude associated with $J_1$, consistent with the flat band found in the corresponding tight-binding model~\cite{MizoguchiKunoHatsugai2021}. In contrast, for the $(\pi,0,0)$ state, the flat band lies at the Fermi energy and is entirely localized on the square-lattice sites. Including further-neighbor hopping amplitudes disperses this flat band, giving rise to Fermi pockets. The $(0,0,\pi)$ state, on the other hand, exhibits two Dirac nodes per spin.

At zero Lagrange multipliers, these states do not satisfy the on-site constraint exactly. For the $(0,0,0)$ and $(0,0,\pi)$ states, the square sites accommodate $1.888$ and $2.106$ fermions per spin per unit cell, respectively, rather than the required value of $2$. This imbalance between the two site orbits necessitates distinct Lagrange multipliers for the two orbits. In the gapped $(\pi,0,\pi)$ state, the two bow-tie sites within a unit cell accommodate $0.33$ and $0.67$ fermions per spin, respectively. Enforcing the on-site constraint therefore requires different Lagrange multipliers on these sites, a distinction permitted by the PSG.

\subsection{Spectral Types}
\label{sec:types}

\begin{table}
\caption{Spectral types at half filling of the realized TRS spin liquids at \textit{Point A} (Sec.~\ref{sec:spectra}). The types are gapped (G), Dirac nodes (D), Dirac nodes together with quadratic band touchings (M), an isolated quadratic band touching (Q), nodal lines (NL), Fermi pockets (FS), and exactly flat bands at $E_F$ (F, Sec.~\ref{sec:flatband}), the only type that holds in the whole amplitude space of the PSG.}
\label{tab:types}
\begin{ruledtabular}
\begin{tabular}{lcccccccc}
IGG & states & G & D & M & Q & NL & FS & F\\
\hline
$\Z_2$ & 160 & 11 & 27 & 4 & 0 & 71 & 7 & 40\\
$U(1)$ & 57 & 1 & 3 & 1 & 0 & 18 & 9 & 25\\
$SU(2)$ & 4 & 0 & 0 & 0 & 0 & 0 & 0 & 4
\end{tabular}
\end{ruledtabular}
\end{table}

Table~\ref{tab:types} lists the spectral types of all $221$ realized TRS states at \textit{Point A}. Apart from the flat-band states, nodal lines are the most common type, and only $12$ states are gapped. The Dirac states have $N_f=4$ to $36$ Dirac cones in the Brillouin zone, counting both spin species, for $\Z_2$ states, and $N_f=16$ or $20$ for $U(1)$ states. The large values come from nodes at general momenta, which the point group copies up to eight times, and from enlarged unit cells. The types are fairly robust. At \textit{Point B} we rechecked the $\Z_2$ and $U(1)$ states that are gapped or have point nodes at \textit{Point A}, together with one $\Z_2$ state with small nodal loops. Of these, $34$ of the $43$ $\Z_2$ states and $4$ of the $5$ $U(1)$ states keep their type. Amplitudes, spectra, and densities of states of all states are given in the SM.

Outside the $\zeta=-$ PSGs, flat bands at the Fermi level need fine-tuned amplitudes. With nearest-neighbor bonds only, $78$ further TRS states have them, among them the $(\pi,0,0)$ state U0111-2. The chiral state $\mathcal{B}$:SU000-000, which is not counted among them, is another example. At \textit{Point A}, no state outside the $\zeta=-$ PSGs keeps a flat band at the Fermi level.

Spin structure factors of all states, computed from the mean-field two-spinon bubble with the actual site positions, are given in Sec.~\ref{sm:sec:sf} of the SM, with representative examples in Fig.~\ref{sm:fig:sq}. Apart from the flat-band states, whose structure factor takes amplitude-independent values in the grand-canonical $T\to0^+$ mean-field ensemble, in which the zero modes are half filled (a state at fixed particle number or a projected state need not share these values), they show broad maxima of similar height, $0.15$ to $0.19$, for all spectral types, so that the static structure factor alone does not distinguish them.

\section{Spin--Orbit Coupling}
\label{sec:soc}

Spin--orbit coupling allows spin-dependent spinon hopping and pairing, the triplet links of Eq.~\eqref{eq:links}. Their allowed directions on each bond are fixed by the PSG through Eq.~\eqref{eq:psg-spatial}. This section states two symmetry results for the mean-field spinon Hamiltonian and the topological spinon bands they allow, and Sec.~\ref{sec:soc-beyond} discusses what happens beyond mean-field theory. Symmetry classes are given in Appendix~\ref{app:soc}, and the full survey and further examples in \SMtext{Sec.~\ref{sm:sec:soc} of the SM}{the SM}.

\subsection{Projective Moriya Rule}
\label{sec:moriya}

Moriya's rules~\cite{Moriya1960} determine the allowed directions of a Dzyaloshinskii--Moriya (DM) vector from the symmetries that leave a bond invariant, using the spin rotations $R_g$ of Appendix~\ref{app:rotations}. For spinon links, each of these symmetries also carries a PSG sign. To see its effect, regard the singlet link $q^0_{ij}=-\ii u^{(0)}_{ij}$ and the triplet links $q^a_{ij}=u^{(a)}_{ij}$ as real quaternions, and form the vector $v^a_{ij}=\langle q^0_{ij},q^a_{ij}\rangle$. The PSG sign multiplies both links, so it cancels in $v^a_{ij}$. This vector therefore transforms exactly like a DM vector and obeys Moriya's rules on every bond, in every PSG.

Apparent violations can occur only on bonds where the PSG forbids the singlet hopping, so that $v^a_{ij}$ vanishes. There the triplet links can point in directions that Moriya's rules would not allow. In the class U0001-0, for example, the bonds $J_d$, $J_{bs}$, $J_\times$, and $J_+$ carry no singlet hopping and an out-of-plane DM-like vector, where Moriya's rule requires an in-plane one (Appendix~\ref{app:soc}). This concerns the spinon ansatz only, since the DM vectors of the physical spin Hamiltonian always obey Moriya's rules.

\subsection{Spinon Quantum Spin Hall States}
\label{sec:qsh}

With spin--orbit coupling, a gapped spinon band structure can be a quantum spin Hall insulator, with a nontrivial $\Z_2$ invariant ($\Z_2$-odd) and helical edge modes. Among the $U(1)$ and $SU(2)$ PSGs, this is possible only in the $32$ $U(1)$ PSGs with $w_\T=1$. These conserve spinon number and have $\T^2=-1$ on the spinons, which places them in symmetry class AII~\cite{AltlandZirnbauer1997,Schnyder2008} (Appendix~\ref{app:az}). The $\Z_2$ PSGs with spin--orbit coupling belong to class DIII and are not studied here.

Two rules restrict where such states can occur.
\begin{itemize}
\item \textit{Charge conjugation.} If spin--orbit coupling conserves $S^z$, the spin Chern number $C_\uparrow$ vanishes whenever a symmetry exchanges spinon particles and holes ($w_T$, $w_{C_4}$, or $w_\sigma$ equal to $1$). Such a symmetry maps the occupied bands of one spin species onto the empty bands of a spin species, which reverses the Chern number twice, through the exchange of occupied and empty bands and through complex conjugation. A translation or $C_4$, which keeps $S^z$, maps spin up onto spin down and gives $C_\downarrow=C_\uparrow$, which together with $C_\downarrow=-C_\uparrow$ from time reversal forces $C_\uparrow=0$. The mirror $\sigma$, whose spin rotation reverses $S^z$, maps spin up onto itself and reverses the orientation of the lattice, which gives $C_\uparrow=-C_\uparrow$ directly. With conserved $S^z$, spinon quantum spin Hall states are therefore possible only in the four classes U0001-$n$.
\item \textit{Forced zero modes.} Some states cannot be gapped at all: the flat-band PSGs, whose zero modes (Sec.~\ref{sec:flatband}) survive any allowed spin--orbit coupling, and nine further $U(1)$ classes in which every ansatz has exact zero modes at a time-reversal-invariant momentum (Appendix~\ref{app:az}).
\end{itemize}

\begin{figure*}
\centering
\includegraphics[width=\textwidth]{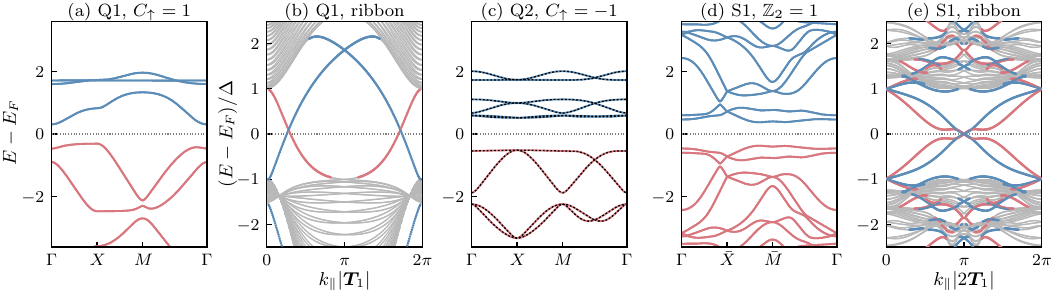}
\caption{Mean-field spinon quantum spin Hall states (\SMtext{Table~\ref{sm:tab:qshshow} of the SM}{SM}), with occupied bands rose and empty bands blue. (a) Bands of state Q1 (U0001-2) with uniform hopping on $J_1$ and $J_{23}$ and an out-of-plane DM amplitude $0.3$ on $J_1$, along $\Gamma\to X\to M\to\Gamma$. (b) Its ribbon along $\bm{T}_1$, $24$ cells wide, in units of the single-spinon gap $\Delta$, half the band gap. States with more than half of their weight in the two cells next to the lower (upper) edge are rose (blue). (c) Bands of state Q2 (U0001-0) with an out-of-plane DM amplitude $0.3$ on $J_{23}$ (solid), and the same bands shifted by $(\pi,\pi)$ (dotted). (d) Bands of state S1 (U1101-3, staggered gauge sector) with generic spin--orbit coupling on all orbits up to $J_+$, on the $2\times2$ magnetic cell ($\bar X$, $\bar M$: points of its zone). (e) Its ribbon along $2\bm{T}_1$, ten supercells wide, in units of $\Delta$, colored as in (b).}
\label{fig:qsh}
\end{figure*}

Gapped $\Z_2$-odd states do occur (Fig.~\ref{fig:qsh}). State Q1 belongs to the class U0001-2. Its singlet parent, uniform hopping on $J_1$ and $J_{23}$, is the flux state $(0,0,0)$ of Table~\ref{tab:flux}. With the multipliers fixed by $\langle n_i\rangle=1$, its highest occupied band lies at the Fermi level along the lines $k_1=0$ and $k_2=0$, and the lowest empty band touches it quadratically at $\Gamma$, where the lines cross. An out-of-plane DM term on the square edges gaps both, linear in its strength, and gives one helical pair per edge [Fig.~\ref{fig:qsh}(a), (b)]. State Q2, in the class U0001-0, is gapped in the same way by a DM term on the bow-tie edges. Its bands satisfy $E(\bm{k})=E(\bm{k}+(\pi,\pi))$ [Fig.~\ref{fig:qsh}(c)], a direct signature of its projective PSG (Appendix~\ref{app:az}).

With generic spin--orbit coupling, $\Z_2$-odd states also occur in the staggered gauge sectors ($w_T=1$). There the translations exchange spinon particles and holes, and a translation combined with time reversal acts as a chiral symmetry that pins the Fermi level at $E=0$. State S1, in U1101-3, has a spectrum that is symmetric under $E\to-E$ and twofold degenerate at every momentum [Fig.~\ref{fig:qsh}(d)]. On a ribbon, one Kramers pair per edge crosses $E=0$ [Fig.~\ref{fig:qsh}(e)].

A survey of all $32$ AII classes (\SMtext{Sec.~\ref{sm:sec:soc} of the SM}{the SM}) confirms these rules. With conserved $S^z$, $\Z_2$-odd states occur only in U0001-0 and U0001-2. With generic spin--orbit coupling, they also occur in the staggered gauge sectors and, rarely, in U0111-2.

\subsection{Beyond Mean-Field Theory}
\label{sec:soc-beyond}

These spinon insulators are not stable spin liquids. With conserved $S^z$, time reversal forbids a Chern--Simons term for the emergent gauge field, but the monopoles of the gauge field carry spin $S^z=C_\uparrow$, as charge is carried by skyrmions in quantum spin Hall insulators~\cite{GroverSenthil2008,RanVishwanathLee2008}. If no other gapless matter is present, the monopoles therefore cannot enter the action, the gauge field becomes the Goldstone mode of broken spin rotation about $\hat{\bm z}$, and the physical state orders. Its order parameter carries $S^z=C_\uparrow$: an in-plane magnetic moment for $|C_\uparrow|=1$, a spin-nematic for $|C_\uparrow|=2$, and $(S^+)^3$ for $|C_\uparrow|=3$. Its spatial pattern depends on the lattice quantum numbers of the monopoles, which we have not determined.

With generic spin--orbit coupling the monopoles are allowed, and the $U(1)$ gauge field confines~\cite{Polyakov1977}. A stable state then requires spinon pairing, which reduces the gauge group to $\Z_2$ and leads to class DIII~\cite{Kitaev2009}, and we do not study these states here. Even then, the protected spinon edges need not survive, because pairing can gap a helical spinon edge. If the state confines instead, the constraint of Sec.~\ref{sec:lsm}, which we expect but do not prove for time reversal alone, would forbid a featureless symmetric paramagnet, so that the confined state breaks a symmetry or is gapless.

\begin{figure*}
\centering
\includegraphics[width=0.86\textwidth]{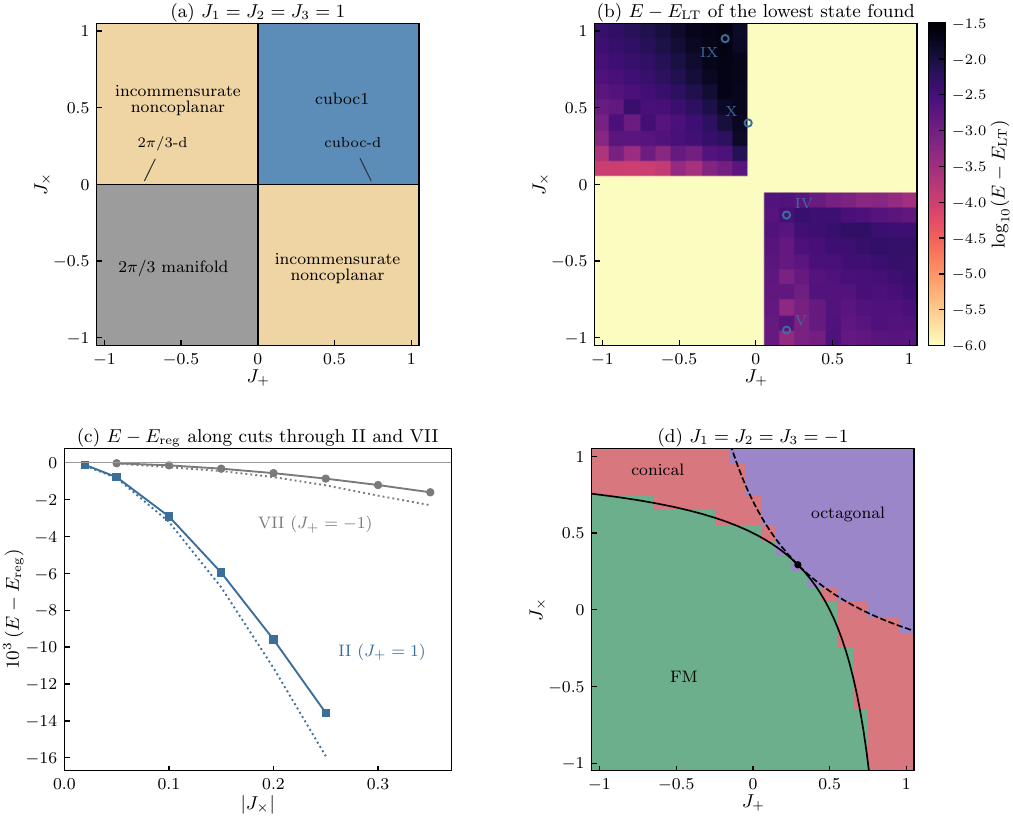}
\caption{Classical ground states of the square-kagome Heisenberg model with couplings $J_+$ (bow-tie sites across an octagon) and $J_\times$ (square sites across an octagon), unit spins, energies per site. (a) Antiferromagnetic nearest neighbors $J_1=J_2=J_3=1$, with the regular orders cuboc-d ($J_+>0$) and $2\pi/3$-d ($J_+<0$) marked on the line $J_\times=0$. Grid spacing $0.1$. (b) Difference $E-E_{\rm LT}$ between the lowest state found and the generalized LT bound (logarithmic scale). Differences below $10^{-6}$, where the bound is reached, are shown at $10^{-6}$ (pale). Circles: points at which Ref.~\cite{Gembe2023} quotes spiral energies for its phases IV, V, IX, and X. (c) Energies along the cuts $J_+=1$ (through phase II into phase III) and $J_+=-1$ (through phase VII) relative to the respective regular order: lowest state found (symbols) and LT bound (dotted). (d) Ferromagnetic nearest neighbors $J_1=J_2=J_3=-1$, with ferromagnet (FM), octagonal, and conical orders. Solid and dashed lines are the two phase boundaries given in the text.}
\label{fig:classical}
\end{figure*}

\section{Noncoplanar Orders and Chiral Spin Liquids}
\label{sec:orders}

We now return to the question: which chiral liquids can the noncoplanar orders of the cross-octagon model~\cite{Gembe2023} produce, and does the energetics favor one? We first find the classical phases (Sec.~\ref{sec:classical}) and the chiral class that each order would pass on to a liquid (Secs.~\ref{sec:regular} and \ref{sec:nematic}). We then compare candidate states in self-consistent mean-field theory (Sec.~\ref{sec:mf}) and with projected wave functions (Sec.~\ref{sec:vmc}).

\subsection{Classical Phases of the Cross-Octagon Model}
\label{sec:classical}

\begin{figure*}
\centering
\includegraphics[width=\textwidth]{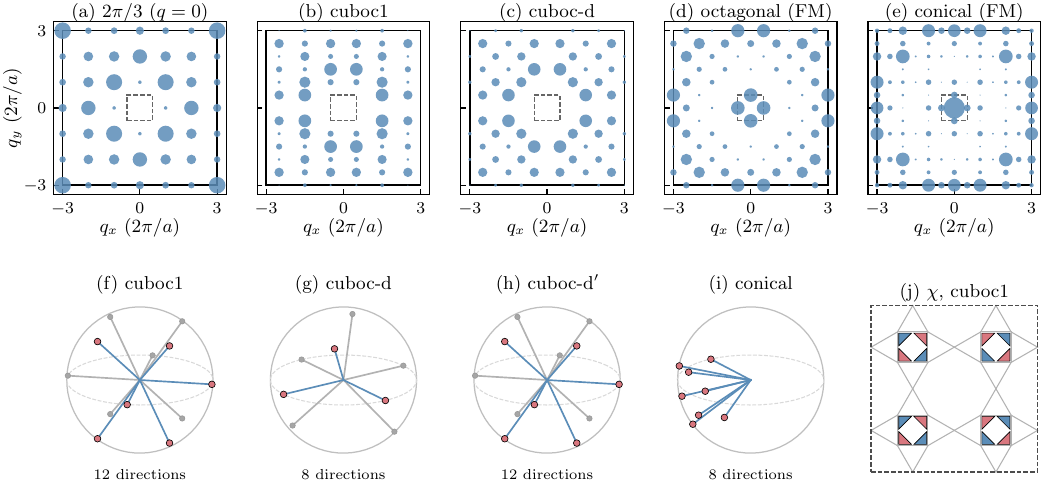}
\caption{Commensurate classical orders of the square-kagome Heisenberg model. (a)--(e) Bragg weights $w(\bm{Q})=|N^{-1}\sum_i\bm{S}_ie^{\ii\bm{Q}\cdot\bm{r}_i}|^2$ in the window $|q_x|,|q_y|\le3$ (units $2\pi/a$, solid square, with the first Brillouin zone dashed), normalized to unit total weight. Disc areas are proportional to $w$. (a) $2\pi/3$ order with $\bm{q}=0$ ($J_+=J_\times=-0.5$), (b) cuboc1 ($J_+=J_\times=0.5$), (c) cuboc-d ($J_+=0.5$, $J_\times=0$), (d) octagonal and (e) conical orders for ferromagnetic nearest neighbors ($J_+=J_\times=0.5$, and $J_+=-0.5$, $J_\times=0.8$). (f)--(i) Spin directions of the noncoplanar orders: cuboc1 ($12$ directions, the vertices of a cuboctahedron), cuboc-d ($8$), cuboc-d$'$ ($12$), and the conical order ($8$ directions on two cones). Front directions are colored, back directions gray. (j) Sign of the scalar chirality $\bm{S}_i\cdot(\bm{S}_j\times\bm{S}_k)$ of cuboc1 on the half-square triangles ($J_1$, $J_1$, $J_d$), drawn on the $2\times2$ magnetic cell (red positive, blue negative). The $J_d$ link only closes the triangle and is not a coupling of the cross-octagon model.}
\label{fig:orders}
\end{figure*}

We study the Heisenberg model with equal nearest-neighbor couplings $J_1=J_2=J_3=\pm1$ and couplings $J_+$ and $J_\times$ across the octagons~\cite{Gembe2023}. To find exact classical ground states, we use the Luttinger--Tisza (LT) method~\cite{LuttingerTisza1946,LyonsKaplan1960,KaplanMenyuk2007} with one Lagrange multiplier per site orbit. It gives a lower bound on the energy, and a spin state that reaches the bound is a ground state.

For antiferromagnetic nearest neighbors, the bound is reached by the order cuboc1 for $J_+,J_\times\ge0$, and by a degenerate family of $2\pi/3$ states, with $120^\circ$ between the three spins of every triangle, for $J_+,J_\times\le0$ [Fig.~\ref{fig:classical}(a), with the energies in Eq.~\eqref{eq:regE}]. For ferromagnetic nearest neighbors, the bound is reached everywhere, by ferromagnetic, octagonal, and conical orders, with the phase boundaries $(J_\times-1)(J_+-1)=1/2$ and $(J_\times+1/\sqrt2)(J_++1/\sqrt2)=1$ [Fig.~\ref{fig:classical}(d)]. These results agree with the closed forms of Ref.~\cite{Gembe2023}.

In the two quadrants with $J_+J_\times<0$ the bound is not reached. There, simulated annealing on large and twisted cells finds incommensurate noncoplanar states [Fig.~\ref{fig:classical}(b)]. They are the lowest states we found, but without a reached bound they carry no certificate of global optimality. This refines Ref.~\cite{Gembe2023} in two ways. First, the regular orders of its phases II and VII, which we call cuboc-d and $2\pi/3$-d, are ground states only on the line $J_\times=0$. Away from it, long-wavelength modulations of them have lower energy [Fig.~\ref{fig:classical}(c)], with periods that grow as $1/|J_\times|$ and are too long to fit on the finite clusters of Ref.~\cite{Gembe2023}. Second, the spiral states of its phases III, IV, V, VIII, IX, and X are not the lowest states: incommensurate states lie slightly lower. Details are given in \SMtext{Sec.~\ref{sm:sec:classicalsm} of the SM}{the SM}.

Phase II has two degenerate regular orders, built from the irreducible blocks $B_2+M_5$ ($8$ spin directions) and $M_4+M_5$ ($12$ directions). We call them cuboc-d and cuboc-d$'$. Figure~\ref{fig:orders} shows the commensurate orders.

\subsection{Regular Orders and Chiral Classes}
\label{sec:regular}
\label{sec:order-classes}

In a regular order, every lattice symmetry $g$ can be undone by a global spin rotation $R_g\in O(3)$~\cite{MessioLhuillierMisguich2011}. If $\det R_g=-1$, this rotation includes a spin inversion, so in a liquid that keeps the symmetry of the order, $g$ appears only together with time reversal. The order thus passes on the chiral class
\begin{equation}
\epsilon(g)=\det R_g
\label{eq:detR}
\end{equation}
of Sec.~\ref{sec:chiral-psg}. This assumes that the liquid keeps the magnetic symmetry of its parent order, a mean-field heuristic rather than a theorem. A liquid that breaks further symmetries is not constrained by Eq.~\eqref{eq:detR}.

\begin{table}[!htb]
\caption{Noncoplanar orders of the cross-octagon model~\cite{Gembe2023}, their irreducible blocks, number of spin directions, and the chiral class of a liquid that keeps their magnetic symmetry. cuboc1 and cuboc3 are regular only under the nematic subgroup $p2mm$, and $X_2^{(1)}$ denotes one arm of the $X$ star. The noncoplanar $2\pi/3$-d order is the $p4mm$-regular member of a degenerate family. None of these orders allows a Chern number.}
\label{tab:order-classes}
\begin{ruledtabular}
\begin{tabular}{lccl}
order & blocks & dir. & class\\
\hline
cuboc1, cuboc3 & $X_2^{(1)}+M_5$ & 12 & nematic, $(T_1\T,T_2)$\\
cuboc-d & $B_2+M_5$ & 8 & $\mathcal{D}$\\
cuboc-d$'$ & $M_4+M_5$ & 12 & $\mathcal{D}'$\\
conical & $A_1+X_1$ & 8 & $\mathcal{D}'$\\
$2\pi/3$-d (noncoplanar) & $A_1+X_2$ & 5 & $\mathcal{D}'$
\end{tabular}
\end{ruledtabular}
\end{table}

Table~\ref{tab:order-classes} gives the result for the orders of the cross-octagon model. None of them falls into class $\mathcal{B}$, so none allows a Chern number. Their scalar chirality alternates in space, under a translation or under the fourfold rotation, and their net chirality is zero. cuboc1 and its partner cuboc3 also break the fourfold rotation and the diagonal mirrors. Their magnetic group is the nematic group $p2mm$, in which $T_1$ is combined with time reversal [Fig.~\ref{fig:orders}(j)].\footnote{Reference~\cite{Gembe2023} remarks that its noncoplanar orders are not regular. This holds for cuboc1 and the spirals, while the cuboc-d and conical orders are regular under the full group $p4mm$.} We state the result as a selection rule: \textit{a phase that keeps the magnetic group of a noncoplanar parent order has $C=0$ and $\kappa_{xy}=0$ whenever that group contains an element $g$ with $\epsilon(g)\det(g)=-1$}, such as a translation or rotation combined with time reversal. It follows from Sec.~\ref{sec:chern} and needs only the magnetic group. Its one assumption is that the liquid inherits this group. It does not constrain a liquid that spontaneously breaks the Hall-reversing element, as a uniformly chiral liquid would.

Uniformly chiral orders of class $\mathcal{B}$ do exist. With magnetic cells of up to $2\times2$ unit cells there are $34$ families of noncoplanar regular orders, and $17$ of them have exact parent Hamiltonians with couplings up to $J_+$ (Appendix~\ref{app:regular}). Three of these are in class $\mathcal{B}$: $A_1+M_5$, $M_1+X_4$, and $M_4+X_1$. They have a uniform chirality on the nearest-neighbor triangles, but their parent Hamiltonians need ferromagnetic bow-tie couplings and sizable further-neighbor couplings, far from the antiferromagnetic regime considered here. We have not studied their quantum melting, nor searched for uniformly chiral orders that are not regular, such as incommensurate ones, so the need for ferromagnetic bow-tie couplings is established only for the regular orders.

For each order of Table~\ref{tab:order-classes} we also identified the PSGs that reproduce its chirality pattern. For cuboc-d these are class-$\mathcal{D}$ PSGs such as $\mathcal{D}$:SU011-001. For the conical order, the natural candidate $\mathcal{D}'$:U100-0 is gapped with $C=0$, as it must be.

\subsection{Nematic PSGs of the Cuboc1 Class}
\label{sec:nematic}

A liquid that keeps the symmetry of cuboc1 is invariant under $T_2$, the two axial mirrors $\sigma_x$ and $\sigma_y$, and $T_1\T$, and breaks the fourfold rotation. Since $T_1$ and $T_2$ are treated differently, this is not one of the chiral classes of $p4mm$. We therefore classified the PSGs of the smaller group $p2mm$ (\SMtext{Sec.~\ref{sm:sec:nematicsm} of the SM}{the SM}). It gives $100$ $\Z_2$, $48$ discrete $U(1)$ and one flux family, and $4$ $SU(2)$ PSGs.

To reproduce the chirality pattern of cuboc1, which sits on the half-square triangles, a state needs the square diagonals $J_d$ in addition to the nearest-neighbor bonds. This is possible in $58$ $\Z_2$ and $7$ $U(1)$ PSGs, and in no $SU(2)$ PSG. A natural candidate is N:U1000-0, which reproduces the chirality of cuboc1 on all loops up to $J_+$. With nearest-neighbor and $J_d$ amplitudes it is gapped, with $\Delta=0.156$, and has $C=0$, as $T_1\T$ requires. These fermionic states complement the nematic spin liquid found in Schwinger-boson theory of a related model~\cite{LuganJaubertRalko2019}.

\subsection{Self-Consistent Mean-Field States}
\label{sec:mf}

\begin{figure}[!htb]
\centering
\includegraphics[width=\columnwidth]{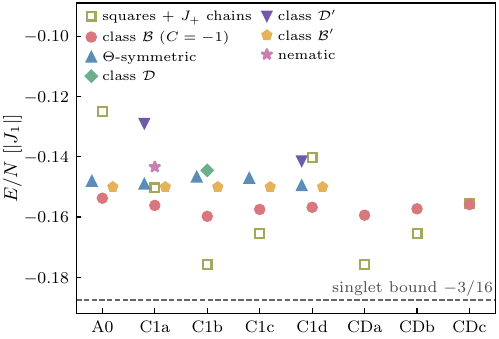}
\caption{Self-consistent mean-field energies per site of symmetric states of the model of Ref.~\cite{Gembe2023} with $J_1=J_2=J_3=1$, at the points of Table~\ref{tab:mf}: the class-$\mathcal{B}$ chiral state (circles), the decoupled state of $\pi$-flux squares and $J_+$ chains (open squares, whose bond graph is not connected), the lowest TRS state with a connected two-dimensional bond graph (triangles), the class-$\mathcal{B}'$ saddle (pentagons), and states of classes $\mathcal{D}$ and $\mathcal{D}'$ and of the nematic class (star). Symbols are offset horizontally for clarity. The dashed line is the singlet bound $-3/16$.}
\label{fig:mf}
\end{figure}

To see which state a Heisenberg model selects at mean-field level, we solved the self-consistency equations of the model of Ref.~\cite{Gembe2023} with $J_1=J_2=J_3=1$ at eight points: the nearest-neighbor point A0, four points C1a--C1d in the cuboc1 phase, and three points CDa--CDc in phase II (Table~\ref{tab:mf} and Fig.~\ref{fig:mf}). The on-site constraints are imposed by site-dependent multipliers.

\begin{table}[!htb]
\caption{Self-consistent mean-field energies per site for $J_1=J_2=J_3=1$. $E_{\mathcal{B}}$ and $\Delta_{\mathcal{B}}$ are the energy and single-spinon gap of the class-$\mathcal{B}$ state, and $E_{\rm dec}$ is the energy of the decoupled state. The last column states which of the two is lower. Energies and $\Delta_{\mathcal{B}}$ are in units of $|J_1|$. The singlet bound is $-3/16=-0.1875$ at every point.}
\label{tab:mf}
\begin{ruledtabular}
\begin{tabular}{lcccccc}
point & $J_+$ & $J_\times$ & $E_{\mathcal{B}}$ & $\Delta_{\mathcal{B}}$ & $E_{\rm dec}$ & lower\\
\hline
A0 & 0 & 0 & $-0.153741$ & 0.0899 & $-0.125000$ & $\mathcal{B}$\\
C1a & 0.5 & 0.5 & $-0.156130$ & 0.1348 & $-0.150330$ & $\mathcal{B}$\\
C1b & 1 & 1 & $-0.159759$ & 0.1285 & $-0.175661$ & dec.\\
C1c & 0.8 & 0.3 & $-0.157464$ & 0.1267 & $-0.165528$ & dec.\\
C1d & 0.3 & 0.8 & $-0.156757$ & 0.1288 & $-0.140198$ & $\mathcal{B}$\\
CDa & 1 & $-0.1$ & $-0.159371$ & 0.0946 & $-0.175661$ & dec.\\
CDb & 0.8 & $-0.075$ & $-0.157268$ & 0.1197 & $-0.165528$ & dec.\\
CDc & 0.6 & $-0.055$ & $-0.155855$ & 0.1261 & $-0.155396$ & $\mathcal{B}$
\end{tabular}
\end{ruledtabular}
\end{table}

Unprojected mean-field theory has a known bias. Each bond contributes $\langle\bm{S}_i\cdot\bm{S}_j\rangle=-\frac38(|\chi_{ij}|^2+|p_{ij}|^2)$, with $\chi_{ij}$ and $p_{ij}$ the hopping and pairing expectation values, and the lowest possible energy, $-3/16$ per site at all eight points, is reached by isolated dimers and four-site clusters. These break the lattice symmetries and say nothing about the classical phases. We therefore compare only symmetric states whose bonds connect the whole lattice, each solved within its PSG. This comparison diagnoses the PSG manifolds, whereas the physical ranking requires projection (Sec.~\ref{sec:vmc}). At C1a--C1d we solved all $SU(2)$ and $U(1)$ PSGs and a sample of $\Z_2$ PSGs in each class, and at the other points fewer candidates. All solved manifolds and their energies are in the SM.

The self-consistent solutions do not inherit the class of the melted cuboc1 order. Among the connected symmetric states, the lowest at A0 and C1a--C1d is the class-$\mathcal{B}$ state $\mathcal{B}$:SU011-001 of Sec.~\ref{sec:kl}, with its flux pattern, $SU(2)$ gauge group, and Chern number unchanged. It is also the lowest of the fewer candidates at CDa--CDc. The self-consistency map flows to it from many starting points, but it is a saddle point: one soft distortion, which breaks the fourfold rotation, lowers the energy towards a dimer covering. The closest connected competitor is a gapless class-$\mathcal{B}'$ state, lying higher in energy by $0.004$–$0.010$. The best TRS states are higher by $0.006$–$0.013$, and the states of the nematic class and of classes $\mathcal{D}$ and $\mathcal{D}'$ lie higher still (Sec.~\ref{sm:sec:mfscans} of the SM). At the nearest-neighbor point, this ranking agrees with Ref.~\cite{Astrakhantsev2021}.

At four of the eight points, a state that is not two-dimensional lies lower: $\pi$-flux squares and decoupled chains of bow-tie sites along the $J_+$ bonds, with energy $E/N=-1/8-J_+/(2\pi^2)$ (Table~\ref{tab:mf}). We list it separately. The mean-field structure factor of the class-$\mathcal{B}$ state at C1b (\SMtext{Fig.~\ref{sm:fig:mfsq} of the SM}{the SM}) has its maxima close to the Bragg peaks of cuboc1, so the static structure factor alone cannot distinguish the chiral liquid from short-range cuboc1 correlations.

These are unprojected results. Already at the nearest-neighbor point, projection places three nonchiral flux states and two valence-bond crystals below the class-$\mathcal{B}$ state~\cite{Astrakhantsev2021}. For the ferromagnetic regimes of cuboc3 and the conical order, singlet mean-field theory is not meaningful.

\subsection{Projected Wave Functions}
\label{sec:vmc}

\begin{figure}[!htb]
\centering
\includegraphics[width=\columnwidth]{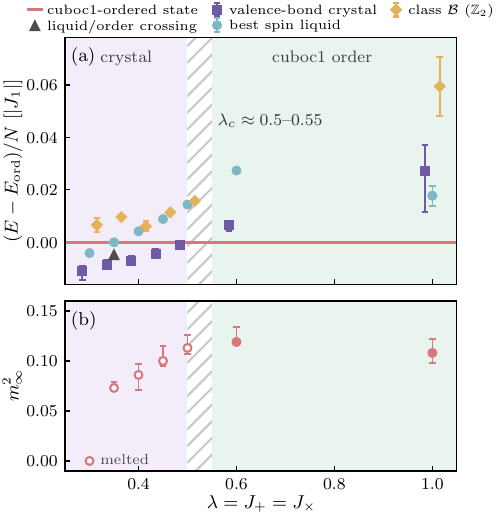}
\caption{Gutzwiller-projected wave functions along $J_+=J_\times=\lambda$ ($J_1=J_2=J_3=1$), extrapolated to $N\to\infty$. (a) Energies per site relative to the cuboc1-ordered state (line): valence-bond crystal (squares), best spin liquid (circles), and chiral $\Z_2$ state of class $\mathcal{B}$ (diamonds). Error bars for the crystal indicate the spread among different fitting forms, while those for the other states represent fit uncertainties, and neither includes the scatter between local optima of the optimization (\SMtext{Sec.~\ref{sm:sec:vmcsm} of the SM}{the SM}). Shading marks the lowest state, and the hatched band is the estimate of the transition range within these variational spaces, $\lambda_c\approx0.5$--$0.55$. (b) Squared cuboc1 order parameter $m^2_\infty$ of the ordered state. Open symbols mark couplings at which it is not the lowest state, and at $\lambda=0.3$ the order has melted.}
\label{fig:vmc}
\end{figure}

\begin{table}
\caption{Extrapolated projected energies per site along $J_+=J_\times=\lambda$, relative to the cuboc1-ordered state: valence-bond crystal (VBC), best spin liquid (SL), and the $\Z_2$ state of class $\mathcal{B}$. The best spin liquid is Z000-011-100(1) for $\lambda\le0.4$ and at C1b, N:Z00000-101 at $\lambda=0.45$ and C1a, and the zero-field ordered ansatz at $\lambda=0.6$. The class-$\mathcal{B}$ $SU(2)$ state lies $0.0318(3)$ (C1a) and $0.062(2)$ (C1b) above the ordered state. At $\lambda=0.3$ the ordered state has melted. Differences on the largest common cluster are given in \SMtext{Table~\ref{sm:tab:vmclargest} of the SM}{the SM}.}
\label{tab:vmc}
\begin{ruledtabular}
\begin{tabular}{lccc}
$\lambda$ & $E_{\rm VBC}-E_{\rm ord}$ & $E_{\rm SL}-E_{\rm ord}$ & $E^{\Z_2}_{\mathcal{B}}-E_{\rm ord}$\\
\hline
$0.30$ & $-0.0108(13)$ & $-0.0041(3)$ & $+0.0066(26)$\\
$0.35$ & $-0.0085(3)$ & $0.0000(2)$ & $+0.0096(6)$\\
$0.40$ & $-0.0070(5)$ & $+0.0042(2)$ & $+0.0061(19)$\\
$0.45$ & $-0.0044(6)$ & $+0.0089(4)$ & $+0.0114(5)$\\
$0.50$ (C1a) & $-0.0010(3)$ & $+0.0144(6)$ & $+0.0158(2)$\\
$0.60$ & $+0.0065(4)$ & $+0.0274(6)$ & --\\
$1.00$ (C1b) & $+0.0273(18)$ & $+0.0178(38)$ & $+0.060(11)$
\end{tabular}
\end{ruledtabular}
\end{table}

\textit{Method.} To rank the physical spin-$1/2$ states, we compared Gutzwiller-projected wave functions~\cite{Gutzwiller1963,Gros1989} along the cuboc1 line $J_+=J_\times=\lambda$, which passes through C1a ($\lambda=0.5$) and C1b ($\lambda=1$). The wave functions are projected Slater determinants, optimized by stochastic reconfiguration~\cite{Sorella1998,Sorella2005,BeccaSorella2017}. We compare:
\begin{itemize}
\item the PSG spin liquids of Sec.~\ref{sec:mf}: the class-$\mathcal{B}$ $SU(2)$ and $\Z_2$ states, the TRS state Z000-011-100(1), and two states of the nematic group, N:Z00000-101 (TRS) and N:Z00000-110 (the nematic chiral state),
\item a cuboc1-ordered state, obtained by adding a field $-h\,\bm{n}_i\cdot\bm{\sigma}$ along the classical cuboc1 directions $\bm{n}_i$,
\item a valence-bond crystal with free hopping on all bonds of the $2\times2$ magnetic cell, and
\item a combination of the crystal and the cuboc1 field.
\end{itemize}
Energies are averaged over boundary twists~\cite{LinZongCeperley2001}, computed on clusters of $96$ to $600$ sites, and extrapolated in $1/N$. For the crystal and the $\Z_2$ states these fits have large $\chi^2$ per degree of freedom, because their energies scatter between clusters beyond the statistical errors. We therefore base the ranking primarily on the differences on the largest common cluster (\SMtext{Table~\ref{sm:tab:vmclargest} of the SM}{the SM}) and treat the extrapolated values of Table~\ref{tab:vmc} as secondary, and both give the same ranking. The quoted errors are statistical. The scatter between local optima of the optimization, which is one-sided because the lowest optimum is kept, and the spread over fit forms are discussed in \SMtext{Sec.~\ref{sm:sec:vmcsm} of the SM}{the SM}. The variational families also differ in flexibility: the crystal has $63$ free parameters, the PSG states $3$ to $11$, the nematic states $18$ and $26$, and the ordered state $4$, and only the ordered state receives a Jastrow factor (\SMtext{Table~\ref{sm:tab:vmcpar} of the SM}{the SM}). The comparison ranks these families and does not determine the exact phase diagram. Two independent codes agree, and the projected energies of Ref.~\cite{Astrakhantsev2021} at the nearest-neighbor point are reproduced. Details are given in \SMtext{Sec.~\ref{sm:sec:vmcsm} of the SM}{the SM}.

\textit{Along the cuboc1 line} (Table~\ref{tab:vmc}, Fig.~\ref{fig:vmc}). For $\lambda\ge0.4$ the cuboc1-ordered state lies below every spin liquid, and its order survives in the thermodynamic limit, with $m^2=0.07$ to $0.13$ (the classical value is $1/4$). For $\lambda\le0.3$ the order melts during optimization. At weaker couplings a valence-bond crystal is lowest. It has a pinwheel pattern, with strong dimers around half of the squares and resonating squares on the other half (\SMtext{Fig.~\ref{sm:fig:vbc} of the SM}{the SM}). Within these variational spaces, the crystal and the ordered state cross near $\lambda_c\approx0.5$, between $0.5$ and $0.55$. Near the crossing, independent starts of the crystal optimization agree within $1.5\times10^{-4}$ on $216$ sites, but the location of the crossing depends on further optimization and on correlation factors. A Jastrow factor for the ordered state lowers its energy by up to $2\times10^{-3}$ per site and makes the two energies equal at C1a. Since both energies are variational upper bounds, $\lambda_c$ can move either way: a better crystal would move it up, and an $SU(2)$-invariant Jastrow factor, which usually favors order, would move it down.

No chiral state is lowest at any $\lambda$. The class-$\mathcal{B}$ states lie above the ordered state at every $\lambda$ computed. The $\Z_2$ state of class $\mathcal{B}$ comes within $0.003$ of the best spin liquid for $0.4\le\lambda\le0.5$, but only by giving up its chirality. Its optimized amplitudes carry a weak time-reversal-odd flux, $|\sin\Phi_\triangle|\le0.1$ on the triangles, compared with $1$ in the $SU(2)$ state, and its projected scalar chirality falls with the cluster size, at $\lambda=0.5$ from $0.053$ on $96$ sites to $0.011$ on $384$ sites (\SMtext{Table~\ref{sm:tab:vmcchi} of the SM}{the SM}). Its energy is therefore no evidence for a nearby chiral phase. The $SU(2)$ state keeps its chirality but lies $0.03$ to $0.06$ above the ordered state. The nematic chiral state N:Z00000-110, the liquid that would inherit the symmetry of cuboc1, lies $0.025$ to $0.05$ above the best spin liquid. This differs from the kagome lattice, where variational and density-matrix renormalization group studies find uniformly chiral liquids near the cuboc orders~\cite{BieriMessioBernuLhuillier2015,BieriLhuillierMessio2016,HuZhuZhangGongBeccaSheng2015,GongZhuBalentsSheng2015,WietekSterdyniakLauchli2015}. At C1b the $\Z_2$ energies depend strongly on the boundary twist, and the ranking of the spin liquids there is not settled.

\textit{Across the phase boundary.} On the kagome lattice, chiral liquids appear where classical phases meet~\cite{WietekSterdyniakLauchli2015}. We therefore also followed a cut at $J_+=1$ from $J_\times=0.5$ to $-0.1$, into phase II, on clusters of up to $384$ sites. The cuboc1 order does not melt along this cut, but it is overtaken by nonchiral states. On $384$ sites a valence-bond crystal is lowest at $J_\times=0.5$, $0.006$ below the ordered state, and for $J_\times\le0.25$ the TRS state Z000-011-100(1), $0.0013$ to $0.0019$ below the crystal, a margin too small to establish a spin-liquid phase there. Both are close to a product of singlet squares and Heisenberg chains along $J_+$, with $\langle\bm{S}_i\cdot\bm{S}_j\rangle\approx-0.47$ on the square edges, $-0.39$ to $-0.41$ on the $J_+$ bonds, and $-0.04$ to $-0.05$ on the bow-tie edges. In that product state the bonds between squares and chains and the $J_\times$ bonds carry no energy, so it has the energy
\begin{equation}
E_{\rm prod}/N=-\big[2+2J_+(\ln2-\tfrac14)\big]/6
\label{eq:eprod}
\end{equation}
for any $J_\times$, $-0.4810$ at $J_+=1$, an upper bound on the ground-state energy that does not depend on our wave functions. The lowest projected states lie below it, at $E/N=-0.506$ to $-0.508$ on $384$ sites, $0.025$ to $0.027$ lower, whereas the ordered state lies above it for $J_\times\le0.25$. Along the line $J_+=J_\times$ the lowest states lie $0.05$ to $0.09$ below $E_{\rm prod}$ (\SMtext{Table~\ref{sm:tab:vmclargest} of the SM}{the SM}). The class-$\mathcal{B}$ $SU(2)$ state falls below the ordered state for $J_\times\le0.25$ but stays $0.015$ to $0.017$ above the lowest state, and the nematic and class-$\mathcal{D}$ chiral states lose their chirality during optimization.

\textit{Through the nearest-neighbor point.} Continued to $\lambda=0.2$, $0.1$, $-0.1$, $-0.2$, and $-0.3$, into the quadrant of the $2\pi/3$ manifold, the line $J_+=J_\times=\lambda$ has the crystal lowest at every point on clusters of up to $384$ sites, with the TRS state Z000-011-100(1) $0.001$ to $0.004$ and the best chiral state, the $\Z_2$ state of class $\mathcal{B}$, $0.014$ to $0.018$ above it. The cuboc1 and coplanar $2\pi/3$ orders melt near $\lambda=0$, and the $2\pi/3$ order that survives at $\lambda=-0.3$ lies $0.009$ above the crystal.

\textit{Limitations.} These comparisons are variational and limited to the states considered. Unbiased methods are not available here: the smallest cluster that fits the $2\times2$ cells has $24$ sites and maps the $J_+$ bonds onto each other. Crystals on larger cells, other starting patterns, $SU(2)$-invariant Jastrow factors, and coexistence of dimerization and order could shift the boundaries.

\section{Stability and the Lieb--Schultz--Mattis-Type Constraint}
\label{sec:lsm}

We end with what symmetry says about the physical state. With six spins per unit cell, the Lieb--Schultz--Mattis theorem~\cite{LiebSchultzMattis1961,Oshikawa2000,Hastings2004} gives no constraint on the square-kagome lattice, but lattice homotopy~\cite{PoWatanabeJianZaletel2017} does. Assume spin-rotation symmetry, or just its $\pi$ rotations. Each bow-tie site is a twofold rotation center carrying a spin $1/2$, and this spin cannot be moved away without breaking a symmetry. The four square sites of a cell, by contrast, can be moved along their mirror lines to the square center, where together they carry integer spin. The lattice is therefore equivalent to the checkerboard lattice, and it has no symmetric, gapped ground state without topological order. This agrees with the valence-bond crystals found numerically, which break lattice symmetries~\cite{Rousochatzakis2013,Schmoll2023,JahromiIqbal2026}. With time reversal alone, as for generic spin--orbit coupling, each bow-tie spin is a Kramers doublet, and similar constraints are expected~\cite{WatanabePoVishwanathZaletel2015,PoWatanabeJianZaletel2017}, although we do not prove them here.

This constraint acts on the projected state. It combines with the stability of the mean-field states as follows. Gapped $U(1)$ and $SU(2)$ states without a Chern--Simons term confine in two dimensions~\cite{Polyakov1977}. Stable spin liquids among our states are therefore (i) the gapped $\Z_2$ states, (ii) the chiral states with $C\neq0$, such as the class-$\mathcal{B}$ state, whose Chern--Simons term gaps the gauge field, and (iii) the Dirac spin liquids, if their fermion bilinears and monopoles are harmless, which we do not check~\cite{Hermele2004,Song2019,Song2020}. Only (i) and (ii) are gapped, symmetric, and compatible with the constraint. A gapped $U(1)$ state with $C=0$ must, when it confines, either break a lattice symmetry or close its gap. This holds for the gapped TRS $U(1)$ states of Sec.~\ref{sec:types}, the gapped $C=0$ states of Sec.~\ref{sec:mf} outside the primed classes, and the nematic candidate N:U1000-0 of Sec.~\ref{sec:nematic}.

The argument relies only on the unitarity of the twofold rotations about the bow-tie sites. The rotation $C_4^2$ about the square center is unitary in every class (Sec.~\ref{sec:chern}), but the bow-tie rotations $T_1C_4^2$ and $T_2C_4^2$ carry the character $\epsilon_T$. In the nematic class of cuboc1, $\epsilon(T_1)=-1$, so the rotation about the sites $\mu=5$ is combined with time reversal. The sites $\mu=6$ still suffice: they are centers of the unitary rotation $T_2C_4^2$ of the subgroup $\{T_1^2,T_2,\sigma_x,\sigma_y\}$, whose unit cell is doubled, and each carries a spin $1/2$. In the other nematic domain the sites $5$ and $6$ swap roles. In the staggered classes $\mathcal{A}'$--$\mathcal{D}'$ both bow-tie rotations are combined with time reversal, and the argument does not apply in this form.

\section{Conclusions and Outlook}
\label{sec:conclusion}

Both site orbits of the square-kagome lattice have nontrivial site symmetry, and this shapes its fermionic spin liquids. The site symmetries reduce the cohomology classes to the algebraic PSGs of Sec.~\ref{sec:algebraic}. The relative sign $\zeta$ of time reversal on the two orbits decides which PSGs are realizable. For $\zeta=+$ no amplitude survives. For $\zeta=-$ only square--bow-tie bonds survive, and the spinon bands are exactly flat at the Fermi level (Sec.~\ref{sec:flatband}). These flat bands are the zero modes of the bipartite square--bow-tie graph, which has twice as many square sites as bow-tie sites~\cite{Sutherland1986,Lieb1989}. Two further features make the chiral states richer than on the kagome lattice. The lattice contains triangles, so chiral states appear already with nearest-neighbor bonds. Its translations can be combined with time reversal, which gives staggered chiral classes that do not exist with sixfold symmetry~\cite{MessioLhuillierMisguich2013}.

Only class $\mathcal{B}$ allows a Chern number. Its nearest-neighbor $SU(2)$ state is gapped with $C=\pm1$, a mean-field Kalmeyer--Laughlin state (Sec.~\ref{sec:kl}). The cuboc and conical orders of the cross-octagon model fall into other classes. A liquid that keeps the magnetic symmetry of such an order therefore has a chirality that alternates in space, and $C=0$. This rule does not forbid chiral liquids; it forbids only a spontaneous thermal Hall response in a liquid inherited from these orders. The energetics also does not favor a chiral liquid. Unprojected mean-field theory ranks the class-$\mathcal{B}$ state lowest among the symmetric two-dimensional states we solved, but projection reverses this. Among the projected states we compared, the cuboc1 order is lowest for strong cross-octagon couplings and a pinwheel valence-bond crystal for weaker ones along $J_+=J_\times$. Across the cuboc1 phase boundary a TRS $\Z_2$ state and the crystal, both close to decoupled squares and chains and $0.025$ below the energy of that product state, are lowest within $0.002$ of each other. The class-$\mathcal{B}$ states and the nematic chiral state lie higher at every coupling we computed, including a cut across the cuboc1 phase boundary and the continuation of the line $J_+=J_\times$ through the nearest-neighbor point (Sec.~\ref{sec:vmc}). This differs from the kagome lattice, where uniformly chiral liquids are found near the cuboc orders~\cite{BieriMessioBernuLhuillier2015,BieriLhuillierMessio2016,HuZhuZhangGongBeccaSheng2015,GongZhuBalentsSheng2015,WietekSterdyniakLauchli2015}.

Uniformly chiral class-$\mathcal{B}$ regular orders do exist on the square-kagome lattice, but those we found need ferromagnetic bow-tie couplings and further-neighbor couplings beyond those of Ref.~\cite{Gembe2023}, and nonregular ones remain to be explored. A spontaneous zero-field thermal Hall signal from spinons would therefore point to a uniformly chiral liquid, not to a melted cuboc or conical order. In experiments, such a signal must be separated from field-induced magnon and phonon contributions~\cite{Watanabe2016Volborthite,Doki2018,Grissonnanche2019} and from averaging over chiral domains.

The flat bands of the $\zeta=-$ states and the symmetry selection rules hold for all amplitudes. Spectral types, Dirac cone numbers, and Chern numbers were determined at chosen amplitudes, and mapping them across the amplitude space is a natural next step. The self-consistent treatment can be extended to all $\Z_2$ PSGs, to magnetic cells larger than $2\times2$, and to magnetically ordered mean-field states, and the projected comparison to Jastrow factors for the spin liquids as well as for the ordered state. Structure factors, edge states, and thermal Hall responses were computed for mean-field spinons, and evaluating them in the projected states will include the effect of gauge fluctuations. Finally, real materials have lower symmetry than $p4mm$, through inequivalent bow-tie edges, buckling, or decorating sites, and the same classification scheme applies to these subgroups.

These results point to several directions. Interactions and gauge fluctuations will lift the large degeneracy of the flat bands, and it will be interesting to see whether the resulting states connect to the valence-bond crystals found numerically~\cite{Rousochatzakis2013,Astrakhantsev2021,Schmoll2023,JahromiIqbal2026}, for instance through the fermion bilinears and monopoles of the Dirac states~\cite{Hermele2004,Song2019,Song2020}. Two refinements will pin down the crystal--cuboc1 boundary at $\lambda_c\approx0.5$--$0.55$: larger crystal cells and the loop-six and crossed-dimer patterns of Refs.~\cite{Astrakhantsev2021,Schmoll2023}, which would move it up, and $SU(2)$-invariant Jastrow factors~\cite{BeccaSorella2017,Iqbal2014}, which usually favor order and would move it down. By the constraint of Sec.~\ref{sec:lsm}, any gapped symmetric state found there must be topologically ordered. Finally, a Schwinger-boson classification, building on Ref.~\cite{LuganJaubertRalko2019}, would cover spin liquids next to the noncoplanar orders, and the nematic and staggered classes found here provide its symmetry framework.

\begin{acknowledgments}
H.R. acknowledges the Council of Scientific and Industrial Research (CSIR), Government of India, for financial support through a Senior Research Fellowship (File No.~09/0084(16360)/2023-EMR-I). The work of Y.I. was performed, in part, at the Aspen Center for Physics, which is supported by a grant from the Simons Foundation (1161654, Troyer). This research was also supported in part by grant NSF PHY-2309135 to the Kavli Institute for Theoretical Physics (KITP). Y.I. acknowledges support from the Abdus Salam International Centre for Theoretical Physics (ICTP) through the Associates Programme, from the Simons Foundation through Grant No.~284558FY19, from IIT Madras through the Institute of Eminence (IoE) program for establishing QuCenDiEM (Project No.~SP22231244CPETWOQCDHOC), and thanks the International Centre for Theoretical Sciences (ICTS), Bengaluru, for participation in the Discussion Meeting ``Fractionalized Quantum Matter'' (code: ICTS/DMFQM2025/07). A.~M. is supported by the Deutsche Forschungsgemeinschaft (DFG, German Research Foundation) through Project-ID 258499086 - SFB 1170 and through the W\"urzburg-Dresden Cluster of Excellence on Complexity, Topology and Dynamics in Quantum Matter - ctd.qmat Project-ID 390858490 - EXC 2147. We acknowledge the use of the AI tools Claude Opus 5.5 and Fable 5.1 (Anthropic) for assistance with code debugging, verifying some of the calculations, and preparing parts of the manuscript. The authors reviewed all results and changes and take full responsibility for the content.
\end{acknowledgments}

\section*{Data availability}
The data that support the findings of this article, including the canonical forms of all projective symmetry groups, the amplitudes, spectra, and structure factors of all realized spin liquids, and the data of the noncoplanar orders and variational calculations, are openly available on Zenodo as part of the SM~\cite{RajMaityIqbalZenodo}.

\makeatletter\let\SM@hangfrom@section\@hangfrom@section\let\SM@sectioncntformat\@sectioncntformat\makeatother
\appendix
\section{Conventions and Fermion Bilinears}
\label{app:bilinears}

\subsection{Conventions}
\label{app:conventions}

A site is labeled $i=(x,y,\mu)$ (Sec.~\ref{sec:lattice}), and the generators act on the labels by Eq.~\eqref{eq:action}, while time reversal $\T$ fixes every site. Words in the generators are read from right to left. By Eq.~\eqref{eq:compose} a word $O_1O_2\cdots O_n$ is represented by
\begin{align}
G_{O_1\cdots O_n}(i)&=G_{O_1}(i_0)\,G_{O_2}(i_1)\cdots G_{O_n}(i_{n-1}),\nonumber\\
i_0&=i,\qquad i_k=O_k^{-1}(i_{k-1}),
\label{eq:psg-word}
\end{align}
where a letter $O_k=O^{-1}$ contributes $G_{O^{-1}}(i_{k-1})=[G_O(O(i_{k-1}))]^{-1}$. Time reversal, which acts linearly on the link matrices in the form of Eq.~\eqref{eq:psg-T}, composes by the same rule. If a word equals the identity of the symmetry group, $G_{O_1\cdots O_n}(i)$ lies in the IGG and, in the canonical gauge, in which the IGG consists of site-independent matrices, does not depend on $i$. We call it the value of the relation. For IGG $\Z_2$ the values of the relations~\eqref{eq:relations} are the signs $\eta$ of Eq.~\eqref{eq:z2labels}, each equal to $\pm\tau^0$ and identified with $\pm1$. Two sets $\{G_O\}$ describe the same PSG if they are related by a gauge transformation
\begin{equation}
G_O(i)\ \to\ W_i\,G_O(i)\,W^\dagger_{O^{-1}(i)},\qquad W_i\in SU(2),
\label{eq:psg-gauge}
\end{equation}
that preserves the IGG, or by a re-lifting $G_O\to s_OG_O$ with $s_O$ in the IGG. Under Eq.~\eqref{eq:psg-gauge} the inner factors of Eq.~\eqref{eq:psg-word} cancel pairwise, and since a relation returns to its starting site its value $\eta$ becomes $W_i\eta W_i^\dagger=\eta$, so relation values are gauge invariant. A re-lifting $G_O\to-G_O$ multiplies the value of a relation by $(-1)^{N_O}$, where $N_O$ is the number of letters $O$ or $O^{-1}$ in the word. Once $C_4T_1C_4^{-1}T_2^{-1}$ is set to $+1$, which ties the re-liftings of $T_1$ and $T_2$ together, the ten signs of Eq.~\eqref{eq:z2labels} are invariant under all remaining re-liftings and label the $\Z_2$ PSGs, while the other relation values are fixed by them (Sec.~\ref{sec:alg-z2}).

Fluxes are oriented counterclockwise, the hopping $t_{ij}$ is the coefficient of $f^\dagger_{i\sigma}f_{j\sigma}$, and the Berry connection is $\bm{\mathcal{A}}=\ii\sum_n\langle u_n|\nabla_{\bm{k}}u_n\rangle$ over the occupied bands (Sec.~\ref{sec:chern}). The hopping flux of a triangle, $\Phi=\arg(t_{12}t_{23}t_{31})$ for $1\to2\to3$ counterclockwise, is minus the Aharonov--Bohm phase of a spinon encircling the triangle counterclockwise. With these conventions a hopping flux $+\pi/2$ through every triangle of the class-$\mathcal{B}$ state gives the Chern number $C=-1$ per spin of the occupied bands ($\nu=+4$), a mode on the lower edge of a ribbon parallel to $\bm{T}_1$ moving along $+x$, and, with this counterclockwise edge, a free-spinon thermal Hall conductivity $\kappa_{xy}/T\to+2$ in units of $\pi k_B^2/(6\hbar)$. Energies and gaps follow the units and the single-spinon gap convention stated in Sec.~\ref{sec:defs}.

\subsection{Spin Rotations}
\label{app:rotations}

With spin--orbit coupling a lattice operation rotates the spin by the rotation part of the operation acting on an axial vector, in which a mirror acts as the twofold rotation about its normal. In the frame $(\hat{\bm{x}},\hat{\bm{y}},\hat{\bm{z}})$ of Eq.~\eqref{eq:T}, with $\hat{\bm{z}}$ normal to the layer, the matrices $R_O$ of Eq.~\eqref{eq:psg-spatial} are
\begin{align}
R_{T_1}=R_{T_2}&=\one,\qquad
R_{C_4}=\begin{pmatrix}0&-1&0\\1&0&0\\0&0&1\end{pmatrix},\nonumber\\
R_\sigma&=\begin{pmatrix}0&-1&0\\-1&0&0\\0&0&-1\end{pmatrix},
\end{align}
where $R_{C_4}$ is the rotation by $\pi/2$ about $\hat{\bm{z}}$ and $R_\sigma$ the twofold rotation about the normal $(1,-1,0)/\sqrt2$ of the mirror line $X=Y$. Time reversal enters through Eq.~\eqref{eq:psg-T} with $R_\T=\one$, and products follow from $R_{O_1O_2}=R_{O_1}R_{O_2}$, for instance $R_{\sigma C_4}$ for the axial mirror.

\subsection{Fermion Bilinears}

This subsection expresses the mean-field Hamiltonian~\eqref{eq:HMF} in terms of the fermions $f_{i\alpha}$. Write $f_i=(f_{i\uparrow},f_{i\downarrow})^{\rm T}$ and $f^{\dagger\rm T}_j=(f^\dagger_{j\uparrow},f^\dagger_{j\downarrow})^{\rm T}$. Inserting the matrix~\eqref{eq:psi} into $\Tr[\sigma^\alpha\psi_iu^{(\alpha)}_{ij}\psi_j^\dagger]$ and multiplying out the $2\times2$ products gives a sum of terms $f^\dagger_{i\alpha}f_{j\beta}$, $f_{i\alpha}f^\dagger_{j\beta}$, $f^\dagger_{i\alpha}f^\dagger_{j\beta}$, and $f_{i\alpha}f_{j\beta}$. Adding the term with $i$ and $j$ exchanged, which carries $u^{(\alpha)}_{ji}=u^{(\alpha)\dagger}_{ij}$, and anticommuting all operators into the order $f^\dagger_i(\cdots)f_j$ and $f^\dagger_i(\cdots)f^{\dagger\rm T}_j$ yields, for each bond,
\begin{align}
&\sum_{\alpha}\Tr\left[\sigma^\alpha\psi_iu^{(\alpha)}_{ij}\psi_j^\dagger\right]+(i\leftrightarrow j)\nonumber\\
&\qquad=f_i^\dagger\,\mathcal{T}_{ij}\,f_j+f_i^\dagger\,\mathcal{P}_{ij}\,f^{\dagger\rm T}_j+\mathrm{h.c.},
\end{align}
with the spin-space hopping and pairing matrices
\begin{align}
\mathcal{T}_{ij}&=2\left(\ii\chi^0_{ij}-\chi^3_{ij}\right)\sigma^0\nonumber\\
&\quad+2\sum_{a}\left(-t^{a0}_{ij}+\ii t^{a3}_{ij}\right)\sigma^a,\nonumber\\
\mathcal{P}_{ij}&=-2\left(\Delta^1_{ij}+\ii\Delta^2_{ij}\right)\ii\sigma^y\nonumber\\
&\quad+2\sum_{a}\left(\ii t^{a1}_{ij}-t^{a2}_{ij}\right)\sigma^a\,\ii\sigma^y,
\end{align}
where the coefficients are those of Eq.~\eqref{eq:links}. The triplet link matrices produce spin-dependent hopping $\propto\sigma^a$ and triplet pairing $\propto\sigma^a\ii\sigma^y$, whose vector of coefficients is the usual $\bm{d}$-vector. The on-site term gives
\begin{align}
\Tr\left[\psi_ia_i\psi_i^\dagger\right]=&-2a^3_i\left(n_i-1\right)\nonumber\\
&-2\left[\left(a^1_i+\ii a^2_i\right)f^\dagger_{i\uparrow}f^\dagger_{i\downarrow}+\mathrm{h.c.}\right],
\end{align}
with $n_i=f^\dagger_{i\alpha}f_{i\alpha}$: $a^3_i$ is a chemical potential for the fermion number and $a^{1,2}_i$ an on-site singlet pairing field, which together impose the three components of the constraint on average.
\section{Derivation of the Projective Symmetry Groups}
\label{app:psg}

\subsection{Canonical Gauge}

We fix the gauge in three steps. A gauge transformation sets $G_{T_1}=\tau^0$ on all sites and $G_{T_2}=\tau^0$ on the line $x=0$, and the relation $T_2T_1T_2^{-1}T_1^{-1}=\eta_{12}$ then gives $G_{T_2}(x,y,\mu)=\eta_{12}^x\tau^0$. The relations that fix the action of the point group on the translations, $C_4T_1C_4^{-1}T_2^{-1}$, $C_4T_2C_4^{-1}T_1$, $\sigma T_1\sigma^{-1}T_2^{-1}$, and $\sigma T_2\sigma^{-1}T_1^{-1}$, and the commutators of the translations with $\T$ then determine the dependence of $G_{C_4}$, $G_\sigma$, and $G_\T$ on $(x,y)$ up to the signs of Eq.~\eqref{eq:z2psg}, which are relation values. The relation $C_4T_1C_4^{-1}T_2^{-1}$ is set to $+1$ by re-lifting $T_1$. Since $C_4^{-1}$ maps the cell $(x,y)$ to $(y,-x)$ up to a shift for the bow-tie sites, which does not change the sign of $G_{T_2}$, the forms hold for all six sublattices. A remaining gauge freedom on the sublattices sets $g_{C_4}(\mu)$ to a fixed value along a spanning tree of the $C_4$ action on the sublattices, $\mu=2,3,4$ and $6$, so that $g_{C_4}(1)$ and $g_{C_4}(5)$ carry the relation $C_4^4$ on the two orbits: $g_{C_4}(\mu)=\tau^0$ for the $\Z_2$ and $SU(2)$ PSGs and for the $U(1)$ PSGs with $w_{C_4}=0$, and $g_{C_4}(\mu)=\ii\tau^1$ for the $U(1)$ PSGs with $w_{C_4}=1$, in which every $G_{C_4}(i)$ reverses the $U(1)$ rotation.

Every site stabilizer of $p4mm\times\Z_2^\T$ is an elementary abelian $2$-group: $\langle T_1C_4^2,\sigma C_4,\T\rangle\cong\Z_2^3$ for the bow-tie site $(0,0,5)$ and $\langle\sigma,\T\rangle\cong\Z_2^2$ for the square site $(0,0,1)$. The images of their projective $SU(2)$ representations lie, after a global rotation on each orbit, in the quaternion group $\{\pm\tau^0,\pm\ii\tau^1,\pm\ii\tau^2,\pm\ii\tau^3\}$, so the search for the sublattice matrices $g_O(\mu)$ can be restricted to this group without loss of generality. For each class the canonical form is the first solution in a fixed ordering of the sublattice matrices. The complete canonical forms, sign parameters and all sublattice matrices, of the PSGs of $p4mm\times\Z_2^\T$ and $p4mm$, the latter being the algebraic PSGs of the chiral classes, are provided in the SM, together with the phase functions of the flux families, and the characters of the chiral classes. A chiral state labeled $\mathcal{K}$:X has the PSG X of the space group alone, with the character of class $\mathcal{K}$. The counting of the PSGs, Eq.~\eqref{eq:countpsg}, and the canonical forms of the $U(1)$ PSGs and flux families are given in \SMtext{Sec.~\ref{sm:sec:counting} of the SM}{the SM}.

\subsection{Chiral PSGs}

For a chiral class the elements with $\epsilon(g)=-1$ are combined with time reversal and act on the ansatz as $u\to-G\,u\,G^\dagger$, with the spin rotation acting on the triplet components. This changes the conditions on the ansatz but not the algebra of the gauge matrices, so the algebraic PSGs of every chiral class are those of $p4mm$ alone: $40$ $\Z_2$, $24$ discrete $U(1)$ and one flux family, and $4$ $SU(2)$ PSGs. The character enters the realization through the allowed amplitudes and on-site fields, and the canonical forms and labels of these PSGs are listed in the SM.
\section{Hierarchy of PSGs}
\label{app:hierarchy}

The realized PSGs form a family tree. Adding new symmetry-allowed amplitudes to a state lowers its IGG while keeping all lattice symmetries: real hopping lowers $SU(2)$ to $U(1)$, and pairing lowers $SU(2)$ or $U(1)$ to $\Z_2$. The new state is a descendant of the old one, and its PSG is a subgroup of the parent PSG. Physically, parent and descendant are connected by a Higgs transition, in which a spinon hopping or pairing channel condenses.

The four $SU(2)$ states sit at the top of the tree. To lower one of them to $U(1)$, each generator must either keep or reverse a common $\tau^3$ axis, which gives one descendant in each of the $16$ sectors $w$ of Table~\ref{tab:u1}. Lowering to $\Z_2$ gives $52$ descendants per parent. Different parents can share descendants, so these numbers do not add up to totals. Every realized $\Z_2$ PSG descends from a realized $U(1)$ or $SU(2)$ PSG, so no state in the classification stands alone.

Since all four $SU(2)$ parents have $\zeta=-$, the tree starts from flat-band states with bonds only between square and bow-tie sites. Bonds within a site orbit, such as the square edge $J_1$, appear only further down the tree, in descendants in which time reversal no longer acts as a plain sign.

\section{Spin--Orbit Coupling: Moriya Rules and Symmetry Classes}
\label{app:soc}

\subsection{Moriya Rules and Inverted Directions}

For the classical Moriya rules of Sec.~\ref{sec:moriya} we assume a layer without the horizontal mirror $z\to-z$, as for buckled or supported layers in the materials. With this mirror every $\bm{D}$ would be along $\hat{\bm{z}}$, and $\bm{D}$ would vanish on the bonds whose stabilizer contains a twofold rotation. On the square-kagome lattice the rules give $\bm{D}$ in the plane of the perpendicular bisecting mirror on $J_1$, $J_{bp}$, and $J_{bb}$, in the plane on $J_{bc}$, only along the in-plane normal $\hat{\bm{e}}_\perp$ of the bond on $J_d$, $J_\times$, $J_{bs}$, and $J_+$, whose stabilizers contain a twofold rotation and two mirrors, and no constraint on $J_{23}$, $J_{sb}$, $J_{o3}$, and $J_t$.

Inverted directions arise on bonds where the PSG forbids the singlet hopping. In the $U(1)$ classes with $w_\T=1$, which are of Altland--Zirnbauer class AII (Appendix~\ref{app:az}), time reversal in the gauge $G_\T=\ii\tau^1$ keeps exactly the channels $\tau^3$ (singlet) and $\ii\tau^3$ (triplet) on every bond, so that the triplet link is always parallel, in quaternion space, to the would-be singlet. Each stabilizer element $g$ of a bond multiplies this channel by a sign $\tilde\epsilon_g$, a character of the bond stabilizer. The singlet hopping is allowed exactly when $\tilde\epsilon=1$ on the whole stabilizer, and the DM directions satisfy $R_g\bm{n}=\tilde\epsilon_g\bm{n}$ for fixing and $R_g\bm{n}=-\tilde\epsilon_g\bm{n}$ for reversing elements, which gives the classical set for $\tilde\epsilon=1$ and an inverted set otherwise. The characters follow in closed form from the relation values and the sector bits (\SMtext{Sec.~\ref{sm:sec:bonds} of the SM}{the SM}): for instance $\tilde\epsilon=(-1)^{w_{C_4}+w_\sigma}$ on $J_1$, $(-1)^{w_T}$ on $J_{bc}$, $\eta_{C_4}$ for the twofold rotation about the square center on $J_d$ and $J_{bs}$, and $\eta_{C_4}\eta_{12}$ for the twofold rotation about the octagon center on $J_\times$ and $J_+$.

The class U0001-0, projective only through $\eta_{\sigma T}$, $\eta_{C_4}$, and $\eta_{\T T}$, illustrates the mechanism. Its sign $\eta_{C_4}=-1$ forbids the singlet hopping on $J_d$ and $J_{bs}$, and $\eta_{C_4}\eta_{12}=-1$ forbids it on $J_\times$ and $J_+$. All four orbits carry instead an out-of-plane DM vector, $\bm{D}\parallel\hat{\bm{z}}$, instead of $\hat{\bm{e}}_\perp$. In classes in which $w_{C_4}+w_\sigma$ is odd, such as U0011-$n$ and U0101-$n$, the square edges carry no singlet hopping and a DM vector along the bond.

\subsection{Altland--Zirnbauer Classes and Momentum Action}
\label{app:az}

In the $U(1)$ classes with $w_\T=0$ time reversal acts as a unitary chiral operator on the Bloch Hamiltonian, class AIII, which in two dimensions has no strong invariant and forces $C=C_\uparrow=0$. Every such PSG has an orbit-wise constant central $G_\T$ and either admits no amplitude or is a flat-band PSG, so that the statement is automatic. The $SU(2)$ PSGs are of class CII and trivial in two dimensions. Spinon quantum spin Hall states can therefore occur only in the AII classes among the $U(1)$ and $SU(2)$ PSGs, whereas $\Z_2$ PSGs with spin--orbit coupling are of class DIII and are not surveyed. The twofold rotation $C_2=C_4^2$ conserves spinon number in every class. Time reversal maps $\bm{k}\to-\bm{k}+(\pi,\pi)$ in the classes with $\eta_{\T T}=-1$, and $C_2$ maps $\bm{k}\to-\bm{k}+(\pi,\pi)$ when $\eta_{C_4T}=-1$. The product $C_2\T$ is therefore local in momentum only when the two shifts coincide, and then $(C_2\T)^2=\eta_{C_4}$. When the shifts differ, as in U0001-0, $C_2\T$ maps $\bm{k}\to\bm{k}+(\pi,\pi)$, and every band satisfies $E(\bm{k})=E(\bm{k}+(\pi,\pi))$. This gauge-invariant bulk property is the projective signature of such classes, whereas the momentum at which edge Kramers pairs cross depends on the gauge of the projective translation on a ribbon and carries no such information. In the staggered sectors the translation combined with time reversal is a chiral operator at every momentum, so that the spectrum is symmetric under $E\to-E$ and $E_F=0$ is pinned.

The flat-band PSGs of Sec.~\ref{sec:flatband} allow triplet amplitudes only on the square--bow-tie bonds, so that the Bloch matrix keeps the block form of Eq.~\eqref{eq:hflat} and at least four exact zero modes per momentum survive any PSG-allowed spin--orbit coupling. Where $B(\bm{k})$ has full rank at every momentum, the flat manifold is the orthogonal complement of the span of its columns and hence a trivial bundle with Chern number zero. Where dispersive bands touch the flat bands, as in SU000-000-000$-$ at $M$, the flat manifold is not separated from them and carries no Chern number of its own. In all our spin--orbit-coupled samples of these PSGs the flat manifold is separated from the dispersive bands, by at least $0.13$. In nine further $U(1)$ classes of $p4mm$, U0011-0, U0011-2, U0101-0, U1001-0, U1011-0, U1011-3, U1101-0, U1111-0, and U1111-1, every ansatz up to $J_+$, with singlet, triplet, and on-site terms, has exact zero modes at a time-reversal-invariant momentum. The mechanism is a splitting of the levels there into sectors of odd dimension, each mapped onto itself by an antiunitary particle--hole operation, and it accounts for all zero modes except four of the twelve in U1011-0.

\section{Regular Magnetic Orders}
\label{app:regular}

In a regular order every element $g$ of $p4mm$ is undone by a global $O(3)$ operation $R_g$, and $g\mapsto R_g$ is a homomorphism~\cite{MessioLhuillierMisguich2011}. For magnetic cells of at most $2\times2$ unit cells, $R_g$ factors through $F=p4mm/\langle T_1^2,T_2^2\rangle$, a group of order $32$ with fourteen real irreducible representations: $A_1$, $A_2$, $B_1$, $B_2$, $E$ (wave vector $\Gamma$), $M_1$--$M_5$ ($M$), and $X_1$--$X_4$ ($X$ and $Y$, induced from the little group $2mm$). A three-dimensional orthogonal representation of $F$ defines a family of regular orders through the spin of a reference site in each orbit, which must be fixed by the image of the site stabilizer. Of the $168$ three-dimensional representations, $84$ allow nonzero spins on both orbits. By wave-vector content, $15$ are at $\Gamma$, $45$ at $\Gamma$ and $M$, $12$ at $\Gamma$ and $X$, $8$ at $M$ and $X$, and $4$ at $M$. Of the irreducible blocks, $A_1$, $M_5$, $X_1$, and $X_4$ carry spins on both orbits, $B_1$ on the bow-tie sites only, and $B_2$, $E$, $M_1$, $M_4$, $X_2$, and $X_3$ on the square sites only, while $A_2$, $M_2$, and $M_3$ carry no spins. Grouping the representations by the space they span gives $46$ families: one collinear, containing the ferromagnet and the Lieb ferrimagnet, $11$ coplanar, and $34$ noncoplanar.

The energy per site of a family is a quadratic form in the amplitudes $x$, $y$ of its blocks on the two orbits, $E/N=\sum(ax^2+by^2-2|c|xy)$, with coefficients linear in all twelve couplings, minimized over unit spin length on every site. For the orders of the cross-octagon model with $J_1=J_2=J_3=1$ this gives
\begin{align}
E_{2\pi/3}&=-1+\tfrac13(J_++J_\times),\nonumber\\
E_{\rm cuboc1}&=-1-\tfrac13(J_++J_\times),\nonumber\\
E_{B_2+M_5}&=\frac{3+J_\times^2+J_+-J_\times(J_++3)}{3(J_\times-1)},\nonumber\\
E_{2\pi/3\text{-d}}&=-\frac{6+J_\times^2-2J_+-J_\times(J_+-4)}{3(J_\times+2)},
\label{eq:regE}
\end{align}
and for $J_1=J_2=J_3=-1$ the ferromagnetic and octagonal energies $E_{\rm FM}=-2+\frac13(J_++J_\times)$ and $E_{\rm oct}=-\frac13(2+2\sqrt2+J_++J_\times)$, in agreement with Ref.~\cite{Gembe2023}. The energies of cuboc1 and cuboc-d$'$ ($M_4+M_5$) coincide with that of $B_2+M_5$ on the line $J_\times=0$. The scalar chiralities of the cross-octagon orders are as follows. The nearest-neighbor triangles are coplanar in all of them. The chirality of cuboc1 is given in Sec.~\ref{sec:order-classes}. In $B_2+M_5$ the chiralities alternate around each square with translation-invariant signs and a magnitude of $1/\sqrt2$ at $J_\times=0$ that decreases with $|J_\times|$. In $M_4+M_5$ they have the same magnitude with a $(\pi,\pi)$ staggering, and in the conical order only loops around the octagons carry chirality, staggered in sign. None has a net chirality.

The $34$ noncoplanar families, with their blocks, chiral classes, and parent Hamiltonians, are provided in the SM. They distribute over the chiral classes as $\mathcal{B}$ $6$, $\mathcal{C}$ $4$, $\mathcal{D}$ $4$, $\mathcal{A}'$ $4$, $\mathcal{B}'$ $6$, $\mathcal{C}'$ $4$, and $\mathcal{D}'$ $6$, and no family falls into class $\mathcal{A}$. For each family we optimized the twelve couplings for the largest margin by which all other modes lie above the Luttinger--Tisza minimum, with one Lagrange multiplier per site orbit, and evaluated the bound on dense momentum grids including rings around the ordering wave vectors, and by annealing on $2\times2$, $4\times4$, and $8\times8$ tori and on twisted cells. Saturation of the generalized Luttinger--Tisza bound proves that the order is a classical ground state, and uniqueness means that every annealed state equals the order up to $O(3)$ and lattice symmetry.

Of the $17$ families with exact parent Hamiltonians and an open stability region in the space of the twelve couplings, $13$ have a unique ground state, and the three class-$\mathcal{B}$ families among them are $A_1+M_5$, $M_1+X_4$, and $M_4+X_1$. Within the couplings of Ref.~\cite{Gembe2023} only the conical order has an exact parent with an open stability region (besides the $C_4$-breaking cuboc1 and cuboc3), and all other parents need further orbits. The noncoplanar member of the $2\pi/3$-d family is exact with the additional coupling $J_{sb}$, where it is degenerate with $C_4$-breaking members of the same family that harmonic fluctuations favor. Three families are exact only on a manifold of codimension one in coupling space: $A_1+B_1+B_2$ and $B_1+M_1+M_4$, where the order is degenerate with coplanar members, and the twelve-direction cuboc-d order $M_4+M_5$, which stays the unique ground state on the manifold $J_\times+J_{bp}-J_{bc}=2J_{o3}(J_1-J_d-J_{bp}+2J_{bc}-J_\times)/|J_{23}-J_{sb}-J_{o3}|$. On this manifold the coupling between its two blocks that is linear in the momentum deviation vanishes. In the model of Ref.~\cite{Gembe2023} the manifold reduces to the line $J_\times=0$, where $M_4+M_5$ is degenerate with $B_2+M_5$ and cuboc1. For the remaining families, which have no Luttinger--Tisza-saturating parent with couplings up to $J_+$, two kinds of obstruction appear: blocks at the same wave vector that couple linearly in the momentum deviation, which makes the LT minimum incommensurate, and degeneracies with coplanar members that hold on manifolds of codimension one or two.

\clearpage
\bibliography{references}

@article{Wen2002,
  author = {Wen, Xiao-Gang},
  title = {{Quantum orders and symmetric spin liquids}},
  journal = {Phys. Rev. B},
  volume = {65},
  pages = {165113},
  year = {2002},
  doi = {10.1103/PhysRevB.65.165113}
}

@article{BieriLhuillierMessio2016,
  author = {Bieri, Samuel and Lhuillier, Claire and Messio, Laura},
  title = {{Projective symmetry group classification of chiral spin liquids}},
  journal = {Phys. Rev. B},
  volume = {93},
  pages = {094437},
  year = {2016},
  doi = {10.1103/PhysRevB.93.094437}
}

@article{MessioLhuillierMisguich2011,
  author = {Messio, L. and Lhuillier, C. and Misguich, G.},
  title = {{Lattice symmetries and regular magnetic orders in classical frustrated antiferromagnets}},
  journal = {Phys. Rev. B},
  volume = {83},
  pages = {184401},
  year = {2011},
  doi = {10.1103/PhysRevB.83.184401}
}

@article{Gembe2023,
  author = {Gemb{\'e}, Martin and Schmidt, Heinz-J{\"u}rgen and Hickey, Ciar{\'a}n and Richter, Johannes and Iqbal, Yasir and Trebst, Simon},
  title = {{Noncoplanar magnetic order in classical square-kagome antiferromagnets}},
  journal = {Phys. Rev. Res.},
  volume = {5},
  pages = {043204},
  year = {2023},
  doi = {10.1103/PhysRevResearch.5.043204},
}

@article{Chauhan2023,
  author = {Chauhan, Aishwarya and Maity, Atanu and Liu, Chunxiao and Sonnenschein, Jonas and Ferrari, Francesco and Iqbal, Yasir},
  title = {{Quantum spin liquids on the diamond lattice}},
  journal = {Phys. Rev. B},
  volume = {108},
  pages = {134424},
  year = {2023},
  doi = {10.1103/PhysRevB.108.134424}
}

@article{Mackey1952,
  author = {Mackey, George W.},
  title = {{Induced Representations of Locally Compact Groups I}},
  journal = {Ann. Math.}, volume = {55}, pages = {101}, year = {1952},
  doi = {10.2307/1969423}
}

@book{Brown1982,
  author = {Brown, Kenneth S.},
  title = {{Cohomology of Groups}},
  publisher = {Springer},
  address = {New York},
  year = {1982},
  series = {Graduate Texts in Mathematics},
  volume = {87},
  doi = {10.1007/978-1-4684-9327-6}
}

@misc{RajIqbalSHD,
  author = {Raj, Harsh and Iqbal, Yasir},
  title = {{Kitaev, semi-Dirac, and spinon quantum spin Hall liquids on the square-hexagon-dodecagon lattice: A parent projective symmetry group classification}},
  year = {2026},
  archivePrefix = {arXiv},
  eprint = {2610.22965},
  primaryClass = {cond-mat.str-el},
  doi = {10.48550/arXiv.2610.22965}
}

@article{Astrakhantsev2021,
  author = {Astrakhantsev, Nikita and Ferrari, Francesco and Niggemann, Nils and M{\"u}ller, Tobias and Chauhan, Aishwarya and Kshetrimayum, Augustine and Ghosh, Pratyay and Regnault, Nicolas and Thomale, Ronny and Reuther, Johannes and Neupert, Titus and Iqbal, Yasir},
  title = {{Pinwheel valence bond crystal ground state of the spin-$\frac{1}{2}$ Heisenberg antiferromagnet on the shuriken lattice}},
  journal = {Phys. Rev. B},
  volume = {104},
  pages = {L220408},
  year = {2021},
  doi = {10.1103/PhysRevB.104.L220408},
}

@article{FukuiHatsugaiSuzuki2005,
  author = {Fukui, Takahiro and Hatsugai, Yasuhiro and Suzuki, Hiroshi},
  title = {{Chern Numbers in Discretized Brillouin Zone: Efficient Method of Computing (Spin) Hall Conductances}},
  journal = {J. Phys. Soc. Jpn.},
  volume = {74},
  pages = {1674},
  year = {2005},
  doi = {10.1143/JPSJ.74.1674}
}

@article{FangGilbertBernevig2012,
  author = {Fang, Chen and Gilbert, Matthew J. and Bernevig, B. Andrei},
  title = {{Bulk topological invariants in noninteracting point group symmetric insulators}},
  journal = {Phys. Rev. B},
  volume = {86},
  pages = {115112},
  year = {2012},
  doi = {10.1103/PhysRevB.86.115112}
}

@article{SumiyoshiFujimoto2013,
  author = {Sumiyoshi, Hiroaki and Fujimoto, Satoshi},
  title = {{Quantum Thermal Hall Effect in a Time-Reversal-Symmetry-Broken Topological Superconductor in Two Dimensions: Approach from Bulk Calculations}},
  journal = {J. Phys. Soc. Jpn.},
  volume = {82},
  pages = {023602},
  year = {2013},
  doi = {10.7566/JPSJ.82.023602}
}

@article{KalmeyerLaughlin1987,
  author = {Kalmeyer, V. and Laughlin, R. B.},
  title = {{Equivalence of the resonating-valence-bond and fractional quantum Hall states}},
  journal = {Phys. Rev. Lett.},
  volume = {59},
  pages = {2095},
  year = {1987},
  doi = {10.1103/PhysRevLett.59.2095}
}

@article{WenWilczekZee1989,
  author = {Wen, X. G. and Wilczek, Frank and Zee, A.},
  title = {{Chiral spin states and superconductivity}},
  journal = {Phys. Rev. B},
  volume = {39},
  pages = {11413},
  year = {1989},
  doi = {10.1103/PhysRevB.39.11413}
}

@article{Kitaev2006,
  author = {Kitaev, Alexei},
  title = {{Anyons in an exactly solved model and beyond}},
  journal = {Ann. Phys.},
  volume = {321},
  pages = {2},
  year = {2006},
  doi = {10.1016/j.aop.2005.10.005}
}

@article{BieriMessioBernuLhuillier2015,
  author = {Bieri, Samuel and Messio, Laura and Bernu, Bernard and Lhuillier, Claire},
  title = {{Gapless chiral spin liquid in a kagome Heisenberg model}},
  journal = {Phys. Rev. B},
  volume = {92},
  pages = {060407},
  year = {2015},
  doi = {10.1103/PhysRevB.92.060407}
}

@article{Rousochatzakis2013,
  author = {Rousochatzakis, Ioannis and Moessner, Roderich and van den Brink, Jeroen},
  title = {{Frustrated magnetism and resonating valence bond physics in two-dimensional kagome-like magnets}},
  journal = {Phys. Rev. B},
  volume = {88},
  pages = {195109},
  year = {2013},
  doi = {10.1103/PhysRevB.88.195109},
}

@article{Schmoll2023,
  author = {Schmoll, Philipp and Kshetrimayum, Augustine and Naumann, Jan and Eisert, Jens and Iqbal, Yasir},
  title = {{Tensor network study of the spin-$\frac{1}{2}$ Heisenberg antiferromagnet on the shuriken lattice}},
  journal = {Phys. Rev. B},
  volume = {107},
  pages = {064406},
  year = {2023},
  doi = {10.1103/PhysRevB.107.064406},
}

@article{JahromiIqbal2026,
  author = {Jahromi, Saeed S. and Iqbal, Yasir},
  title = {{Quantum phase diagram of the spin-$\frac{1}{2}$ Heisenberg antiferromagnet on the square-kagome lattice: A tensor network study}},
  journal = {Phys. Rev. B},
  volume = {113},
  pages = {L140402},
  year = {2026},
  doi = {10.1103/76vf-63zr},
}

@article{Hermele2004,
  author = {Hermele, Michael and Senthil, T. and Fisher, Matthew P. A. and Lee, Patrick A. and Nagaosa, Naoto and Wen, Xiao-Gang},
  title = {{Stability of U(1) spin liquids in two dimensions}},
  journal = {Phys. Rev. B},
  volume = {70},
  pages = {214437},
  year = {2004},
  doi = {10.1103/PhysRevB.70.214437}
}

@article{Song2019,
  author = {Song, Xue-Yang and Wang, Chong and Vishwanath, Ashvin and He, Yin-Chen},
  title = {{Unifying description of competing orders in two-dimensional quantum magnets}},
  journal = {Nat. Commun.},
  volume = {10},
  pages = {4254},
  year = {2019},
  doi = {10.1038/s41467-019-11727-3}
}

@article{Song2020,
  author = {Song, Xue-Yang and He, Yin-Chen and Vishwanath, Ashvin and Wang, Chong},
  title = {{From Spinon Band Topology to the Symmetry Quantum Numbers of Monopoles in Dirac Spin Liquids}},
  journal = {Phys. Rev. X},
  volume = {10},
  pages = {011033},
  year = {2020},
  doi = {10.1103/PhysRevX.10.011033}
}

@book{BeccaSorella2017,
  author = {Becca, Federico and Sorella, Sandro},
  title = {{Quantum Monte Carlo Approaches for Correlated Systems}},
  publisher = {Cambridge University Press},
  address = {Cambridge},
  year = {2017},
  doi = {10.1017/9781316417041}
}

@article{Iqbal2014,
  author = {Iqbal, Yasir and Poilblanc, Didier and Becca, Federico},
  title = {{Vanishing spin gap in a competing spin-liquid phase in the kagome Heisenberg antiferromagnet}},
  journal = {Phys. Rev. B},
  volume = {89},
  pages = {020407},
  year = {2014},
  doi = {10.1103/PhysRevB.89.020407}
}

@article{LuganJaubertRalko2019,
  author = {Lugan, Tristan and Jaubert, L. D. C. and Ralko, Arnaud},
  title = {{Topological nematic spin liquid on the square kagome lattice}},
  journal = {Phys. Rev. Res.},
  volume = {1},
  pages = {033147},
  year = {2019},
  doi = {10.1103/PhysRevResearch.1.033147},
}

@article{MessioLhuillierMisguich2013,
  author = {Messio, Laura and Lhuillier, Claire and Misguich, Gr{\'e}goire},
  title = {{Time reversal symmetry breaking chiral spin liquids: Projective symmetry group approach of bosonic mean-field theories}},
  journal = {Phys. Rev. B},
  volume = {87},
  pages = {125127},
  year = {2013},
  doi = {10.1103/PhysRevB.87.125127}
}

@article{SiddharthanGeorges2001,
  author = {Siddharthan, Rahul and Georges, Antoine},
  title = {{Square kagome quantum antiferromagnet and the eight-vertex model}},
  journal = {Phys. Rev. B},
  volume = {65},
  pages = {014417},
  year = {2001},
  doi = {10.1103/PhysRevB.65.014417},
}

@article{TomczakRichter2003,
  author = {Tomczak, Piotr and Richter, Johannes},
  title = {{Specific heat of the spin-$\frac{1}{2}$ Heisenberg antiferromagnet on squagome lattice}},
  journal = {J. Phys. A: Math. Gen.},
  volume = {36},
  pages = {5399--5403},
  year = {2003},
  doi = {10.1088/0305-4470/36/20/304}
}

@article{NakanoSakai2013,
  author = {Nakano, Hiroki and Sakai, T{\^o}ru},
  title = {{The Two-Dimensional $S=1/2$ Heisenberg Antiferromagnet on the Shuriken Lattice -- A Lattice Composed of Vertex-Sharing Triangles --}},
  journal = {J. Phys. Soc. Jpn.},
  volume = {82},
  pages = {083709},
  year = {2013},
  doi = {10.7566/JPSJ.82.083709}
}

@article{Richter2009CMP,
  author = {Richter, J. and Schulenburg, J. and Tomczak, P. and Schmalfu{\ss}, D.},
  title = {{The Heisenberg antiferromagnet on the square-kagom{\'e} lattice}},
  journal = {Condens. Matter Phys.},
  volume = {12},
  pages = {507--517},
  year = {2009},
  doi = {10.5488/CMP.12.3.507}
}

@article{RichterDerzhkoSchnack2022,
  author = {Richter, Johannes and Derzhko, Oleg and Schnack, J{\"u}rgen},
  title = {{Thermodynamics of the spin-half square kagome lattice antiferromagnet}},
  journal = {Phys. Rev. B},
  volume = {105},
  pages = {144427},
  year = {2022},
  doi = {10.1103/PhysRevB.105.144427},
}

@article{RalkoRousochatzakis2015,
  author = {Ralko, Arnaud and Rousochatzakis, Ioannis},
  title = {{Resonating-Valence-Bond Physics Is Not Always Governed by the Shortest Tunneling Loops}},
  journal = {Phys. Rev. Lett.},
  volume = {115},
  pages = {167202},
  year = {2015},
  doi = {10.1103/PhysRevLett.115.167202},
}

@article{MoritaTohyama2018,
  author = {Morita, Katsuhiro and Tohyama, Takami},
  title = {{Magnetic Phase Diagrams and Magnetization Plateaus of the Spin-1/2 Antiferromagnetic Heisenberg Model on a Square-Kagome Lattice with Three Nonequivalent Exchange Interactions}},
  journal = {J. Phys. Soc. Jpn.},
  volume = {87},
  pages = {043704},
  year = {2018},
  doi = {10.7566/JPSJ.87.043704},
}

@article{HasegawaNakanoSakai2018,
  author = {Hasegawa, Yasumasa and Nakano, Hiroki and Sakai, T{\^o}ru},
  title = {{Metamagnetic jump in the spin-$\frac{1}{2}$ antiferromagnetic Heisenberg model on the square kagome lattice}},
  journal = {Phys. Rev. B},
  volume = {98},
  pages = {014404},
  year = {2018},
  doi = {10.1103/PhysRevB.98.014404},
}

@article{RichterDerzhkoSchulenburg2004,
  author = {Richter, Johannes and Derzhko, Oleg and Schulenburg, J{\"o}rg},
  title = {{Magnetic-Field Induced Spin-Peierls Instability in Strongly Frustrated Quantum Spin Lattices}},
  journal = {Phys. Rev. Lett.},
  volume = {93},
  pages = {107206},
  year = {2004},
  doi = {10.1103/PhysRevLett.93.107206},
}

@article{Derzhko2013,
  author = {Derzhko, Oleg and Richter, Johannes and Krupnitska, Olesia and Krokhmalskii, Taras},
  title = {{Frustrated quantum Heisenberg antiferromagnets at high magnetic fields: Beyond the flat-band scenario}},
  journal = {Phys. Rev. B},
  volume = {88},
  pages = {094426},
  year = {2013},
  doi = {10.1103/PhysRevB.88.094426},
}

@article{PohleBentonJaubert2016,
  author = {Pohle, Rico and Benton, Owen and Jaubert, L. D. C.},
  title = {{Reentrance of disorder in the anisotropic shuriken Ising model}},
  journal = {Phys. Rev. B},
  volume = {94},
  pages = {014429},
  year = {2016},
  doi = {10.1103/PhysRevB.94.014429},
}

@article{SchmidtRichter2024,
  author = {Schmidt, Heinz-J{\"u}rgen and Richter, Johannes},
  title = {{Exotic magnetization curves in classical square-kagom{\'e} spin lattices}},
  journal = {J. Phys. A: Math. Theor.},
  volume = {57},
  pages = {185001},
  year = {2024},
  doi = {10.1088/1751-8121/ad3ca5},
}

@article{RichterSchmidtSchnack2025,
  author = {Richter, Johannes and Schmidt, Heinz-J{\"u}rgen and Schnack, J{\"u}rgen},
  title = {{The magnetization process of classical Heisenberg magnets with non-coplanar cuboc ground states}},
  journal = {J. Phys.: Condens. Matter},
  volume = {37},
  pages = {015804},
  year = {2025},
  doi = {10.1088/1361-648X/ad7fb2},
}

@article{Fujihala2020,
  author = {Fujihala, Masayoshi and Morita, Katsuhiro and Mole, Richard and Mitsuda, Setsuo and Tohyama, Takami and Yano, Shin-ichiro and Yu, Dehong and Sota, Shigetoshi and Kuwai, Tomohiko and Koda, Akihiro and Okabe, Hirotaka and Lee, Hua and Itoh, Shinichi and Hawai, Takafumi and Masuda, Takatsugu and Sagayama, Hajime and Matsuo, Akira and Kindo, Koichi and Ohira-Kawamura, Seiko and Nakajima, Kenji},
  title = {{Gapless spin liquid in a square-kagome lattice antiferromagnet}},
  journal = {Nat. Commun.},
  volume = {11},
  pages = {3429},
  year = {2020},
  doi = {10.1038/s41467-020-17235-z}
}

@article{Goto2024,
  author = {Goto, Takayuki and Fujihala, Masayoshi and Mitsuda, Setsuo},
  title = {{$^{27}$Al-NMR study on a square-kagome lattice antiferromagnet}},
  journal = {Interactions},
  volume = {245},
  pages = {40},
  year = {2024},
  doi = {10.1007/s10751-024-01882-5},
}

@article{Yakubovich2021,
  author = {Yakubovich, Olga V. and Shvanskaya, Larisa V. and Kiriukhina, Galina V. and Volkov, Anatoly S. and Dimitrova, Olga V. and Vasiliev, Alexander N.},
  title = {{Hydrothermal Synthesis and a Composite Crystal Structure of Na$_6$Cu$_7$BiO$_4$(PO$_4$)$_4$[Cl,(OH)]$_3$ as a Candidate for Quantum Spin Liquid}},
  journal = {Inorg. Chem.},
  volume = {60},
  pages = {11450--11457},
  year = {2021},
  doi = {10.1021/acs.inorgchem.1c01459}
}

@article{Liu2022,
  author = {Liu, Bo and Zeng, Zhenyuan and Xu, Aini and Sun, Yili and Yakubovich, Olga and Shvanskaya, Larisa and Li, Shiliang and Vasiliev, Alexander},
  title = {{Low-temperature specific-heat studies on two square-kagome antiferromagnets}},
  journal = {Phys. Rev. B},
  volume = {105},
  pages = {155153},
  year = {2022},
  doi = {10.1103/PhysRevB.105.155153},
}

@article{Niggemann2023,
  author = {Niggemann, Nils and Astrakhantsev, Nikita and Ralko, Arnaud and Ferrari, Francesco and Maity, Atanu and M{\"u}ller, Tobias and Richter, Johannes and Thomale, Ronny and Neupert, Titus and Reuther, Johannes and Iqbal, Yasir and Jeschke, Harald O.},
  title = {{Quantum paramagnetism in the decorated square-kagome antiferromagnet Na$_6$Cu$_7$BiO$_4$(PO$_4$)$_4$Cl$_3$}},
  journal = {Phys. Rev. B},
  volume = {108},
  pages = {L241117},
  year = {2023},
  doi = {10.1103/PhysRevB.108.L241117},
}

@article{Taran2026,
  author = {Taran, Leonid S. and Ralko, Arnaud and Temnikov, Fedor V. and Mazurenko, Vladimir V. and Streltsov, Sergey V. and Iqbal, Yasir},
  title = {{Dzyaloshinskii-Moriya-driven instabilities in square-kagome quantum antiferromagnets}},
  journal = {Phys. Rev. B},
  volume = {114},
  pages = {074414},
  year = {2026},
  doi = {10.1103/9p42-vt6m},
}

@article{Murtazoev2023,
  author = {Murtazoev, Alisher F. and Lyssenko, Konstantin A. and Markina, Maria M. and Dolgikh, Valery A. and Vasiliev, Alexander N. and Berdonosov, Peter S.},
  title = {{New Nabokoite-like Phases $A$Cu$_7$TeO$_4$(SO$_4$)$_5$Cl ($A$ = Na, K, Rb, Cs) with Decorated and Distorted Square Kagome Lattices}},
  journal = {ChemPhysChem},
  volume = {24},
  pages = {e202300111},
  year = {2023},
  doi = {10.1002/cphc.202300111}
}

@article{Markina2024,
  author = {Markina, M. M. and Berdonosov, P. S. and Vasilchikova, T. M. and Zakharov, K. V. and Murtazoev, A. F. and Dolgikh, V. A. and Moskin, A. V. and Glazkov, V. N. and Smirnov, A. I. and Vasiliev, A. N.},
  title = {{Static and resonant properties of decorated square kagom{\'e} lattice KCu$_7$(TeO$_4$)(SO$_4$)$_5$Cl}},
  journal = {Mater. Chem. Phys.},
  volume = {319},
  pages = {129348},
  year = {2024},
  doi = {10.1016/j.matchemphys.2024.129348},
}

@article{Jing2026,
  author = {Jing, Jingjing and Eich, Andreas and Liu, Yiqiu and Wang, Liran and He, Lunhua and Wang, Aifeng and Chai, Yisheng and Sun, Young and Cui, Yi and Hardy, Frederic and Meingast, Christoph and Yu, Weiqiang and Mi, Xinrun and Merz, Michael and He, Mingquan},
  title = {{Single-crystal growth, structural characterization, and physical properties of a decorated square-kagome antiferromagnet KCu$_7$TeO$_4$(SO$_4$)$_5$Cl}},
  journal = {Phys. Rev. B},
  volume = {113},
  pages = {134447},
  year = {2026},
  doi = {10.1103/zr7m-l3dk},
}

@article{BaskaranZouAnderson1987,
  author = {Baskaran, G. and Zou, Z. and Anderson, P. W.},
  title = {{The resonating valence bond state and high-$T_c$ superconductivity---A mean field theory}},
  journal = {Solid State Commun.},
  volume = {63},
  pages = {973},
  year = {1987},
  doi = {10.1016/0038-1098(87)90642-9}
}

@article{BaskaranAnderson1988,
  author = {Baskaran, G. and Anderson, P. W.},
  title = {{Gauge theory of high-temperature superconductors and strongly correlated Fermi systems}},
  journal = {Phys. Rev. B},
  volume = {37},
  pages = {580},
  year = {1988},
  doi = {10.1103/PhysRevB.37.580}
}

@article{AffleckZouHsuAnderson1988,
  author = {Affleck, Ian and Zou, Z. and Hsu, T. and Anderson, P. W.},
  title = {{SU(2) gauge symmetry of the large-$U$ limit of the Hubbard model}},
  journal = {Phys. Rev. B},
  volume = {38},
  pages = {745},
  year = {1988},
  doi = {10.1103/PhysRevB.38.745}
}

@book{Wen2004,
  author = {Wen, Xiao-Gang},
  title = {{Quantum Field Theory of Many-Body Systems: From the Origin of Sound to an Origin of Light and Electrons}},
  publisher = {Oxford University Press},
  address = {Oxford},
  year = {2004},
  isbn = {9780198530947},
  series = {Oxford Graduate Texts}
}

@article{ReutherLeeAlicea2014,
  author = {Reuther, Johannes and Lee, Shu-Ping and Alicea, Jason},
  title = {{Classification of spin liquids on the square lattice with strong spin-orbit coupling}},
  journal = {Phys. Rev. B},
  volume = {90},
  pages = {174417},
  year = {2014},
  doi = {10.1103/PhysRevB.90.174417}
}

@misc{ZhengMeiQi2015,
  author = {Zheng, Wayne and Mei, Jia-Wei and Qi, Yang},
  title = {{Classification and Monte Carlo study of symmetric $Z_2$ spin liquids on the triangular lattice}},
  year = {2015},
  archivePrefix = {arXiv},
  eprint = {1505.05351},
  primaryClass = {cond-mat.str-el}
}

@article{Lu2016,
  author = {Lu, Yuan-Ming},
  title = {{Symmetric $Z_2$ spin liquids and their neighboring phases on triangular lattice}},
  journal = {Phys. Rev. B},
  volume = {93},
  pages = {165113},
  year = {2016},
  doi = {10.1103/PhysRevB.93.165113}
}

@article{LuRan2011,
  author = {Lu, Yuan-Ming and Ran, Ying},
  title = {{$\mathbb{Z}_2$ spin liquid and chiral antiferromagnetic phase in the Hubbard model on a honeycomb lattice}},
  journal = {Phys. Rev. B},
  volume = {84},
  pages = {024420},
  year = {2011},
  doi = {10.1103/PhysRevB.84.024420},
}

@article{LuRanLee2011,
  author = {Lu, Yuan-Ming and Ran, Ying and Lee, Patrick A.},
  title = {{$\mathbb{Z}_2$ spin liquids in the $S=\frac{1}{2}$ Heisenberg model on the kagome lattice: A projective symmetry-group study of Schwinger fermion mean-field states}},
  journal = {Phys. Rev. B},
  volume = {83},
  pages = {224413},
  year = {2011},
  doi = {10.1103/PhysRevB.83.224413}
}

@article{ChoyKim2009,
  author = {Choy, Ting-Pong and Kim, Yong Baek},
  title = {{Classification of quantum phases for the star-lattice antiferromagnet via a projective symmetry group analysis}},
  journal = {Phys. Rev. B},
  volume = {80},
  pages = {064404},
  year = {2009},
  doi = {10.1103/PhysRevB.80.064404}
}

@article{YangParamekantiKim2010,
  author = {Yang, Bohm-Jung and Paramekanti, Arun and Kim, Yong Baek},
  title = {{Competing quantum paramagnetic ground states of the Heisenberg antiferromagnet on the star lattice}},
  journal = {Phys. Rev. B},
  volume = {81},
  pages = {134418},
  year = {2010},
  doi = {10.1103/PhysRevB.81.134418}
}

@article{MaitySamajdarIqbal2024,
  author = {Maity, Atanu and Samajdar, Rhine and Iqbal, Yasir},
  title = {{Gapped and gapless quantum spin liquids on the ruby lattice}},
  journal = {Phys. Rev. B},
  year = {2026},
  doi = {10.1103/h3d9-mmx2},
}

@article{MaityIqbalSamajdar2026,
  author = {Maity, Atanu and Iqbal, Yasir and Samajdar, Rhine},
  title = {{Fermionic parton theory of Rydberg $\mathbb{Z}_2$ quantum spin liquids}},
  journal = {Nat. Commun.},
  volume = {17},
  pages = {8588},
  year = {2026},
  doi = {10.1038/s41467-026-75245-9},
}

@article{SahooFlint2020,
  author = {Sahoo, Jyotisman and Flint, Rebecca},
  title = {{Symmetric spin liquids on the stuffed honeycomb lattice}},
  journal = {Phys. Rev. B},
  volume = {101},
  pages = {115103},
  year = {2020},
  doi = {10.1103/PhysRevB.101.115103}
}

@article{Maity2023,
  author = {Maity, Atanu and Ferrari, Francesco and Thomale, Ronny and Mandal, Saptarshi and Iqbal, Yasir},
  title = {{Projective symmetry group classification of Abrikosov fermion mean-field ans{\"a}tze on the square-octagon lattice}},
  journal = {Phys. Rev. B},
  volume = {107},
  pages = {134438},
  year = {2023},
  doi = {10.1103/PhysRevB.107.134438}
}

@article{Sonnenschein2024,
  author = {Sonnenschein, Jonas and Maity, Atanu and Liu, Chunxiao and Thomale, Ronny and Ferrari, Francesco and Iqbal, Yasir},
  title = {{Candidate quantum spin liquids on the maple-leaf lattice}},
  journal = {Phys. Rev. B},
  volume = {110},
  pages = {014414},
  year = {2024},
  doi = {10.1103/PhysRevB.110.014414}
}

@article{Chatterjee2026,
  author = {Chatterjee, Sourin and Maity, Atanu and Potten, Janik and M{\"u}ller, Tobias and Feuerpfeil, Andreas and Thomale, Ronny and Penc, Karlo and Jeschke, Harald O. and Samajdar, Rhine and Iqbal, Yasir},
  title = {{Semi-Dirac spin liquids and frustrated quantum magnetism on the trellis lattice}},
  journal = {Phys. Rev. Res.},
  volume = {8},
  pages = {013191},
  year = {2026},
  doi = {10.1103/xmtj-4kj9},
}

@misc{Maity2025SS,
  author = {Maity, Atanu and Ferrari, Francesco and Capponi, Sylvain and Penc, Karlo and Potten, Janik and M{\"u}ller, Tobias and Feuerpfeil, Andreas and Thomale, Ronny and Lee, Jong Yeon and Samajdar, Rhine and Iqbal, Yasir},
  title = {{Evidence for a $\mathbb{Z}_2$ Dirac spin liquid in the generalized Shastry-Sutherland model}},
  year = {2025},
  archivePrefix = {arXiv},
  eprint = {2501.00096},
  primaryClass = {cond-mat.str-el},
  doi = {10.48550/arXiv.2501.00096}
}

@article{Sonnenschein2020,
  author = {Sonnenschein, Jonas and Chauhan, Aishwarya and Iqbal, Yasir and Reuther, Johannes},
  title = {{Projective symmetry group classifications of quantum spin liquids on the simple cubic, body centered cubic, and face centered cubic lattices}},
  journal = {Phys. Rev. B},
  volume = {102},
  pages = {125140},
  year = {2020},
  doi = {10.1103/PhysRevB.102.125140}
}

@article{LiuHalaszBalents2019,
  author = {Liu, Chunxiao and Hal{\'a}sz, G{\'a}bor B. and Balents, Leon},
  title = {{Competing orders in pyrochlore magnets from a $\mathbb{Z}_2$ spin liquid perspective}},
  journal = {Phys. Rev. B},
  volume = {100},
  pages = {075125},
  year = {2019},
  doi = {10.1103/PhysRevB.100.075125}
}

@article{LiuHalaszBalents2021,
  author = {Liu, Chunxiao and Hal{\'a}sz, G{\'a}bor B. and Balents, Leon},
  title = {{Symmetric U(1) and $\mathbb{Z}_2$ spin liquids on the pyrochlore lattice}},
  journal = {Phys. Rev. B},
  volume = {104},
  pages = {054401},
  year = {2021},
  doi = {10.1103/PhysRevB.104.054401}
}

@article{SchneiderHalimehPunk2022,
  author = {Schneider, Benedikt and Halimeh, Jad C. and Punk, Matthias},
  title = {{Projective symmetry group classification of chiral $\mathbb{Z}_2$ spin liquids on the pyrochlore lattice: Application to the spin-$\frac{1}{2}$ XXZ Heisenberg model}},
  journal = {Phys. Rev. B},
  volume = {105},
  pages = {125122},
  year = {2022},
  doi = {10.1103/PhysRevB.105.125122}
}

@article{ChernKimCastelnovo2022,
  author = {Chern, Li Ern and Kim, Yong Baek and Castelnovo, Claudio},
  title = {{Competing quantum spin liquids, gauge fluctuations, and anisotropic interactions in a breathing pyrochlore lattice}},
  journal = {Phys. Rev. B},
  volume = {106},
  pages = {134402},
  year = {2022},
  doi = {10.1103/PhysRevB.106.134402}
}

@article{LiuBalentsIqbal2026,
  author = {Liu, Chunxiao and Balents, Leon and Iqbal, Yasir},
  title = {{Classification and correlation signatures of chiral spin liquids on the pyrochlore lattice}},
  journal = {Rep. Prog. Phys.},
  volume = {89},
  pages = {078003},
  year = {2026},
  doi = {10.1088/1361-6633/ae8572},
}

@article{HuangKimLu2017,
  author = {Huang, Biao and Kim, Yong Baek and Lu, Yuan-Ming},
  title = {{Interplay of nonsymmorphic symmetry and spin-orbit coupling in hyperkagome spin liquids: Applications to Na$_4$Ir$_3$O$_8$}},
  journal = {Phys. Rev. B},
  volume = {95},
  pages = {054404},
  year = {2017},
  doi = {10.1103/PhysRevB.95.054404}
}

@article{JinZhou2020,
  author = {Jin, Hui-Ke and Zhou, Yi},
  title = {{Classical and quantum order in hyperkagome antiferromagnets}},
  journal = {Phys. Rev. B},
  volume = {101},
  pages = {054408},
  year = {2020},
  doi = {10.1103/PhysRevB.101.054408}
}

@article{HuangChoiKimLu2018,
  author = {Huang, Biao and Choi, Wonjune and Kim, Yong Baek and Lu, Yuan-Ming},
  title = {{Classification and properties of quantum spin liquids on the hyperhoneycomb lattice}},
  journal = {Phys. Rev. B},
  volume = {97},
  pages = {195141},
  year = {2018},
  doi = {10.1103/PhysRevB.97.195141}
}

@article{ChernKim2021,
  author = {Chern, Li Ern and Kim, Yong Baek},
  title = {{Theoretical study of quantum spin liquids in $S=\frac{1}{2}$ hyper-hyperkagome magnets: Classification, heat capacity, and dynamical spin structure factor}},
  journal = {Phys. Rev. B},
  volume = {104},
  pages = {094413},
  year = {2021},
  doi = {10.1103/PhysRevB.104.094413}
}

@article{LiBiswasParameswaran2025,
  author = {Li, Ming-Hao and Biswas, Sounak and Parameswaran, S. A.},
  title = {{Classification of spin-$\frac{1}{2}$ fermionic quantum spin liquids on the trillium lattice}},
  journal = {Phys. Rev. B},
  volume = {112},
  pages = {104429},
  year = {2025},
  doi = {10.1103/3nmp-1vt2}
}

@article{WangVishwanath2006,
  author = {Wang, Fa and Vishwanath, Ashvin},
  title = {{Spin-liquid states on the triangular and Kagom{\'e} lattices: A projective-symmetry-group analysis of Schwinger boson states}},
  journal = {Phys. Rev. B},
  volume = {74},
  pages = {174423},
  year = {2006},
  doi = {10.1103/PhysRevB.74.174423}
}

@article{DoddsBhattacharjeeKim2013,
  author = {Dodds, Tyler and Bhattacharjee, Subhro and Kim, Yong Baek},
  title = {{Quantum spin liquids in the absence of spin-rotation symmetry: Application to herbertsmithite}},
  journal = {Phys. Rev. B},
  volume = {88},
  pages = {224413},
  year = {2013},
  doi = {10.1103/PhysRevB.88.224413}
}

@article{Gros1989,
  author = {Gros, Claudius},
  title = {{Physics of projected wavefunctions}},
  journal = {Ann. Phys.},
  volume = {189},
  pages = {53},
  year = {1989},
  doi = {10.1016/0003-4916(89)90077-8}
}

@article{Sorella2005,
  author = {Sorella, Sandro},
  title = {{Wave function optimization in the variational Monte Carlo method}},
  journal = {Phys. Rev. B},
  volume = {71},
  pages = {241103},
  year = {2005},
  doi = {10.1103/PhysRevB.71.241103}
}

@article{LinZongCeperley2001,
  author = {Lin, C. and Zong, F. H. and Ceperley, D. M.},
  title = {{Twist-averaged boundary conditions in continuum quantum Monte Carlo algorithms}},
  journal = {Phys. Rev. E},
  volume = {64},
  pages = {016702},
  year = {2001},
  doi = {10.1103/PhysRevE.64.016702}
}

@article{PoWatanabeJianZaletel2017,
  author = {Po, Hoi Chun and Watanabe, Haruki and Jian, Chao-Ming and Zaletel, Michael P.},
  title = {{Lattice Homotopy Constraints on Phases of Quantum Magnets}},
  journal = {Phys. Rev. Lett.},
  volume = {119},
  pages = {127202},
  year = {2017},
  doi = {10.1103/PhysRevLett.119.127202}
}

@article{Polyakov1977,
  author = {Polyakov, A. M.},
  title = {{Quark confinement and topology of gauge theories}},
  journal = {Nucl. Phys. B},
  volume = {120},
  pages = {429},
  year = {1977},
  doi = {10.1016/0550-3213(77)90086-4}
}

@article{Issing2026,
  author = {Issing, Jonas and Seufert, Jannis and Klett, Michael and Mazumdar, Sarbajit and Iqbal, Yasir and Thomale, Ronny and Maity, Atanu},
  title = {{Engineering altermagnetic orders on the square-kagome lattice through Wyckoff-selective sublattice polarization}},
  journal = {Phys. Rev. B},
  volume = {114},
  pages = {115128},
  year = {2026},
  doi = {10.1103/yr4y-bw4n},
}

@article{LuttingerTisza1946,
  author = {Luttinger, J. M. and Tisza, L.},
  title = {{Theory of Dipole Interaction in Crystals}},
  journal = {Phys. Rev.},
  volume = {70},
  pages = {954},
  year = {1946},
  doi = {10.1103/PhysRev.70.954}
}

@article{KaplanMenyuk2007,
  author = {Kaplan, T. A. and Menyuk, N.},
  title = {{Spin ordering in three-dimensional crystals with strong competing exchange interactions}},
  journal = {Philos. Mag.},
  volume = {87},
  pages = {3711},
  year = {2007},
  doi = {10.1080/14786430601080229}
}

@article{Gutzwiller1963,
  author = {Gutzwiller, Martin C.},
  title = {{Effect of Correlation on the Ferromagnetism of Transition Metals}},
  journal = {Phys. Rev. Lett.},
  volume = {10},
  pages = {159},
  year = {1963},
  doi = {10.1103/PhysRevLett.10.159}
}

@article{Sorella1998,
  author = {Sorella, Sandro},
  title = {{Green Function Monte Carlo with Stochastic Reconfiguration}},
  journal = {Phys. Rev. Lett.},
  volume = {80},
  pages = {4558},
  year = {1998},
  doi = {10.1103/PhysRevLett.80.4558}
}

@article{Abrikosov1965,
  author = {Abrikosov, A. A.},
  title = {{Electron scattering on magnetic impurities in metals and anomalous resistivity effects}},
  journal = {Phys. Phys. Fiz.},
  volume = {2},
  pages = {5},
  year = {1965},
  doi = {10.1103/PhysicsPhysiqueFizika.2.5}
}

@article{Dzyaloshinsky1958,
  author = {Dzyaloshinsky, I.},
  title = {{A thermodynamic theory of ``weak'' ferromagnetism of antiferromagnetics}},
  journal = {J. Phys. Chem. Solids},
  volume = {4},
  pages = {241},
  year = {1958},
  doi = {10.1016/0022-3697(58)90076-3}
}

@article{Moriya1960,
  author = {Moriya, T{\^o}ru},
  title = {{Anisotropic Superexchange Interaction and Weak Ferromagnetism}},
  journal = {Phys. Rev.},
  volume = {120},
  pages = {91},
  year = {1960},
  doi = {10.1103/PhysRev.120.91}
}

@article{LiebSchultzMattis1961,
  author = {Lieb, Elliott and Schultz, Theodore and Mattis, Daniel},
  title = {{Two soluble models of an antiferromagnetic chain}},
  journal = {Ann. Phys.},
  volume = {16},
  pages = {407},
  year = {1961},
  doi = {10.1016/0003-4916(61)90115-4}
}

@article{Oshikawa2000,
  author = {Oshikawa, Masaki},
  title = {{Commensurability, Excitation Gap, and Topology in Quantum Many-Particle Systems on a Periodic Lattice}},
  journal = {Phys. Rev. Lett.},
  volume = {84},
  pages = {1535},
  year = {2000},
  doi = {10.1103/PhysRevLett.84.1535}
}

@article{Hastings2004,
  author = {Hastings, M. B.},
  title = {{Lieb-Schultz-Mattis in higher dimensions}},
  journal = {Phys. Rev. B},
  volume = {69},
  pages = {104431},
  year = {2004},
  doi = {10.1103/PhysRevB.69.104431}
}

@article{WatanabePoVishwanathZaletel2015,
  author = {Watanabe, Haruki and Po, Hoi Chun and Vishwanath, Ashvin and Zaletel, Michael},
  title = {{Filling constraints for spin-orbit coupled insulators in symmorphic and nonsymmorphic crystals}},
  journal = {Proc. Natl. Acad. Sci. U.S.A.},
  volume = {112},
  pages = {14551},
  year = {2015},
  doi = {10.1073/pnas.1514665112}
}

@article{AltlandZirnbauer1997,
  author = {Altland, Alexander and Zirnbauer, Martin R.},
  title = {{Nonstandard symmetry classes in mesoscopic normal-superconducting hybrid structures}},
  journal = {Phys. Rev. B},
  volume = {55},
  pages = {1142},
  year = {1997},
  doi = {10.1103/PhysRevB.55.1142}
}

@article{Schnyder2008,
  author = {Schnyder, Andreas P. and Ryu, Shinsei and Furusaki, Akira and Ludwig, Andreas W. W.},
  title = {{Classification of topological insulators and superconductors in three spatial dimensions}},
  journal = {Phys. Rev. B},
  volume = {78},
  pages = {195125},
  year = {2008},
  doi = {10.1103/PhysRevB.78.195125}
}

@article{SoluyanovVanderbilt2011,
  author = {Soluyanov, Alexey A. and Vanderbilt, David},
  title = {{Computing topological invariants without inversion symmetry}},
  journal = {Phys. Rev. B},
  volume = {83},
  pages = {235401},
  year = {2011},
  doi = {10.1103/PhysRevB.83.235401}
}

@article{FukuiHatsugai2007,
  author = {Fukui, Takahiro and Hatsugai, Yasuhiro},
  title = {{Quantum Spin Hall Effect in Three Dimensional Materials: Lattice Computation of $Z_2$ Topological Invariants and Its Application to Bi and Sb}},
  journal = {J. Phys. Soc. Jpn.},
  volume = {76},
  pages = {053702},
  year = {2007},
  doi = {10.1143/JPSJ.76.053702}
}

@article{KaneMele2005Z2,
  author = {Kane, C. L. and Mele, E. J.},
  title = {{$Z_2$ Topological Order and the Quantum Spin Hall Effect}},
  journal = {Phys. Rev. Lett.},
  volume = {95},
  pages = {146802},
  year = {2005},
  doi = {10.1103/PhysRevLett.95.146802}
}

@article{FuKane2006,
  author = {Fu, Liang and Kane, C. L.},
  title = {{Time reversal polarization and a $Z_2$ adiabatic spin pump}},
  journal = {Phys. Rev. B},
  volume = {74},
  pages = {195312},
  year = {2006},
  doi = {10.1103/PhysRevB.74.195312}
}

@article{GroverSenthil2008,
  author = {Grover, Tarun and Senthil, T.},
  title = {{Topological Spin Hall States, Charged Skyrmions, and Superconductivity in Two Dimensions}},
  journal = {Phys. Rev. Lett.},
  volume = {100},
  pages = {156804},
  year = {2008},
  doi = {10.1103/PhysRevLett.100.156804}
}

@article{RanVishwanathLee2008,
  author = {Ran, Ying and Vishwanath, Ashvin and Lee, Dung-Hai},
  title = {{Spin-Charge Separated Solitons in a Topological Band Insulator}},
  journal = {Phys. Rev. Lett.},
  volume = {101},
  pages = {086801},
  year = {2008},
  doi = {10.1103/PhysRevLett.101.086801}
}

@article{Kitaev2009,
  author = {Kitaev, Alexei},
  title = {{Periodic table for topological insulators and superconductors}},
  journal = {AIP Conf. Proc.},
  volume = {1134},
  pages = {22},
  year = {2009},
  doi = {10.1063/1.3149495}
}

@article{Lieb1989,
  author = {Lieb, Elliott H.},
  title = {{Two theorems on the Hubbard model}},
  journal = {Phys. Rev. Lett.},
  volume = {62},
  pages = {1201},
  year = {1989},
  doi = {10.1103/PhysRevLett.62.1201}
}

@article{MizoguchiKunoHatsugai2021,
  author = {Mizoguchi, Tomonari and Kuno, Yoshihito and Hatsugai, Yasuhiro},
  title = {{Flat band, spin-1 Dirac cone, and Hofstadter diagram in the fermionic square kagome model}},
  journal = {Phys. Rev. B},
  volume = {104},
  pages = {035161},
  year = {2021},
  doi = {10.1103/PhysRevB.104.035161},
}

@misc{RajMaityIqbalZenodo,
  author       = {Raj, Harsh and Maity, Atanu and Iqbal, Yasir},
  title        = {Supplemental {M}aterial and data for: {M}elting noncoplanar order on the square-kagome lattice: {C}hiral, flat-band, and staggered spin liquids from projective symmetry},
  year         = {2026},
  howpublished = {Zenodo},
  doi          = {10.5281/zenodo.23723060},
  note         = {\href{https://doi.org/10.5281/zenodo.23723060}{doi:10.5281/zenodo.23723060}}
}

@article{HickeyCincioPapicParamekanti2017,
  author = {Hickey, Ciar{\'a}n and Cincio, Lukasz and Papi{\'c}, Zlatko and Paramekanti, Arun},
  title = {{Emergence of chiral spin liquids via quantum melting of noncoplanar magnetic orders}},
  journal = {Phys. Rev. B},
  volume = {96},
  pages = {115115},
  year = {2017},
  doi = {10.1103/PhysRevB.96.115115}
}

@article{MessioBernuLhuillier2012,
  author = {Messio, Laura and Bernu, Bernard and Lhuillier, Claire},
  title = {{Kagome Antiferromagnet: A Chiral Topological Spin Liquid?}},
  journal = {Phys. Rev. Lett.},
  volume = {108},
  pages = {207204},
  year = {2012},
  doi = {10.1103/PhysRevLett.108.207204}
}

@article{GongZhuSheng2014,
  author = {Gong, Shou-Shu and Zhu, Wei and Sheng, D. N.},
  title = {{Emergent Chiral Spin Liquid: Fractional Quantum Hall Effect in a Kagome Heisenberg Model}},
  journal = {Sci. Rep.},
  volume = {4},
  pages = {6317},
  year = {2014},
  doi = {10.1038/srep06317}
}

@article{HeShengChen2014,
  author = {He, Yin-Chen and Sheng, D. N. and Chen, Yan},
  title = {{Chiral Spin Liquid in a Frustrated Anisotropic Kagome Heisenberg Model}},
  journal = {Phys. Rev. Lett.},
  volume = {112},
  pages = {137202},
  year = {2014},
  doi = {10.1103/PhysRevLett.112.137202}
}

@article{WietekSterdyniakLauchli2015,
  author = {Wietek, Alexander and Sterdyniak, Antoine and L{\"a}uchli, Andreas M.},
  title = {{Nature of chiral spin liquids on the kagome lattice}},
  journal = {Phys. Rev. B},
  volume = {92},
  pages = {125122},
  year = {2015},
  doi = {10.1103/PhysRevB.92.125122}
}

@article{HuZhuZhangGongBeccaSheng2015,
  author = {Hu, Wen-Jun and Zhu, Wei and Zhang, Yi and Gong, Shoushu and Becca, Federico and Sheng, D. N.},
  title = {{Variational Monte Carlo study of a chiral spin liquid in the extended Heisenberg model on the kagome lattice}},
  journal = {Phys. Rev. B},
  volume = {91},
  pages = {041124},
  year = {2015},
  doi = {10.1103/PhysRevB.91.041124}
}

@article{GongZhuBalentsSheng2015,
  author = {Gong, Shou-Shu and Zhu, Wei and Balents, Leon and Sheng, D. N.},
  title = {{Global phase diagram of competing ordered and quantum spin-liquid phases on the kagome lattice}},
  journal = {Phys. Rev. B},
  volume = {91},
  pages = {075112},
  year = {2015},
  doi = {10.1103/PhysRevB.91.075112}
}

@article{Sutherland1986,
  author = {Sutherland, Bill},
  title = {{Localization of electronic wave functions due to local topology}},
  journal = {Phys. Rev. B},
  volume = {34},
  pages = {5208},
  year = {1986},
  doi = {10.1103/PhysRevB.34.5208}
}

@article{DomengeSindzingreLhuillierPierre2005,
  author = {Domenge, J.-C. and Sindzingre, P. and Lhuillier, C. and Pierre, L.},
  title = {{Twelve sublattice ordered phase in the $J_1-J_2$ model on the kagom{\'e} lattice}},
  journal = {Phys. Rev. B},
  volume = {72},
  pages = {024433},
  year = {2005},
  doi = {10.1103/PhysRevB.72.024433}
}

@article{LyonsKaplan1960,
  author = {Lyons, D. H. and Kaplan, T. A.},
  title = {{Method for Determining Ground-State Spin Configurations}},
  journal = {Phys. Rev.},
  volume = {120},
  pages = {1580},
  year = {1960},
  doi = {10.1103/PhysRev.120.1580}
}

@article{Watanabe2016Volborthite,
  author = {Watanabe, Daiki and Sugii, Kaori and Shimozawa, Masaaki and Suzuki, Yoshitaka and Yajima, Takeshi and Ishikawa, Hajime and Hiroi, Zenji and Shibauchi, Takasada and Matsuda, Yuji and Yamashita, Minoru},
  title = {{Emergence of nontrivial magnetic excitations in a spin-liquid state of kagom{\'e} volborthite}},
  journal = {Proc. Natl. Acad. Sci. U.S.A.},
  volume = {113},
  pages = {8653},
  year = {2016},
  doi = {10.1073/pnas.1524076113}
}

@article{Doki2018,
  author = {Doki, Hayato and Akazawa, Masatoshi and Lee, Hyun-Yong and Han, Jung Hoon and Sugii, Kaori and Shimozawa, Masaaki and Kawashima, Naoki and Oda, Migaku and Yoshida, Hiroyuki and Yamashita, Minoru},
  title = {{Spin Thermal Hall Conductivity of a Kagome Antiferromagnet}},
  journal = {Phys. Rev. Lett.},
  volume = {121},
  pages = {097203},
  year = {2018},
  doi = {10.1103/PhysRevLett.121.097203}
}

@article{Grissonnanche2019,
  author = {Grissonnanche, G. and Legros, A. and Badoux, S. and Lefran{\c{c}}ois, E. and Zatko, V. and Lizaire, M. and Lalibert{\'e}, F. and Gourgout, A. and Zhou, J.-S. and Pyon, S. and Takayama, T. and Takagi, H. and Ono, S. and Doiron-Leyraud, N. and Taillefer, L.},
  title = {{Giant thermal Hall conductivity in the pseudogap phase of cuprate superconductors}},
  journal = {Nature},
  volume = {571},
  pages = {376},
  year = {2019},
  doi = {10.1038/s41586-019-1375-0}
}

\FloatBarrier
\clearpage
\onecolumngrid
\FloatBarrier
\makeatletter\let\@hangfrom@section\SM@hangfrom@section\let\@sectioncntformat\SM@sectioncntformat\@removefromreset{equation}{section}\makeatother
\setcounter{figure}{0}\setcounter{table}{0}\setcounter{equation}{0}\setcounter{section}{0}
\renewcommand{\thefigure}{S\arabic{figure}}\renewcommand{\theHfigure}{S\arabic{figure}}
\renewcommand{\thetable}{S\arabic{table}}\renewcommand{\theHtable}{S\arabic{table}}
\renewcommand{\theequation}{S\arabic{equation}}\renewcommand{\theHequation}{S\arabic{equation}}
\renewcommand{\thesection}{S\Roman{section}}\renewcommand{\theHsection}{S\Roman{section}}
\renewcommand{\thesubsection}{\Alph{subsection}}
\begin{center}
{\large\bfseries Supplemental Material for ``Melting noncoplanar order on the square-kagome lattice: Chiral, flat-band, and staggered spin liquids from projective symmetry''}\\[1ex]
Harsh Raj\textsuperscript{1}, Atanu Maity\textsuperscript{2,1}, and Yasir Iqbal\textsuperscript{1}\\[0.5ex]
{\small\itshape \textsuperscript{1}Department of Physics, Indian Institute of Technology Madras, Chennai 600036, India\\
\textsuperscript{2}Institut f\"ur Theoretische Physik und Astrophysik and W\"urzburg-Dresden Cluster of Excellence ctd.qmat, Julius-Maximilians-Universit\"at W\"urzburg, Am Hubland, Campus S\"ud, W\"urzburg 97074, Germany}
\end{center}
\phantomsection\addcontentsline{toc}{section}{\protect\numberline{}Supplemental Material}
\vspace{1ex}
This Supplemental Material (SM) collects derivations and method details that support the main text. The data files that accompany it~\cite{RajMaityIqbalZenodo} contain the canonical forms of all projective symmetry groups, the amplitudes, spectra, densities of states, and structure factors of all realized spin liquids, the representative chiral and spin--orbit-coupled states, the self-consistent PSG-manifold scans, the catalog of regular orders, and the projected energies.

{\small\noindent\begin{tabular}{@{}l@{\quad}l@{}}
\multicolumn{2}{@{}l}{\textbf{Contents}}\\[0.5ex]
\ref{sm:sec:bonds} & Bond Orbits and Reference-Bond Amplitudes\\
\ref{sm:sec:counting} & Counting and $U(1)$ Canonical Forms of the PSGs\\
\ref{sm:sec:relationsm} & PSGs of Other $p4mm$ Lattices\\
\ref{sm:sec:sf} & Spectra and Structure Factors\\
\ref{sm:sec:soc} & Spin--Orbit Coupling: Survey and Representative States\\
\ref{sm:sec:params} & Parameters of the States in the Figures of the Main Text\\
\ref{sm:sec:classicalsm} & Classical Modulated and Spiral States\\
\ref{sm:sec:mfscans} & Self-Consistent Solutions\\
\ref{sm:sec:vmcsm} & Projected Wave Functions: Methods and Data\\
\ref{sm:sec:nematicsm} & Projective Symmetry Groups of the Nematic Class\\
\end{tabular}}

\vspace*{-0.6\baselineskip}

\section{Bond Orbits and Reference-Bond Amplitudes}
\label{sm:sec:bonds}

Table~\ref{sm:tab:bonds} lists the twelve bond orbits up to the length $1+\sqrt3$ of $J_\times$ and $J_+$, drawn in Fig.~\ref{fig:lattice}(b) and (c) of the main text, with their reference bonds and stabilizers. The orbits $J_{23}$, $J_{sb}$, $J_{o3}$, and $J_t$ have eight bonds per cell, as many as the order of the point group, so their bonds have trivial stabilizers. Several further-neighbor bonds close triangles with nearest-neighbor bonds: $J_d$ with two square edges, $J_{bp}$ and $J_{bb}$ with two bow-tie edges, and $J_{sb}$ with one edge of each kind. The bond $J_{bc}$ is the straight diagonal of a bow-tie and closes no triangle. The next shell, at length $(5+2\sqrt3)^{1/2}$, contains two further orbits at graph distance three, which are not included.

\begin{table}[H]
\caption{The twelve symmetry-inequivalent bond orbits up to the length $1+\sqrt3$ of $J_+$. The reference bond $i\to j$ starts at site $\mu_i$ of the unit cell $(0,0)$ and ends at $(x,y,\mu_j)$. Lengths are in units of the nearest-neighbor bond length. ``Graph'' is the graph distance on the nearest-neighbor network. The column ``fixing'' gives the element of $p4mm$ that maps the reference bond onto itself with fixed endpoints, and ``reversing'' the element that exchanges its endpoints. Orbits with eight bonds per cell have neither. The names $J_\times$ and $J_+$ are those of the main text.}
\label{sm:tab:bonds}
\begin{ruledtabular}
\begin{tabular}{llcccccc}
bond & description & $i\to j$ & length & per cell & graph & fixing & reversing\\
\hline
$J_1$ & square edge & $1\to(0,0,2)$ & $1$ & 4 & 1 & -- & $\sigma C_4^{-1}$\\
$J_{23}$ & bow-tie edge & $1\to(0,0,5)$ & $1$ & 8 & 1 & -- & --\\
$J_d$ & square diagonal & $1\to(0,0,3)$ & $\sqrt2$ & 2 & 2 & $\sigma$ & $C_4^2$\\
$J_{bp}$ & square sites beside a bow-tie & $1\to(1,0,2)$ & $\sqrt3$ & 4 & 2 & -- & $T_1\sigma C_4^{-1}$\\
$J_{sb}$ & square site to next bow-tie & $1\to(-1,0,5)$ & $2\cos(\pi/12)$ & 8 & 2 & -- & --\\
$J_{bb}$ & bow-tie sites on an octagon & $5\to(0,0,6)$ & $2\cos(\pi/12)$ & 4 & 2 & -- & $\sigma$\\
$J_{bc}$ & across a bow-tie & $1\to(0,1,3)$ & $2$ & 4 & 2 & -- & $T_2C_4^2$\\
$J_{o3}$ & third neighbor on an octagon & $1\to(0,1,5)$ & $(4+\sqrt3)^{1/2}$ & 8 & 3 & -- & --\\
$J_\times$ & square sites across an octagon & $1\to(1,1,3)$ & $\sqrt6$ & 2 & 4 & $\sigma$ & $T_1T_2C_4^2$\\
$J_t$ & translate of a square site & $1\to(-1,0,1)$ & $1+\sqrt3$ & 8 & 3 & -- & --\\
$J_{bs}$ & bow-tie sites across a square & $5\to(-1,0,5)$ & $1+\sqrt3$ & 2 & 3 & $\sigma C_4$ & $C_4^2$\\
$J_+$ & bow-tie sites across an octagon & $5\to(0,-1,5)$ & $1+\sqrt3$ & 2 & 4 & $T_1\sigma C_4^{-1}$ & $T_1T_2^{-1}C_4^2$
\end{tabular}
\end{ruledtabular}
\end{table}

Once the amplitudes on the reference bond $i\to j$ of an orbit are fixed, Eq.~\eqref{eq:psg-spatial} of the main text determines them on every other bond of the orbit. The reference amplitudes are restricted by the elements of $p4mm$ that map the unoriented bond onto itself and by time reversal. Two kinds of spatial element occur (Table~\ref{sm:tab:bonds}). A reversing element $\bar O$, with $\bar O(i)=j$ and $\bar O(j)=i$, imposes
\begin{equation}
u^{(0)\dagger}_{ij}=G_{\bar O}(j)\,u^{(0)}_{ij}\,G_{\bar O}^\dagger(i),
\label{sm:eq:rev}
\end{equation}
and a fixing element $F$, with $F(i)=i$ and $F(j)=j$, imposes
\begin{equation}
u^{(0)}_{ij}=G_F(i)\,u^{(0)}_{ij}\,G_F^\dagger(j),
\label{sm:eq:fix}
\end{equation}
together with the corresponding conditions on the triplet links, which carry the spin rotation $R_{\bar O}$ or $R_F$. Fixing elements occur for $J_d$ and $J_\times$ (the diagonal mirror $\sigma$), $J_{bs}$ (the axial mirror $\sigma C_4$), and $J_+$ (the axial mirror $T_1\sigma C_4^{-1}$ through the octagon center). Because the sublattice matrices of Eq.~\eqref{eq:z2psg} of the main text are quaternions rather than signs, $G_{\bar O}(i)$ and $G_F(i)$ are in general not proportional to $\tau^0$, and the conditions mix the channels of Eq.~\eqref{eq:links}. The allowed singlet spaces are then not always spanned by basis matrices. In the data files of the SM such spaces are marked $d^\ast$. Time reversal imposes Eq.~\eqref{eq:psg-T} with $G_\T(i)=\eta_{\T T}^{x+y}g_\T(\mu_i)$. For a central $G_\T$ it reduces to the selection rule of Sec.~\ref{sec:alg-zeta}.

In a chiral class an element $O$ with $\epsilon(O)=-1$ is combined with time reversal and imposes Eq.~\eqref{sm:eq:rev} or \eqref{sm:eq:fix} with an additional minus sign, and the time-reversal condition is dropped. For the $U(1)$ PSGs the factors $(\ii\tau^1)^{w_O}$ of the transforming elements reverse the channel $\tau^3$ and leave $\ii\tau^0$ unchanged, and for the $SU(2)$ PSGs only the channel $\ii\tau^0$ remains. The allowed spaces, the common null spaces of all these linear conditions, were computed for every PSG and every orbit, with a fixed sign convention for their basis vectors. For the nearest-neighbor $SU(2)$ states the result is simple: a singlet amplitude $\ii\chi\tau^0$ is allowed on a bond only if it is odd under time reversal, $h_ih_j=-1$, which for the four realized $SU(2)$ PSGs selects the square--bow-tie orbits $J_{23}$, $J_{sb}$, and $J_{o3}$.

\textit{Triplet directions.} The triplet links carry the spin rotation $R_g$ of each stabilizer element, the axial-vector action of its rotation part, in which a mirror acts as the twofold rotation about its normal. In the AII classes ($U(1)$, $w_\T=1$) time reversal in the gauge $G_\T=\ii\tau^1$ keeps only the channels $\tau^3$ and $\ii\tau^3$, and each stabilizer element multiplies them by a common sign $\tilde\epsilon_g$ (Sec.~\ref{sec:moriya}). Table~\ref{sm:tab:moriya} lists, for the orbits of $p4mm$ with a nontrivial stabilizer, the classical DM directions, the closed forms of $\tilde\epsilon$ in terms of the relation values and sector bits, and the inverted directions. In every AII class $\eta_\sigma=(-1)^{w_\sigma}$. The orbits $J_{23}$, $J_{sb}$, $J_{o3}$, and $J_t$ have trivial stabilizers and allow the singlet hopping and all three DM components in every AII class. In the $U(1)$ classes with $w_\T=0$ and in the flat-band PSGs, time reversal allows every channel on bonds with $h_ih_j=-1$ and none on bonds with $h_ih_j=+1$, so that the flat-band PSGs carry spin--orbit coupling only on $J_{23}$, $J_{sb}$, and $J_{o3}$.

\begin{table}[H]
\caption{Moriya rules for the triplet (DM-like) spinon links on the orbits of $p4mm$ with a nontrivial stabilizer. ``Classical'' gives the DM directions allowed by Moriya's rules, in the bond frame of $\hat{\bm{e}}_\parallel$ (along the reference bond), $\hat{\bm{e}}_\perp=\hat{\bm{z}}\times\hat{\bm{e}}_\parallel$, and $\hat{\bm{z}}$. In the AII classes the stabilizer elements multiply the channels $\tau^3$ and $\ii\tau^3$ by the signs $\tilde\epsilon$ given in the third column ($w$: sector bits, $\eta$: relation values). The singlet hopping is allowed only if all signs are $+1$, and the last column gives the DM directions otherwise. For the orbits with stabilizer $\Z_2\times\Z_2$ the three nontrivial characters give $\hat{\bm{e}}_\parallel$, $\hat{\bm{z}}$, or no allowed direction.}
\label{sm:tab:moriya}
\begin{ruledtabular}
\begin{tabular}{llll}
orbit & classical & $\tilde\epsilon$ in the AII classes & inverted\\
\hline
$J_1$, $J_{bp}$, $J_{bb}$ & $\hat{\bm{e}}_\perp$, $\hat{\bm{z}}$ & \parbox[t]{8.4cm}{\raggedright $(-1)^{w_{C_4}+w_\sigma}$, $(-1)^{w_T+w_{C_4}+w_\sigma}$, $(-1)^{w_\sigma}$\strut} & $\hat{\bm{e}}_\parallel$\\
$J_{bc}$ & in plane & $(-1)^{w_T}$ & $\hat{\bm{z}}$\\
$J_d$, $J_{bs}$ & $\hat{\bm{e}}_\perp$ & \parbox[t]{8.4cm}{\raggedright $C_2$: $\eta_{C_4}$; mirrors: $(-1)^{w_\sigma}$ and $\eta_{C_4}(-1)^{w_\sigma}$ ($J_d$), $(-1)^{w_{C_4}+w_\sigma}$ and $\eta_{C_4}(-1)^{w_{C_4}+w_\sigma}$ ($J_{bs}$)\strut} & $\hat{\bm{e}}_\parallel$, $\hat{\bm{z}}$, or none\\
$J_\times$, $J_+$ & $\hat{\bm{e}}_\perp$ & \parbox[t]{8.4cm}{\raggedright $C_2$: $\eta_{C_4}\eta_{12}$; mirrors: $(-1)^{w_\sigma}$ and $\eta_{C_4}\eta_{12}(-1)^{w_\sigma}$ ($J_\times$), $(-1)^{w_T+w_{C_4}+w_\sigma}$ and $\eta_{C_4}\eta_{12}(-1)^{w_T+w_{C_4}+w_\sigma}$ ($J_+$)\strut} & $\hat{\bm{e}}_\parallel$, $\hat{\bm{z}}$, or none
\end{tabular}
\end{ruledtabular}
\end{table}

\textit{Realization by range.} Table~\ref{sm:tab:realized} gives the number of realized PSGs of $p4mm\times\Z_2^\T$ as a function of the range of the ansatz (main text, Sec.~\ref{sec:ansatz}).

\begin{table}[H]
\caption{Number of realized PSGs of $p4mm\times\Z_2^\T$ as a function of the longest bond orbit included, for singlet amplitudes and, in parentheses, for singlet and triplet amplitudes. Ranges are cumulative. GD2 is reached at $+J_{bc}$. The last row is the coupling set $\{J_1,J_{23},J_\times,J_+\}$ of Ref.~\cite{Gembe2023}. $U(1)$ entries include the realized flux family.}
\label{sm:tab:realized}
\begin{ruledtabular}
\begin{tabular}{lccc}
range & $\Z_2$ & $U(1)$ & $SU(2)$\\
\hline
$J_1,J_{23}$ & 40 (160) & 21 (57) & 4 (4)\\
$+J_d$ & 56 (160) & 25 (57) & 4 (4)\\
$+J_{bp}$ & 82 (160) & 31 (57) & 4 (4)\\
$+J_{sb},J_{bb}$ & 152 (160) & 53 (57) & 4 (4)\\
$+J_{bc}$ & 156 (160) & 57 (57) & 4 (4)\\
$+J_{o3}$ & 160 (160) & 57 (57) & 4 (4)\\
$+J_\times$ & 160 (160) & 57 (57) & 4 (4)\\
$+J_t,J_{bs},J_+$ & 160 (160) & 57 (57) & 4 (4)\\
\hline
$J_1,J_{23},J_\times,J_+$ & 66 (160) & 27 (57) & 4 (4)
\end{tabular}
\end{ruledtabular}
\end{table}

\section{Counting and $U(1)$ Canonical Forms of the PSGs}
\label{sm:sec:counting}

This section completes the derivation of the projective symmetry groups outlined in Appendix~\ref{app:psg} of the main text, whose canonical gauge and relation values, Eq.~\eqref{eq:z2psg} of the main text, it uses.

\subsection{Counting}

The $\Z_2$ PSGs follow from the cohomology classes and the on-site representations of the site stabilizers. For the cohomology, the group acts freely on a set of points consisting of the general-position orbit of $p4mm$ and two copies of it exchanged by $\T$. On such a set every cohomology class is realized by sign-valued gauge transformations, and eliminating the signs $G_O(x)$ on a patch of $(2R+1)^2$ cells over $\Z_2$, without gauge fixing, leaves the linear constraints on the thirteen relation values. Modulo re-liftings this is $H^2(p4mm\times\Z_2^\T,\Z_2)=\Z_2^{10}$, stable for $R=1$, $2$, $3$. Carrying along the relators of a presentation of each site stabilizer, written as words in the generators, expresses the extension of the stabilizer as a linear function of the class $\eta$. For each class we enumerated the two-dimensional representations of the extended stabilizers that map the nontrivial element of the IGG to $-\tau^0$, in the dicyclic group of order $48$ and in the binary octahedral group, and identified them by their characters. For the elementary abelian stabilizers of the square-kagome lattice the quaternion group suffices. The binary octahedral group, which contains elements of order $8$, is needed for the $1a$ positions of the Lieb and square lattices compared in Sec.~\ref{sm:sec:relationsm}, where $C_4$ can be represented by $e^{\ii\pi\tau^3/4}$. A PSG is a class together with one on-site representation per orbit, and the re-liftings, the characters of $p4mm\times\Z_2^\T$, act on the representations of both orbits simultaneously. The number of PSGs is
\begin{equation}
N=\sum_\eta\big|\big(R_b(\eta)\times R_s(\eta)\big)/\mathrm{Hom}(p4mm\times\Z_2^\T,\Z_2)\big|,
\label{eq:countpsg}
\end{equation}
where $R_b(\eta)$ and $R_s(\eta)$ are the inequivalent on-site representations of the class on the bow-tie and square orbits. Because the twist is simultaneous, $N$ is not a product of single-orbit counts: the relative sign $\zeta$ of a central time-reversal matrix on the two orbits survives it. Equation~\eqref{eq:countpsg} gives the $200$ PSGs of $p4mm\times\Z_2^\T$. Equivalently, the PSGs are the solutions of the PSG equations on a $2\times2$ torus, with the matrices $G_O(i)$ of all generators on all sites and the thirteen relation values as unknowns, gauge-fixed only on a spanning tree of each orbit, reduced to their relation values and the traces of their on-site stabilizer matrices and taken modulo the re-lifting group. A depth-first search with constraint propagation enumerates them.

\textit{Independent check.} The script \texttt{check\_psg\_classification.py} of the data files uses only the tabulated forms, the symmetry action, and the bond orbits. It reconstructs every PSG on a patch and confirms that all thirteen relations take the tabulated values on every site and that each $G_O$ of a $U(1)$ PSG lies in its sector. For the $\Z_2$ PSGs, the relation values together with the traces of $G_h$ at the fixed sites of the site stabilizers $h$ and of $h\T$ are gauge invariant up to the sign changes $G_g\to-G_g$ of the generators. Minimized over these sign changes, they differ for all $200$ PSGs, which are therefore pairwise inequivalent. Finally, it computes the allowed singlet amplitudes of every PSG on all twelve orbits, builds a random ansatz on a patch, and obtains its IGG as the commutant of the links. This reproduces the realized PSGs, $160$, $57$, and $4$ up to $J_+$ and $40$, $21$, and $4$ with nearest-neighbor bonds, with the same labels as the tables.

\subsection{$U(1)$ Canonical Forms}

For the IGG $U(1)$ every PSG element has the form $G_O(i)=e^{\ii\phi_O(i)\tau^3}(\ii\tau^1)^{w_O}$, and the sector $w$ is a character of $p4mm\times\Z_2^\T$. In the staggered sectors, $w_T=1$, the gauge transformation $W(x,y)=(\ii\tau^1)^{w_T(x+y)}$, an integer matrix power (not reduced mod $2$), maps the translations to sign forms, and we write the canonical forms $G^\prime$ in this gauge, in which the IGG is $\{e^{\ii\theta(-1)^{w_T(x+y)}\tau^3}\}$. The PSG acting on the ansatz is $G_O(i)=W(i)^{-1}G^\prime_O(i)\,W(O^{-1}i)$, with IGG $\{e^{\ii\theta\tau^3}\}$. All tabulated amplitudes and multipliers refer to this gauge. Within each sector the cohomology classes were computed by an exact integer elimination on a free set of sites, without on-site conditions, and the PSGs by an elimination on the sites of the square-kagome lattice that includes the on-site conditions of both orbits (Table~\ref{tab:u1} of the main text). The canonical forms are structured as in Eq.~\eqref{eq:z2psg}, with $g_O(\mu)$ in the normalizer of the IGG, diagonal for $w_O=0$ and off-diagonal for $w_O=1$. In particular, $g_{C_4}(\mu)=\ii\tau^1$ on the tree sublattices $\mu=2,3,4,6$ for $w_{C_4}=1$. Two PSGs are counted as distinct if they differ in their relation values or in the gauge-invariant on-site values of the stabilizer elements with $w_O=0$, modulo re-liftings and the global transformation $\ii\tau^1$. In the sector $(0,0,1,0)$ the flux family takes the form
\begin{align}
G_{T_2}(x,y,\mu)&=e^{\ii\Phi x\tau^3},\nonumber\\
G_{C_4}(x,y,\mu)&=e^{\ii\Phi xy\tau^3}\,g_{C_4}(\mu),\nonumber\\
G_\sigma(x,y,\mu)&=e^{\ii\Phi xy\tau^3}\,g_\sigma(\mu),
\end{align}
with $[T_2,T_1]=e^{\ii\Phi\tau^3}$, while $G_{T_1}=\tau^0$ and $G_\T(x,y,\mu)=\eta_{\T T}^{x+y}g_\T(\mu)$ carry no $\Phi$-dependent factor (the data files of the SM list the same phase functions $\Phi xy$ for $G_{C_4}$ and $G_\sigma$ in all three flux families), and the sublattice matrices of the canonical form at $\Phi=0$, in which $g_\sigma(\mu)\propto\ii\tau^1$ reverses the $U(1)$ rotation. All relations are uniform elements of the IGG for every $\Phi$, and $\Phi$ and $-\Phi$ are identified by $\ii\tau^1$.

\section{PSGs of Other $p4mm$ Lattices}
\label{sm:sec:relationsm}

The square-kagome lattice is built from two simpler lattices with the same symmetry. The bow-tie sites alone form a checkerboard lattice (Wyckoff position $2c$), and the square sites alone form a lattice of squares on the diagonal mirror lines ($4d$). The PSGs of these and other $p4mm$ lattices follow from the same $1024$ classes by the same on-site condition. Their counts are compared below.

A square-kagome PSG restricted to one site orbit gives a valid PSG of that sublattice, with the same class $\eta$. Conversely, a checkerboard PSG $c$ and a square-site PSG $d$ combine into a square-kagome PSG exactly when they share the same class. When time reversal acts as a plain sign on both orbits, the relative sign $\zeta$ gives two ways to combine them:
\begin{align}
\mathrm{PSG}_{\rm sqk}&=\{(c,d):\eta(c)=\eta(d)\}\nonumber\\
&\quad\times\{\zeta=\pm\ \text{if}\ G_\T\ \text{acts as a sign on both orbits}\}.
\end{align}
Of the $240$ checkerboard and $640$ square-site PSGs, $160$ of each form matching pairs. $120$ pairs give one square-kagome PSG and $40$ give two, which reproduces the $200$ $\Z_2$ PSGs of Sec.~\ref{sec:alg-z2} of the main text. Similarly, the $8$ $SU(2)$ PSGs come from $4$ pairs with two choices of $\zeta$ each. The PSGs of a lattice with two site orbits are thus built from those of its sublattices, and are not simply the cohomology classes.

The construction of the main text applies to any $p4mm$ lattice: the PSGs follow from the cohomology classes of $p4mm\times\Z_2^\T$ through the on-site representations of the site stabilizers. Table~\ref{sm:tab:relation} compares the counts for the single orbits of the square-kagome lattice and for other $p4mm$ lattices. For the square lattice the construction gives the $272$ $\Z_2$ PSGs of Wen's classification.

\begin{table}[H]
\caption{Algebraic PSGs of $p4mm$ lattices with time reversal and, in brackets, of the space group alone. ``F'' marks one-parameter $U(1)$ families. The last column gives the lattice-homotopy class of the spin-$1/2$ configuration with spin-rotation symmetry (main text, Sec.~\ref{sec:lsm}), where ``obstructed'' means that no symmetric, gapped state without topological order exists. Sites: checkerboard $2c$, square-site set $4d$, square-octagon $4e$ or $4f$, Lieb $1a+2c$, square $1a$. The square-kagome lattice ($2c+4d$), with the counts of Table~\ref{tab:counts} of the main text, is obstructed.}
\label{sm:tab:relation}
\begin{ruledtabular}
\footnotesize\setlength{\tabcolsep}{1pt}
\begin{tabular}{lllll}
lattice & $\Z_2$ & $U(1)$ & $SU(2)$ & homotopy\\
\hline
checkerboard & 240 [40] & $128+4$F [$32+2$F] & 16 [8] & obstructed\\
square-site set & 640 [64] & $576+16$F [$80+4$F] & 128 [32] & trivial\\
square-octagon & 640 [64] & $576+16$F [$80+4$F] & 128 [32] & trivial\\
Lieb & 192 [40] & $72+2$F [$20+1$F] & 8 [4] & obstructed\\
square & 272 [48] & -- & 16 [8] & obstructed
\end{tabular}
\end{ruledtabular}
\end{table}

The square-site set is not the square-octagon lattice, whose vertices lie on the axial mirror lines, at $4e$ or $4f$. No automorphism of $p4mm$ exchanges the two classes of mirrors. The two lattices have the same totals, because their on-site conditions differ only by the relabeling $n_\sigma\leftrightarrow n_{\sigma C_4}$ together with $\eta_{\T\sigma}\to\eta_{\T\sigma}\eta_{\T C_4}$, a bijection of the label space, while their $U(1)$ sector distributions are related by $w_\sigma\to w_\sigma+w_{C_4}$. The square-octagon classification of Maity \textit{et al.} cited in the main text fixes the on-site time-reversal matrix such that nearest-neighbor amplitudes are allowed: its $384$ $\Z_2$ PSGs are exactly the classes of our $640$ with $G_\T(i)^2=-1$, and its $32$ $SU(2)$ PSGs are those of one of our four time-reversal sectors, the one in which $G_\T$ is the bipartite sign of the lattice. The four realized $SU(2)$ states are the same. Its $U(1)$ entries count the parameter choices of its canonical forms, a different counting rule, and we compare only the $\Z_2$ and $SU(2)$ counts.

\section{Spectra and Structure Factors}
\label{sm:sec:sf}

The ans\"atze of the spectral survey (main text, Sec.~\ref{sec:types}) contain singlet amplitudes on all orbits allowed by their PSG up to $J_+$, with the magnitudes stated there. On the reference bond of each orbit (Table~\ref{sm:tab:bonds}) the link is $u=|u|\,\hat{\bm{n}}\cdot(\ii\tau^0,\tau^1,\tau^2,\tau^3)$ with $\hat{\bm{n}}$ a fixed random unit vector in the allowed space, including its sign, since for a one-dimensional allowed space the sign selects a flux sector. The amplitudes on the other bonds follow from the PSG. $U(1)$ ans\"atze are written in the gauge in which the IGG is $\{e^{\ii\theta\tau^3}\}$ (Sec.~\ref{sm:sec:counting}), and the flux families are taken at $\Phi=\pi/2$. The multipliers allowed by the PSG are solved on both site orbits at half filling. The second amplitude point B uses independently drawn directions.

The reciprocal vectors of the Bravais lattice, Eq.~\eqref{eq:T} of the main text, are $\bm{b}_1=(2\pi/a)(1,0)$ and $\bm{b}_2=(2\pi/a)(0,1)$, and the first Brillouin zone (FBZ) is the square $|k_x|,|k_y|\le\pi/a$. Its high-symmetry points are $\Gamma=(0,0)$, with little group $4mm$, $X=(1/2,0)$ and the equivalent $Y=(0,1/2)$, with little group $2mm$, and $M=(1/2,1/2)$, with little group $4mm$, in units of $2\pi/a$. For PSGs with $\eta_{12}=-1$ the ansatz is periodic on the doubled cell spanned by $\bm{T}_1$ and $2\bm{T}_2$, for the staggered $U(1)$ sectors on the cell spanned by $2\bm{T}_1$ and $2\bm{T}_2$, and for the flux families at $\Phi=\pi/2$ on a cell of four unit cells.

A state is classified as gapped if its indirect and direct gaps are nonzero, as having Fermi pockets if the bands overlap, and as having nodal lines if the direct gap vanishes along lines across the zone. Isolated nodes are classified by the powers of the gap along the principal axes of its quadratic form, linear in both directions for a Dirac node and quadratic in both for a quadratic band touching. Exactly flat bands at the Fermi level define a separate type. A count of the zero levels at all momenta, which also detects flat bands crossed by dispersive bands, finds such a level at every momentum in the $69$ states of Sec.~\ref{sec:flatband} of the main text and in no other state at point A. Restricted to their nearest-neighbor amplitudes, with the directions of point A and self-consistent multipliers, $78$ further states have flat bands at $E_F$. Small closed nodal loops, such as those around $\Gamma$ and $M$ in Z000-101-111(1), count as nodal lines. For each Dirac state we count $N_f$ as the number of two-component cones including spin. In states whose bands are twofold degenerate at every momentum a fourfold linear point counts as two cones. Quadratic band touchings in mixed states are listed separately and not counted in $N_f$. Where the cones are strongly anisotropic, the density of states is linear only below a crossover energy, and fitted density-of-states prefactors are not cone prefactors. For these states we quote velocities and crossover energies.

The spin structure factors depend on $\bm{q}$ through the phases $e^{\ii\bm{q}\cdot(\bm{r}_i-\bm{r}_j)}$. A momentum $\bm{G}$ is a period of $S(\bm{q})$ for all correlations if and only if $\bm{G}\cdot\bm{d}\in2\pi\Z$ for every site difference $\bm{d}$. In the ideal geometry the site coordinates lie in $\frac12\Z[\sqrt3]$, and the site differences contain $(1,0)$, a square edge, and $(\sqrt3,0)=\bm{T}_1-(1,0)$. They generate a group of rank four, which is dense in the plane, so that $S(\bm{q})$ has no period other than $\bm{G}=0$. It is nevertheless defined at every momentum, and we compute it with the ideal positions throughout.

\begin{figure}[H]
\centering
\includegraphics[width=0.86\textwidth]{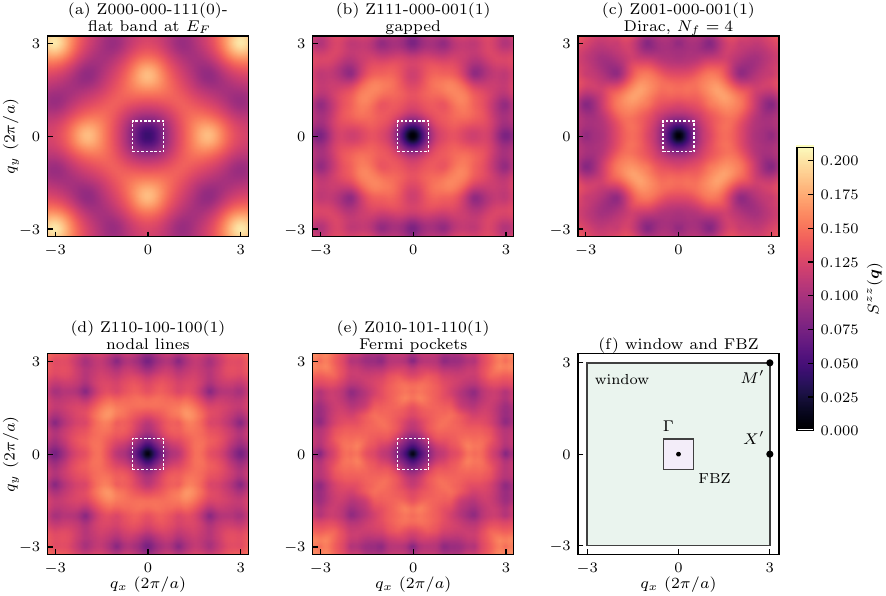}
\caption{Mean-field static structure factor $S^{zz}(\bm{q})$ at amplitude point A of one realized state per spectral type, labeled by its PSG: (a) flat band at $E_F$ ($\zeta=-$), (b) gapped, (c) Dirac nodes, (d) nodal lines, and (e) Fermi pockets. The dashed square is the first Brillouin zone (FBZ). (f) Display window $|q_{x,y}|\le3$ (units $2\pi/a$), the extended Brillouin zone of a commensurate realization of the same bond graph, with the FBZ drawn to scale at its center. Here $X'=6X$ and $M'=6M$.}
\label{sm:fig:sq}
\end{figure}

As a display window [Fig.~\ref{sm:fig:sq}(f)] we use the extended Brillouin zone of a commensurate realization of the same bond graph. Placing the square sites at $(\pm1,\pm1)$ and the bow-tie sites at $(3,0)$ and $(0,3)$, in units of a fine square lattice, with lattice constant $6$, keeps the nearest-neighbor graph and the group $p4mm$, makes all site differences integer vectors of the fine lattice, and distorts the bonds by $12\%$ (bow-tie edges $\sqrt5$, square edges $2$). Its extended zone is the square $|q_x|,|q_y|\le3\times2\pi/a$, six times larger than the FBZ in linear size, with the points $X'=6X=(3,0)$ and $M'=6M=(3,3)$ in units of $2\pi/a$. A realization with square sites at $(\pm3,\pm3)$ and bow-tie sites at $(8,0)$, $(0,8)$ distorts the bonds by only $3\%$, at the price of an extended zone with $128$ times the area of the FBZ. In the ideal geometry the window $|q_{x,y}|\le3$ is not a period cell.

For an ansatz with singlet amplitudes only, the mean-field Hamiltonian conserves the number of the fermions $\Psi_{i1}=f_{i\uparrow}$ and $\Psi_{i2}=f^\dagger_{i\downarrow}$. Indeed, singlet hopping $f^\dagger_{i\uparrow}f_{j\uparrow}+f^\dagger_{i\downarrow}f_{j\downarrow}=\Psi^\dagger_{i1}\Psi_{j1}-\Psi^\dagger_{j2}\Psi_{i2}$ and singlet pairing $f^\dagger_{i\uparrow}f^\dagger_{j\downarrow}=\Psi^\dagger_{i1}\Psi_{j2}$ are both of the form $\Psi^\dagger\Psi$, so that
\begin{equation}
H_{\rm MF}=\sum_{i,j}\Psi_i^\dagger M_{ij}\Psi_j,
\end{equation}
with a Hermitian $2N\times2N$ matrix $M$. Because $\Psi_i^\dagger\Psi_i=n_{i\uparrow}-n_{i\downarrow}+1$, the spin component along $z$ is $S^z_i=\frac12(\Psi^\dagger_i\Psi_i-1)$, and its Fourier transform with the site positions $\bm{r}_i$ of the ideal lattice is, for $\bm{q}\neq0$,
\begin{equation}
S^z_{\bm{q}}=\sum_ie^{-\ii\bm{q}\cdot\bm{r}_i}S^z_i=\frac12\sum_{\bm{k}}\sum_{\mu}\Psi^\dagger_{\bm{k},\mu}\Psi_{\bm{k}+\bm{q},\mu},
\end{equation}
where $\Psi_{\bm{k},\mu}=N_c^{-1/2}\sum_{\bm{R}}e^{-\ii\bm{k}\cdot(\bm{R}+\bm{d}_\mu)}\Psi_{\bm{R},\mu}$ is defined with the phase of the actual site position (atomic gauge) and $N_c$ is the number of unit cells. In this gauge the Bloch matrix $M(\bm{k})$ is not periodic in the reciprocal lattice, but the operator $S^z_{\bm{q}}$ contains no form factors, and the expressions below hold at every momentum $\bm{q}$, not only in the first Brillouin zone. The structure factor is therefore defined at every $\bm{q}$. It is periodic only under those reciprocal vectors $\bm{G}$ with $\bm{G}\cdot(\bm{r}_i-\bm{r}_j)\in2\pi\mathbb{Z}$ for all pairs of sites, of which the ideal geometry has none besides $\bm{G}=0$ (see above).

Let $v_n(\bm{k})$ and $E_n(\bm{k})$ be the eigenvectors and eigenvalues of $M(\bm{k})$. The mean-field ground state fills all states with $E_n(\bm{k})<0$, and $S^z_{\bm{q}}$ creates a particle-hole pair. Inserting a complete set of states in $S^{zz}(\bm{q},\omega)=N^{-1}\sum_f|\langle f|S^z_{\bm{q}}|0\rangle|^2\delta(\omega-E_f+E_0)$ gives
\begin{align}
S^{zz}(\bm{q},\omega)=\frac{1}{4N}\sum_{\bm{k}}{\sum_{a,b}}'&\left|v_b^\dagger(\bm{k})\,v_a(\bm{k}+\bm{q})\right|^2\nonumber\\
&\times\delta\big(\omega-E_b(\bm{k})+E_a(\bm{k}+\bm{q})\big),
\label{sm:eq:sqw}
\end{align}
where the primed sum runs over occupied states $a$ with $E_a(\bm{k}+\bm{q})<0$ and empty states $b$ with $E_b(\bm{k})>0$, and $N=6N_c$ is the number of sites. The overlaps contain the full coherence factors of the Bogoliubov quasiparticles of the original fermions, so Eq.~\eqref{sm:eq:sqw} applies to $\Z_2$ as well as to $U(1)$ and $SU(2)$ ans\"atze. Spin-rotation symmetry gives $S^{xx}=S^{yy}=S^{zz}$, and we plot $S^{zz}(\bm{q},\omega)$ and its frequency integral $S^{zz}(\bm{q})$. The trace over the spin components is three times larger. Its average over large momentum regions is the on-site value $N^{-1}\sum_i\langle(S^z_i)^2\rangle_{\rm MF}$. With the $2\times2$ density matrix $\rho_i$ of $\Psi_i$, Wick's theorem gives $\langle(S^z_i)^2\rangle_{\rm MF}=\frac14\mathrm{Tr}[\rho_i(1-\rho_i)]=\frac18-\frac18|\bm{t}_i|^2$, where $t^k_i=\langle\Psi^\dagger_i\tau^k\Psi_i\rangle$ are the deviations from the on-site constraints and $\mathrm{Tr}\,\rho_i=1$ by spin-rotation symmetry. The value $1/8$, half the value $1/4$ of the physical spin because the mean-field state is singly occupied only with probability $1/2$, therefore holds when the constraints are satisfied on every site. For ans\"atze with triplet amplitudes the number of $\Psi$ fermions is not conserved, and the same calculation is carried out with the four-component Nambu spinor $(f_{i\uparrow},f_{i\downarrow},f^\dagger_{i\uparrow},f^\dagger_{i\downarrow})$, which yields the full tensor $S^{ab}(\bm{q},\omega)$.

\textit{Flat-band states.} The flat-band states are evaluated in the limit $T\to0^+$ of the grand-canonical ensemble. Their zero modes are then maximally mixed: the one-body density matrix is $\rho=\frac12(1-\operatorname{sgn}h)$ with $\operatorname{sgn}0=0$, and because $h$ of Eq.~\eqref{eq:hflat} of the main text is off-diagonal between the orbits, so is $\operatorname{sgn}h$, and correlations between distinct sites of the same orbit vanish. Hence $S(\bm{q})=\frac18+\frac2N\sum_{i,j}\cos[\bm{q}\cdot(\bm{r}_i-\bm{r}_j)]\,\langle S^z_iS^z_j\rangle$, with $i$ on the square and $j$ on the bow-tie sites. The fourfold rotation, which exchanges the two bow-tie sites of a cell and permutes the square sites, gives $S(X')=1/8$. The two half-filled zero modes per spin and cell give $S(\Gamma)=1/24$ and the elastic weight $S_0(\bm{q})\le S_0(\Gamma)=1/24$, and together with the fourfold rotation $S(M')$. These values, $S(M')=\frac16+\frac1{24}\cos[6\pi/(1+\sqrt3)]\approx0.2007$ among them, do not depend on the amplitudes. Where dispersive bands touch the flat bands, they hold up to small mesh effects. These values are properties of the grand-canonical $T\to0^+$ ensemble, which is not a total-spin singlet and has total-spin fluctuations. A state at fixed particle number or a projected state would give other values of $S(\Gamma)$, $S_0$, and $S(M')$. In gapless states without flat bands, zero levels that fall on mesh points give a small weight at $\Gamma$ and at $\omega=0$ that decreases with the mesh and is not a property of the states.

\textit{Representative states.} Figure~\ref{sm:fig:sq} shows $S^{zz}(\bm{q})$ at point A for one state of each spectral type, the state whose maximum lies closest to the median of its type, together with the display window. The maps of all $221$ states are in the data files of the SM. Two properties hold in every state. First, the static weight per spin component averages to $1/8$ (see above). Second, $S(\bm{q},\omega)=S(-\bm{q},\omega)$, chiral states included, because the twofold rotation about the square center maps $\bm{q}\to-\bm{q}$ and is unitary in every chiral class (main text, Sec.~\ref{sec:chern}). On the square-hexagon-dodecagon lattice some chiral classes show an asymmetry instead~\cite{RajIqbalSHD}. The $69$ flat-band states have the amplitude-independent values derived above, with the maximum $S(M')\approx0.2007$ in $65$ of them. In the other $152$ states the maximum lies between $0.145$ and $0.189$, inside the window in $139$ of them, and its range is nearly the same for all spectral types: $0.145$ to $0.180$ for the gapped states, $0.150$ to $0.189$ for the Dirac states, $0.152$ to $0.183$ for the states with nodal lines, and $0.152$ to $0.180$ for those with Fermi pockets. The static structure factor alone therefore does not distinguish the spectral types.

\section{Spin--Orbit Coupling: Survey and Representative States}
\label{sm:sec:soc}

This section gives the survey of the $32$ AII classes and the representative spinon quantum spin Hall states of Sec.~\ref{sec:qsh} of the main text. The symmetry classes are given in Appendix~\ref{app:soc} there.

In the AII classes with $\eta_{12}=-1$, three spinons per primitive cell in a gapped band set require an even $C_\uparrow$. In U0001-0 and U0001-2 the parity of $C_\uparrow$ equals the product $\xi$ of the $C_2$ eigenvalues of the occupied spin-up bands at the time-reversal-invariant momenta~\cite{FangGilbertBernevig2012}, so that a singlet parent with $\xi=-1$ is gapless. The charge-conjugation theorem says nothing about spin--orbit coupling that breaks $S^z$ conservation.

We surveyed all $32$ AII classes. For each class, singlet parents with amplitudes on all allowed orbits up to $J_+$ were combined with random triplet amplitudes, either the $S^z$-conserving out-of-plane part or the full allowed space, of two strengths, with the multipliers fixed by $\langle n_i\rangle=1$. The survey comprises $1280$ samples with $S^z$-conserving and $1280$ with generic spin--orbit coupling. Gaps were determined by separate refinement of the band extrema, minima of the direct gap, and line scans through all special points, and $\Z_2$ invariants from the Wannier-center flow~\cite{SoluyanovVanderbilt2011} and with the Fukui--Hatsugai method~\cite{FukuiHatsugai2007}. In classes with $\eta_{\T T}=-1$ time reversal maps $\bm{k}\to-\bm{k}+(\pi,\pi)$ in the gauge of the tabulated forms, and both methods are evaluated in the zone shifted by $(\pi/2,\pi/2)$. Edge modes are counted from levels localized on one edge, and gaps are also evaluated along the shifted time-reversal-invariant lines and at generic points. For $S^z$-conserving samples the spin Chern number $C_\uparrow$ is computed as well, and $(-1)^{C_\uparrow}=\xi$ holds in all $25$ gapped $S^z$-conserving samples of the classes where it applies. Of the $S^z$-conserving samples, $120$ are gapped and $21$ of them are $\Z_2$-odd, all in U0001-0 and U0001-2, as the charge-conjugation theorem requires. Of the samples with generic spin--orbit coupling, $43$ are gapped and $9$ are $\Z_2$-odd. A further survey of $1320$ samples with generic spin--orbit coupling in the staggered and particle--hole-mirror classes found $11$ gapped states, all $\Z_2$-odd, in U1001-1, U1001-2, U1011-2, U1101-2, and U1101-3. In the non-staggered particle--hole classes an extended survey of $684$ samples, with all orbits up to $J_+$, all allowed multipliers, and two prescriptions for the singlet parents, found $36$ gapped states and one $\Z_2$-odd state, P1 in U0111-2, with single-spinon gap $\Delta=0.067$ (band gap $0.134$) and one helical pair per edge on ribbons along $\bm{T}_1$ and $\bm{T}_1+\bm{T}_2$. Such states are rare in our samples, and the absence of $\Z_2$-odd samples in the other non-staggered particle--hole classes is a sampled observation, not an exclusion. With conserved $S^z$ the charge-conjugation theorem excludes them analytically. The absence of gapped samples in U0011-1, U0011-3, U0101-1, U0101-2, U0101-3, U1001-3, U1011-1, U1101-1, U1111-2, and U1111-3 is likewise a sampled observation, not an exclusion. Only the nine classes with forced zero modes and the flat-band PSGs are excluded analytically.

\begin{table}[H]
\caption{Representative spinon quantum spin Hall states: single-spinon gap $\Delta$ and, in parentheses, the band gap $2\Delta$, spin Chern number $C_\uparrow$ where $S^z$ is conserved (spin--orbit coupling, SOC: out-of-plane DM on the stated orbit, $S^z$-conserving on all orbits, or generic), $\Z_2$ invariant ($1$ nontrivial, $0$ trivial), and number of helical edge pairs per edge. The ribbons run along $\bm{T}_1$, in addition along $\bm{T}_1+\bm{T}_2$ for Q1, Q2, and the $C_\uparrow\neq\pm1$ states, along $2\bm{T}_1$ for S1, and along both supercell directions for S2.}
\label{sm:tab:qshshow}
\begin{ruledtabular}
\footnotesize\setlength{\tabcolsep}{3pt}
\begin{tabular}{lllcccc}
state & class & SOC & $\Delta$ & $C_\uparrow$ & $\Z_2$ & pairs\\
\hline
Q1 & U0001-2 & DM, $J_1$ & 0.307 (0.614) & $+1$ & 1 & 1\\
Q2 & U0001-0 & DM, $J_{23}$ & 0.515 (1.031) & $-1$ & 1 & 1\\
 & U0001-0 & $S^z$, all & 0.411 (0.823) & $-2$ & 0 & 2\\
 & U0001-2 & $S^z$, all & 0.247 (0.493) & $+3$ & 1 & 3\\
S1 & U1101-3 & generic & 0.373 (0.745) & -- & 1 & 1\\
S2 & U1011-2 & generic & 0.655 (1.310) & -- & 1 & 1
\end{tabular}
\end{ruledtabular}
\end{table}

Table~\ref{sm:tab:qshshow} lists the representative states of Fig.~\ref{fig:qsh} of the main text and further examples. The singlet parent of Q1 has zero-energy lines $k_1=0$ and $k_2=0$ through the touching and $\xi=-1$, and a DM amplitude on the bow-tie edges alone does not open the touching. Ribbons along $\bm{T}_1$ and $\bm{T}_1+\bm{T}_2$ carry one helical pair per edge. The parent of Q2 has band touchings at $X$ and $Y$. Both states remain gapped with $\Z_2=1$ when random spin--orbit amplitudes of size $0.1$ on all orbits up to $J_+$ are added. Random searches with all orbits up to $J_+$ find spin Chern numbers of magnitude $2$ and $3$ in addition to $1$. The states with $C_\uparrow=\pm2$ are $\Z_2$-trivial, and their two helical pairs per edge are protected only by $S^z$ conservation, whereas those with $C_\uparrow=\pm3$ have three pairs per edge on both terminations.

In the staggered gauge sector U1101-3 of S1 the element $C_2\T$ is not local in momentum, and we have not derived the origin of the twofold degeneracy, which all staggered sectors U$w$-3 share. On a ribbon along $2\bm{T}_1$, levels localized on one edge cross reference energies inside the gap an odd number of times between the two Kramers points, the criterion for one helical pair~\cite{KaneMele2005Z2,FuKane2006}. The crossing momentum depends on the gauge chosen for the ribbon translation. The state S2 of U1011-2 has the largest gap among the staggered-sector samples and the same edge structure on both supercell terminations. The free-spinon spin thermal Hall conductivity $(\kappa^\uparrow_{xy}-\kappa^\downarrow_{xy})/2$ of the states with conserved $S^z$ approaches $-C_\uparrow$ in units of $\pi k_B^2/(6\hbar)$ well below the gap, as a mean-field property.

\section{Parameters of the States in the Figures of the Main Text}
\label{sm:sec:params}

\textit{Flux states.} The states of Figs.~\ref{fig:chiral}(e)--(g) and \ref{fig:nn-spectra}(a)--(d) of the main text have nearest-neighbor hopping $t_{ij}=e^{\ii\theta_{ij}}$ on the four square edges and eight bow-tie edges of each cell, translation invariant, with $t_{ij}$ the coefficient of $f^\dagger_{i\uparrow}f_{j\uparrow}$ (and of $f^\dagger_{i\downarrow}f_{j\downarrow}$). The phases $\theta_{ij}$ are the least-squares solution of the linear equations that fix the counterclockwise fluxes of the six faces of a cell. The face fluxes of a cell must add to zero modulo $2\pi$, and we lift them accordingly (for the class-$\mathcal{B}$ state $\mathcal{B}$:SU011-001 $4\times\pi/2+\pi-3\pi=0$). The spectrum depends only on the fluxes. With these conventions the $(\pi/2,\pi/2,\pi)$ state has the band gap $4-2\sqrt2$ ($\Delta=2-\sqrt2$) and $C=-1$ per spin, the $(\pi,0,\pi)$ state the band gap $2(\sqrt3-1)$ ($\Delta=\sqrt3-1$), and the $(0,0,\pi)$ state Dirac nodes at the Fermi energy $1-\sqrt3$. Chern numbers are computed with the Fukui--Hatsugai--Suzuki link variables. The plaquette angle sum is multiplied by $-1/2\pi$ to give $C=(2\pi)^{-1}\int F$ with the orientation $k_x\to k_y$, the sign of Sec.~\ref{sec:kl}.

\textit{Ribbon.} The ribbon of Fig.~\ref{fig:chiral}(f) contains $30$ unit cells along $\bm{T}_2$, all six sites of each cell, and the bonds within it. It is periodic along $\bm{T}_1$. A state is assigned to an edge when more than half of its weight lies within $0.1$ of the ribbon width from that edge.

\textit{Thermal Hall conductivity.} The curve of Fig.~\ref{fig:chiral}(g) is Eq.~\eqref{eq:kxy} evaluated with the Kubo Berry curvature.

\textit{Self-consistent states.} The self-consistent states of Sec.~\ref{sec:mf} were obtained with exact on-site constraints in the limit $T\to0^+$, and gapless states at a small Fermi smearing. The structure factors of Fig.~\ref{sm:fig:mfsq} include a small Gaussian broadening.

\section{Classical Modulated and Spiral States}
\label{sm:sec:classicalsm}

The long-wavelength modulations of the regular orders of phases II and VII (main text, Sec.~\ref{sec:classical}) lie below them by $2.2\times10^{-4}$ per site at $(J_+,J_\times)=(0.6,-0.02)$, $1.3\times10^{-3}$ at $(0.6,-0.05)$, and $1.5\times10^{-4}$ at $(-1,0.1)$, with a period of $98$, $42$, and $35$ unit cells, respectively. These energies are obtained on twisted Hermite-normal-form supercells and are therefore upper bounds. They lie within $10^{-5}$--$10^{-4}$ of the generalized Luttinger--Tisza bound evaluated with refined wave vectors, whose minimum lies at $|\delta\bm{q}|\sim10^{-2}$. The modulation period times $|J_\times|$ is $1.9$ to $2.1$ at $J_+=0.6$ for $|J_\times|=0.005$, $0.02$, and $0.05$ (periods of about $380$, $98$, and $42$ cells), consistent with a period proportional to $1/|J_\times|$. A Bloch-Hessian analysis of the regular orders shows that the $12$-direction order cuboc-d$'$ at $(0.6,-0.05)$ and the coplanar $2\pi/3$-d at $(-1,0.1)$ have negative eigenvalues at small wave vectors, so that they are saddle points, whereas cuboc-d is a local minimum at $(0.6,-0.05)$ that the modulated state undercuts only globally. For the spiral phases we have not computed the Hessian and show only that lower states exist. In phase III, for example, the energy at $(J_+,J_\times)=(1,-0.2)$ is $-1.35403$ per site, below the value $-1.34515$ of the $12\times12$ approximant, with a modulation period of $22.6$ cells.

Two details of the cross-octagon study of Gemb\'e \textit{et al.} cited in the main text were compared with our calculations. The energy quoted there for phase IV agrees with our value at $(J_+,J_\times)=(0.2,-0.2)$. The point $(-0.2,-0.2)$ given with it lies in the $2\pi/3$ phase. The magnetization formula numbered 16 there gives imaginary values at some points of the conical phase as printed, and real values at these points with the radicand of the formula numbered 15.

\section{Self-Consistent Solutions}
\label{sm:sec:mfscans}

For the self-consistent solutions of Sec.~\ref{sec:mf} of the main text, in each PSG manifold the minimum of $\langle H\rangle$ within the manifold and the fixed point of the self-consistency map projected onto the manifold were computed. Symmetries were identified from gauge-invariant loop operators. The decoupled state lies in $224$ of the $955$ PSG manifolds of $p4mm$, its chiral classes, and the nematic class: in those that allow a $\pi$-flux $J_1$ square and $J_+$ amplitudes on both chain families.

\textit{Competing states.} The lowest TRS states with a connected bond graph have bond IGG $\Z_2$ at A0, C1a, and C1b and $U(1)$ at C1c and C1d, with energies $-0.148099$ (A0), $-0.149019$ (C1a), $-0.146657$ (C1b), $-0.147092$ (C1c), and $-0.149527$ (C1d). At C1b any gap is below $10^{-3}$. The lowest class-$\mathcal{D}$ state at C1b, at $-0.144515$, is gapped with single-spinon gap $0.0265$, has $C=0$, and breaks time reversal only weakly, with a chirality of $9\times10^{-4}$ on square triples. A gapless class-$\mathcal{C}'$ state, in the PSG $\mathcal{C}'$:U100-1, lies at $-0.143691$. In the nematic class of cuboc1, a gapless state at C1a, at $-0.143359$, breaks time reversal only on loops through the $J_+$ and $J_\times$ chords, with zero chirality on the nearest-neighbor triangles and on square triples, whereas the lowest state in the nematic manifolds at C1a, at $-0.148814$, is TRS, breaks the fourfold rotation, and is nearly decoupled. Self-consistent class-$\mathcal{D}'$ states exist at C1a ($-0.1289$) and C1d ($-0.1414$). They are gapless, and time reversal is broken only on loops through the $J_\times$ and $J_+$ bonds. Further self-consistent time-reversal-symmetric states include the fully symmetric gapped $U(1)$ state at $-0.142434$, with no $J_+$ or $J_\times$ amplitude, which is the nearest-neighbor $(\pi,0,\pi)$ flux state with the on-site constraints imposed exactly (without them its Wick energy is $-0.1531$, but its bow-tie densities are unequal, as discussed in Sec.~\ref{sec:nn-u1} of the main text), and, at C1d, a symmetric state at $-0.145852$ with single-spinon gap $0.0612$, half its band gap $0.1225$.

\begin{figure}[H]
\centering
\includegraphics[width=0.62\textwidth]{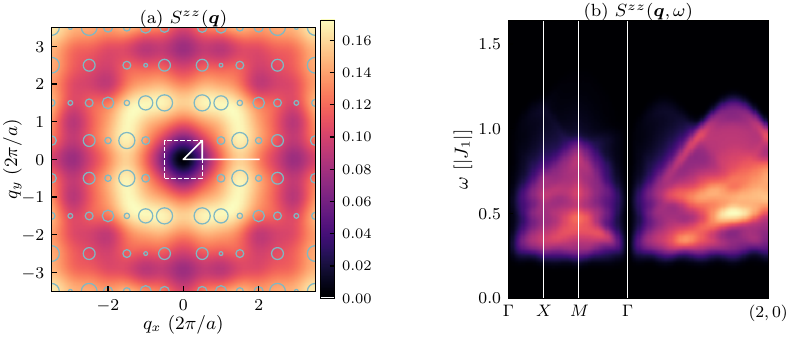}
\caption{Mean-field structure factors of the class-$\mathcal{B}$ state $\mathcal{B}$:SU011-001 at C1b ($J_+=J_\times=1$, $J_1=J_2=J_3=1$). (a) Static structure factor $S^{zz}(\bm{q})$, with the first Brillouin zone (dashed), the path of (b) (white), and the Bragg weights of cuboc1 (circles). (b) Dynamical structure factor $S^{zz}(\bm{q},\omega)$ along $\Gamma\to X\to M\to\Gamma\to(2,0)$ (units $2\pi/a$), with Gaussian broadening $0.03$. The continuum starts at $2\Delta_{\mathcal{B}}$.}
\label{sm:fig:mfsq}
\end{figure}

\textit{Structure factors of the class-$\mathcal{B}$ state.} Figure~\ref{sm:fig:mfsq} shows the static and dynamical structure factors of the class-$\mathcal{B}$ state $\mathcal{B}$:SU011-001 at C1b, computed from the mean-field two-spinon bubble, Eq.~\eqref{sm:eq:sqw}. The maxima of $S^{zz}(\bm{q})$ pass close to the Bragg peaks of cuboc1, so that the static structure factor alone cannot distinguish the chiral liquid from short-range cuboc1 correlations (main text, Sec.~\ref{sec:mf}). The two-spinon continuum starts at $2\Delta_{\mathcal{B}}$.

\textit{Projected energies at the nearest-neighbor point.} In the study of Astrakhantsev \textit{et al.} cited in the main text, the projected energy of the chiral state $(\pi/2,\pi/2,\pi)$, $-0.40489(3)J_1$ per site on a $96$-site cluster, lies above those of the $(0,0,0)$ spinon Fermi-surface state ($-0.42714(2)J_1$), the $(0,0,\pi)$ Dirac state ($-0.41720(3)J_1$), the $(\pi,0,\pi)$ state ($-0.41362(3)J_1$), and the pinwheel and loop-six valence-bond crystals ($-0.43333(1)J_1$ and $-0.43009(1)J_1$).

\section{Projected Wave Functions: Methods and Data}
\label{sm:sec:vmcsm}

\textit{Clusters and wave functions.} The projected states of Sec.~\ref{sec:vmc} of the main text are computed on $L\times L$ unit cells, $N=6L^2$ sites with $L=4$, $6$, $8$, and $10$, with periodic boundary conditions for the spins. Number-conserving states (the $U(1)$ and $SU(2)$ states and the crystal) and the $\Z_2$ states are written as determinants in the basis $(c_{i\uparrow},c^\dagger_{i\downarrow})$. The cuboc1-ordered state, whose field mixes the spin components, uses a determinant without $S^z$ conservation, sampled with exchanges and single and pair spin flips. The ordered state is the hopping ansatz with zero flux on the nearest-neighbor bonds, square hopping $1$, bow-tie hopping $r$, hoppings $t_\times$ and $t_+$ on the cross-octagon bonds, and the field $h$ along the classical cuboc1 directions. Table~\ref{sm:tab:vmcpar} lists the free parameters of each state. One amplitude of each PSG state fixes the normalization.

\begin{table}[H]
\caption{Free variational parameters of the projected states.}
\label{sm:tab:vmcpar}
\begin{ruledtabular}
\begin{tabular}{lc}
state & parameters\\
\hline
class-$\mathcal{B}$ $SU(2)$, $\mathcal{B}$:SU011-001 & 3\\
class-$\mathcal{B}$ $\Z_2$, $\mathcal{B}$:Z000-111 & 11\\
TRS $\Z_2$, Z000-011-100(1) & 9\\
nematic manifolds, N:Z00000-101 / N:Z00000-110 & 26 / 18\\
cuboc1-ordered ($r$, $t_\times$, $t_+$, $h$) & 4\\
valence-bond crystal, all bonds of the $2\times2$ cell & 63
\end{tabular}
\end{ruledtabular}
\end{table}

\textit{Boundary twists.} The spinons are given boundary twists, while the spin Hamiltonian stays periodic. Number-conserving states are averaged over the $4\times4$ grid of complex charge twists $\theta_{1,2}=2\pi(k+\tfrac12)/4$. For $\Z_2$ states a complex twist is not a consistent boundary condition, since a pairing term changes with the choice of the wrapped end of the bond, and they are averaged over the four real twists. Twists with an open mean-field shell are skipped. The energies of every state per cluster and per twist are given in the data files of the SM. The $\Z_2$ states depend strongly on the twist at C1b, by up to $0.068$ per site on $96$ sites and up to $0.05$ on the larger clusters. With the best twist instead of the average, the TRS $\Z_2$ state lies $0.028$ above the twist-averaged ordered state on $384$ sites, and the state N:Z00000-101 lies $0.015$ and $0.020$ above it on $384$ and $600$ sites, so that the ranking at C1b does not depend on the averaging.

\textit{Optimization and measurement.} The parameters are optimized by stochastic reconfiguration with the twist cycled through its set, so that the parameters are twist averaged, in chunks of $10$ steps at one twist with $200$, $200$, $120$, and $80$ sweeps per step for $L=4$, $6$, $8$, and $10$, until the energy drift between the two halves of the last $16$ chunks falls below three standard deviations. The ordered state is optimized by the second code with four twists per step, $40$, $25$, $15$, and $8$ steps for $L=4$ to $10$, each $L$ starting from the optimum at $L-2$. Energies are measured at all twists, for the states of the first code with $4000$, $2500$, $1600$, and $1000$ sweeps per twist, together with all correlations $\langle\bm{S}_i\cdot\bm{S}_j\rangle$. The order parameter is $m^2=\sum_{i\neq j}\langle\bm{S}_i\cdot\bm{S}_j\rangle\,\bm{n}_i\cdot\bm{n}_j/\sum_{i\neq j}(\bm{n}_i\cdot\bm{n}_j)^2$, equal to $1/4$ for the classical cuboc1 order of spins $1/2$.

\textit{Checks.} The two codes agree to $10^{-7}$ in exact enumeration of projected states on $24$ sites, and their twist-averaged energies agree within $1.5$ standard deviations for all $22$ crystal entries compared. Exact diagonalization of the same $24$-site Hamiltonians gives $E_0/N=-0.448329$, $-0.538598$, and $-0.712710$ at A0, C1a, and C1b, below every projected energy on that cluster. On this torus, the smallest compatible with the $2\times2$ cells of cuboc1 and of the crystal, the $J_+$ bonds coincide pairwise, so that these energies do not represent the cross-octagon model. The next compatible cluster has $48$ sites.

\textit{Fits.} Energies are extrapolated linearly in $1/N$ by weighted least squares, with the error scaled by $\sqrt{\chi^2/\nu}$ where it exceeds one. For the ordered state $\chi^2/\nu$ lies between $0.1$ and $1$, whereas for the crystal and the $\Z_2$ states it reaches $10^2$ to $10^3$: their energies scatter between clusters by $10^{-3}$ to $10^{-2}$, beyond their statistical errors, which reflects the several local minima of their optimization rather than a finite-size law. We distinguish three uncertainties. The statistical errors of the twist-averaged energies are $0.5$ to $3\times10^{-4}$ per site. Optimizations from different starting points can end in different local minima. We keep the lowest, so that this uncertainty is one-sided. Where two starts were run, they agree within $1.5\times10^{-4}$ for the crystal at $\lambda=0.4$, $0.45$, and $0.5$ on $216$ sites, but differ by $5\times10^{-3}$ to $9\times10^{-3}$ at $\lambda=0.3$ and $0.35$, and by up to $0.04$ on the cut at $J_+=1$. The extrapolation adds the spread over fit forms. Table~\ref{sm:tab:vmclargest} therefore lists the differences on the largest common cluster next to the extrapolations. Alternative fit forms, linear in $1/L$, the mean of the two largest clusters, and the largest cluster, give the ranges shown in Fig.~\ref{fig:vmc} of the main text.

\begin{table}[H]
\caption{Energies per site relative to the cuboc1-ordered state on the largest cluster on which both states were computed ($L$, $N=6L^2$), along the line $J_+=J_\times=\lambda$, to be compared with the extrapolations of Table~\ref{tab:vmc} of the main text, and on the cut $J_+=1$ across the boundary of the cuboc1 phase. $E_{\rm low}$ is the absolute energy of the lowest state on its largest cluster, and $E_{\rm prod}$ the energy of the product of singlet squares and Heisenberg chains along $J_+$, Eq.~\eqref{eq:eprod} of the main text, an upper bound on the ground-state energy. TRS: $\Z_2$ state Z000-011-100(1), NEM: nematic state N:Z00000-101, ord: cuboc1-ordered state.}
\label{sm:tab:vmclargest}
\begin{ruledtabular}
\begin{tabular}{lccccccc}
 & VBC & TRS & NEM & class $\mathcal{B}$ $\Z_2$ & class $\mathcal{B}$ $SU(2)$ & $E_{\rm low}$ & $E_{\rm prod}$\\
\hline
\multicolumn{8}{l}{line $J_+=J_\times=\lambda$, first column $\lambda$}\\
$0.30$ & $-0.0107$ ($8$) & $-0.0045$ ($8$) & $-0.0011$ ($6$) & $+0.0069$ ($8$) & -- & $-0.4410$ (VBC) & $-0.3776$\\
$0.35$ & $-0.0091$ ($10$) & $-0.0008$ ($10$) & $+0.0004$ ($6$) & $+0.0098$ ($8$) & -- & $-0.4439$ (VBC) & $-0.3850$\\
$0.40$ & $-0.0068$ ($10$) & $+0.0037$ ($10$) & $+0.0044$ ($6$) & $+0.0070$ ($8$) & -- & $-0.4470$ (VBC) & $-0.3924$\\
$0.45$ & $-0.0049$ ($8$) & $+0.0089$ ($8$) & $+0.0082$ ($8$) & $+0.0114$ ($8$) & -- & $-0.4519$ (VBC) & $-0.3998$\\
$0.50$ (C1a) & $-0.0014$ ($10$) & $+0.0158$ ($8$) & $+0.0141$ ($10$) & $+0.0166$ ($8$) & $+0.0312$ ($8$) & $-0.4572$ (VBC) & $-0.4072$\\
$0.60$ & $+0.0059$ ($8$) & -- & -- & -- & -- & $-0.4751$ (ord) & $-0.4220$\\
$1.00$ (C1b) & $+0.0234$ ($10$) & $+0.0293$ ($8$) & $+0.0336$ ($10$) & $+0.0537$ ($8$) & $+0.0549$ ($8$) & $-0.5738$ (ord) & $-0.4810$\\
\hline
\multicolumn{8}{l}{cut $J_+=1$, first column $J_\times$}\\
$0.50$ & $-0.0056$ ($8$) & $-0.0018$ ($8$) & $+0.0377$ ($6$) & $+0.0197$ ($6$) & $+0.0095$ ($8$) & $-0.5077$ (VBC) & $-0.4810$\\
$0.25$ & $-0.0280$ ($8$) & $-0.0299$ ($8$) & $+0.0320$ ($6$) & $-0.0060$ ($6$) & $-0.0145$ ($8$) & $-0.5062$ (TRS) & $-0.4810$\\
$0.10$ & $-0.0423$ ($8$) & $-0.0440$ ($8$) & $+0.0181$ ($6$) & $-0.0219$ ($6$) & $-0.0279$ ($8$) & $-0.5060$ (TRS) & $-0.4810$\\
$0.00$ & $-0.0532$ ($8$) & $-0.0549$ ($8$) & $+0.0153$ ($6$) & $-0.0361$ ($6$) & $-0.0377$ ($8$) & $-0.5059$ (TRS) & $-0.4810$\\
$-0.10$ (CDa) & $-0.0633$ ($8$) & $-0.0647$ ($8$) & $+0.0060$ ($6$) & $-0.0471$ ($6$) & $-0.0487$ ($8$) & $-0.5057$ (TRS) & $-0.4810$
\end{tabular}
\end{ruledtabular}
\end{table}

\textit{Chirality of the class-$\mathcal{B}$ $\Z_2$ state.} Table~\ref{sm:tab:vmcchi} gives the scalar chirality of $\mathcal{B}$:Z000-111 near the crossing. The projected chirality $\langle\bm{S}_i\cdot(\bm{S}_j\times\bm{S}_k)\rangle$ on the nearest-neighbor triangles is uniform, with staggered components below $2\times10^{-3}$, and decreases with the cluster size at each $\lambda$. The optimized amplitudes themselves are only weakly chiral: the time-reversal-odd part $\sin\Phi_\triangle=\operatorname{Im}\operatorname{tr}P/(2|\det P|^{1/2})$ of the loop operator $P=u_{ij}u_{jk}u_{ki}$ of a triangle is at most $0.1$, compared with $1$ for $\mathcal{B}$:SU011-001 and $0$ for the TRS states, and at $\lambda=0.5$ it decreases with the cluster size as well.

\begin{table}[H]
\caption{Scalar chirality of the class-$\mathcal{B}$ $\Z_2$ state $\mathcal{B}$:Z000-111 along $J_+=J_\times=\lambda$: magnitude of the projected chirality on the nearest-neighbor triangles (counterclockwise, four real twists, $800$ configurations, statistical error in parentheses) and magnitude of $\sin\Phi_\triangle$ of the optimized, unprojected ansatz.}
\label{sm:tab:vmcchi}
\begin{ruledtabular}
\begin{tabular}{lccc}
$\lambda$ & $N$ & $|\langle\chi_\triangle\rangle|$ & $|\sin\Phi_\triangle|$\\
\hline
$0.40$ & $96$ & $0.060(4)$ & $0.010$\\
 & $216$ & $0.056(1)$ & $0.024$\\
 & $384$ & $0.033(1)$ & $0.022$\\
$0.45$ & $96$ & $0.039(2)$ & $0.060$\\
 & $216$ & $0.032(2)$ & $0.041$\\
 & $384$ & $0.009(1)$ & $0.029$\\
$0.50$ (C1a) & $96$ & $0.053(2)$ & $0.098$\\
 & $216$ & $0.028(1)$ & $0.046$\\
 & $384$ & $0.011(1)$ & $0.015$
\end{tabular}
\end{ruledtabular}
\end{table}

\textit{Jastrow factor of the ordered state.} A Jastrow factor $\exp(\sum_cv_c\sum_{\langle ij\rangle\in c}S'^z_iS'^z_j)$ in the local frame of the order, in which the quantization axis of site $i$ is $\bm{n}_i$, with one parameter on each of the bond orbits $J_1$, $J_{23}$, $J_\times$, and $J_+$, was optimized on top of the ordered determinant. It lowers the ordered energy by $(5,3,1)\times10^{-4}$ at $\lambda=0.35$, $(6,9,5)\times10^{-4}$ at $\lambda=0.4$, $(10,9,8)\times10^{-4}$ at $\lambda=0.45$, $(11,9,10)\times10^{-4}$ at C1a, and $(2,23,17)\times10^{-4}$ at C1b, on $96$, $216$, and $384$ sites, and $m^2$ by $0.001$ to $0.014$. The extrapolated difference between the best spin liquid and the ordered state at $\lambda=0.35$ remains compatible with zero, $+0.0005(3)$, so that their crossing is unchanged, and the extrapolated energies of the crystal and the ordered state at C1a become equal within the errors. An $S^z S^z$ factor breaks the spin-rotation symmetry of the singlet states and was not applied to them.

\textit{Nematic chiral state.} The time-reversal-breaking state N:Z00000-110 of the nematic class of cuboc1 was optimized on $216$ sites at C1a, C1c ($J_+=0.8$, $J_\times=0.3$), and C1d ($J_+=0.3$, $J_\times=0.8$) and measured on $384$ sites with one boundary condition: $E/N=-0.41871(11)$, $-0.42015(12)$, and $-0.41989(10)$, against $-0.44453(9)$, $-0.46593(11)$, and $-0.46678(12)$ for the lowest projected spin liquid at these points in the same protocol.

\begin{figure}[H]
\centering
\includegraphics[width=0.6\textwidth]{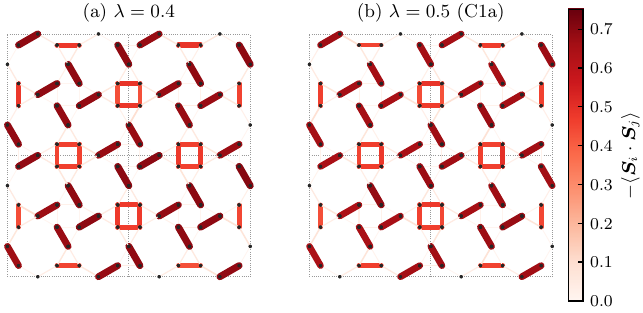}
\caption{Nearest-neighbor correlations $-\langle\bm{S}_i\cdot\bm{S}_j\rangle$ of the projected valence-bond crystal on $384$ sites, averaged over $16$ twists, at (a) $\lambda=0.4$ and (b) C1a ($\lambda=0.5$), $J_+=J_\times=\lambda$, $J_1=J_2=J_3=1$. Line width and color increase with $-\langle\bm{S}_i\cdot\bm{S}_j\rangle$. A window of $4\times4$ unit cells is shown, with the $2\times2$ magnetic cells outlined (dotted).}
\label{sm:fig:vbc}
\end{figure}

\textit{Valence-bond crystal.} Figure~\ref{sm:fig:vbc} shows the nearest-neighbor correlations of the crystal at $\lambda=0.4$ and at C1a. Strong square--bow-tie dimers, with $\langle\bm{S}_i\cdot\bm{S}_j\rangle\approx-0.68$ to $-0.70$, form pinwheels around half of the squares, and the other squares resonate. The pattern is the same at both couplings.

\section{Projective Symmetry Groups of the Nematic Class}
\label{sm:sec:nematicsm}

The nematic subgroup $p2mm$ of $p4mm$ is generated by $T_1$, $T_2$, and the axial mirrors $\sigma_x:(X,Y)\to(-X,Y)$ and $\sigma_y:(X,Y)\to(X,-Y)$ through the square center, $\sigma_x=\sigma C_4^{-1}$ and $\sigma_y=\sigma C_4$ in terms of the generators of Sec.~\ref{sec:lattice} of the main text. Their product is the twofold rotation $C_4^2$. The group $p2mm\times\Z_2^\T$ is defined by the thirteen relations
\begin{align}
&T_2T_1T_2^{-1}T_1^{-1}=\sigma_xT_1\sigma_x^{-1}T_1=\sigma_xT_2\sigma_x^{-1}T_2^{-1}=e,\nonumber\\
&\sigma_yT_1\sigma_y^{-1}T_1^{-1}=\sigma_yT_2\sigma_y^{-1}T_2=e,\nonumber\\
&\sigma_x^2=\sigma_y^2=(\sigma_x\sigma_y)^2=\T^2=e,\nonumber\\
&O\,\T\,O^{-1}\T^{-1}=e\quad(O=T_1,T_2,\sigma_x,\sigma_y).
\end{align}
Every generator occurs an even number of times in every relator, so that all $16$ sign assignments to $(T_1,T_2,\sigma_x,\sigma_y)$ are characters, and re-liftings never change a relation value. We label a $\Z_2$ PSG Z$n_{T_2}n_{xT_1}n_{xT_2}n_{yT_1}n_{yT_2}$-$n_xn_yn_{xy}$, where a digit $1$ marks the value $-1$ of, in this order, the relations $T_2T_1T_2^{-1}T_1^{-1}$, $\sigma_xT_1\sigma_x^{-1}T_1$, $\sigma_xT_2\sigma_x^{-1}T_2^{-1}$, $\sigma_yT_1\sigma_y^{-1}T_1^{-1}$, $\sigma_yT_2\sigma_y^{-1}T_2$, $\sigma_x^2$, $\sigma_y^2$, and $(\sigma_x\sigma_y)^2$. The $SU(2)$ PSGs carry the same digits after SU, and a $U(1)$ PSG is labeled U$w_{T_1}w_{T_2}w_{\sigma_x}w_{\sigma_y}$-$n$, where $w_O=1$ marks the operations that reverse the $U(1)$ rotation and $n$ is an index, as for $p4mm$. The character of cuboc1, read off from its spin configuration, is $\epsilon(T_1,T_2,\sigma_x,\sigma_y)=(-,+,+,+)$: the translation $T_1$ is combined with time reversal, and the mirror $x\to a-X$, $T_1\sigma_x$, also has $\epsilon=-1$.

Under $p2mm$ the four square sites form one orbit at the general position, and the bow-tie sites $5$ and $6$ form two separate orbits, at the positions $(a/2,0)$ and $(0,a/2)$, each with site symmetry $2mm$. The twelve bond orbits of $p4mm$ up to $J_+$ split into $21$: $J_d$, $J_{bb}$, and $J_\times$ do not split, and each of the other nine splits into two orbits related by $C_4$, with bonds along $x$ and along $y$. In particular the horizontal and vertical square edges are inequivalent, and the bow-tie edges split into the legs to site $5$ and those to site $6$, each mixing the orbits $J_2$ and $J_3$ of the model with inequivalent bow-tie edges.

The $\Z_2$ PSGs follow as in Sec.~\ref{sm:sec:counting}. Of the $256$ classes of $H^2(p2mm,\Z_2)$, $100$ are realizable on the three site orbits, each with exactly one PSG, since the characters act transitively on the four sign representations of each $2mm$ bow-tie orbit and the square orbit has trivial site symmetry. There are $4$ $SU(2)$ PSGs and $48$ discrete $U(1)$ PSGs, with one flux family in the sector in which both mirrors reverse the $U(1)$ rotation and the translations do not. As for $p4mm$, the algebraic PSGs are the same for all $16$ characters, which enter only through the realization. Table~\ref{sm:tab:nematic} gives the realized counts for the character of cuboc1. The half-square chirality of cuboc1 needs the $J_d$ bonds and is reproduced by $58$ $\Z_2$ and $7$ $U(1)$ PSGs. The full chirality pattern at graph distance two, over $11$ chiral loop orbits, is reproduced by $24$ $\Z_2$ and $3$ $U(1)$ PSGs, $17$ of which also reproduce its $6$ vanishing loop orbits. The pattern over all $19$ chiral loop orbits up to $J_+$ is reproduced by $24$ $\Z_2$ and $2$ $U(1)$ PSGs. The combination $T_1\T$ reverses the Hall response, so every state of this class has $C=0$ and $\kappa_{xy}=0$.

\begin{table}[H]
\caption{Realized PSGs of $p2mm$ with the character $(-,+,+,+)$ of cuboc1, singlet amplitudes and, in parentheses, singlet and triplet amplitudes. NN: square and bow-tie edges. NN$+J_d$: with the square diagonals. CO: the cross-octagon couplings $J_1$, $J_{23}$, $J_\times$, $J_+$ (Sec.~\ref{sec:classical} of the main text). GD2: all orbits at graph distance two. $J_+$: all $21$ orbits.}
\label{sm:tab:nematic}
\begin{ruledtabular}
\begin{tabular}{lccc}
range & $\Z_2$ & $U(1)$ & $SU(2)$\\
\hline
NN & 64 (100) & 6 (49) & 0 (4)\\
NN$+J_d$ & 92 (100) & 31 (49) & 2 (4)\\
CO & 92 (100) & 29 (49) & 2 (4)\\
GD2 & 100 (100) & 49 (49) & 4 (4)\\
$J_+$ & 100 (100) & 49 (49) & 4 (4)
\end{tabular}
\end{ruledtabular}
\end{table}

\end{document}